\documentclass[12pt,a4paper,JHEP3]{article}
\let\counterwithout\relax
\let\counterwithin\relax
\pdfoutput=1
\usepackage{hhline}
\usepackage[table,dvipsnames]{xcolor}
\usepackage{jheppub}
\usepackage{subcaption}
\usepackage[utf8]{inputenc}
\usepackage[T1]{fontenc}
\usepackage{amsmath}
\usepackage{nicefrac}  % for 1/2 
\usepackage{mathtools}
\usepackage{multirow}
\usepackage{booktabs} %nice tables 
\usepackage{chngcntr}
\usepackage{nicefrac}
\usepackage{cleveref}
\usepackage[normalem]{ulem}
\usepackage{setspace} %%% for padding around boxes
\usepackage{makecell} %%% for multiline cells without using the \tabular{p{width}} option
\usepackage{mathbbol} % lower-case mathbb letters
\usepackage[colorlinks=true]{hyperref}
\usepackage{orcidlink}
\usepackage{refcount} % used for footnote referencing
\usepackage{enumitem}

\newcommand{\iO}{\i 0^{+}}

\newcommand{\vct}[1]{{\boldsymbol{#1}}}
\newcommand{\lad}{\mathcal{L}}
\newcommand{\integral}{I}
\newcommand{\integralspatial}{\mathbf{I}}
\newcommand{\integrand}{\mathcal{I}}
\newcommand{\cint}{\mathcal{C}}
\newcommand{\mR}{\mathcal{R}}
\newcommand{\mP}{\mathcal{P}}
\newcommand{\mT}{\mathcal{T}}

\newcommand{\abn}[1]{{\alpha_{#1}\beta_{#1}}}
\newcommand{\an}[1]{{\alpha_{#1}}}
\newcommand{\bn}[1]{{\beta_{#1}}}
\newcommand{\mascheroni}{{\gamma_{\rm E}}}

\newcommand{\drawResponseBlob}[1]{
  \begin{pgfonlayer}{foreground}
    \draw [fill=gray, thick] #1 circle (.3);
  \end{pgfonlayer}
}
\newcommand{\drawResummedBlob}[1]{
  \begin{pgfonlayer}{foreground}
    \draw [fill=white, thick] #1 circle (.3);
  \end{pgfonlayer}
}
\newcommand{\drawPerturbativeVertex}[1]{
  \begin{pgfonlayer}{foreground}
    \draw [fill=black, thick] #1 circle (.055);
  \end{pgfonlayer}
}
\newcommand{\drawDottedBHLine}[2]{
    \draw [dotted, thick] #1 -- #2 ;
}
\newcommand{\drawShockwave}[2]{
    \draw [black, line width=.6pt,decoration={ticks, segment length=2.8pt, amplitude=1pt}, decorate] #1 -- #2 ;
}
\newcommand{\drawSolidBHLine}[2]{
    \draw [black,thick] #1 -- #2 ;
}
\newcommand{\drawCurvedSolidBHLine}[4]{
    \draw [black,thick] #1 to[in=#3,out=#4] #2 ;
}
\newcommand{\drawSolidBHLineDirected}[2]{
    \draw [zParticle] #1 -- #2 ;
}
\newcommand{\drawGravitonLine}[2]{
    \draw [photonTest] #1 -- #2 ;
}
\newcommand{\drawGravitonLineDirected}[2]{
  \draw [photonTest] #1 -- ($#2!4.5pt!#1$) ;
  \path [
    decoration={
      markings,
      mark=at position 1 with {
        \arrow{Latex[length=5.5pt,width=4pt]}}},
    decorate] #1 -- #2 ;
}
\newcommand{\drawShadowLine}[2]{
    \draw [line width=7pt,white] #1 -- #2 ;
}
\newcommand{\drawCurvedGravitonLine}[4]{
    \draw [photonTest] #1 to[in=#3,out=#4,looseness=1.5] #2 ;
}
\newcommand{\drawSecondaryVertex}[1]{
  \begin{pgfonlayer}{foreground}
    \draw [fill=white, thick] #1 circle (.15);
    \draw [thick] ($#1+(-0.11,-0.11)$) -- ($#1+(0.11,0.11)$);
    \draw [thick] ($#1+(-0.11,0.11)$) -- ($#1+(0.11,-0.11)$);
  \end{pgfonlayer}
}
\newcommand{\drawDottedSecondaryLine}[2]{
    \draw [dotted, thick] #1 -- #2 ;
}
\newcommand{\drawSolidSecondaryLine}[2]{
    \draw [fatzParticle] #1 -- #2 ;
}

\newcommand{\drawSolidSecondaryLineDirected}[2]{
    \draw [fatzParticleDirected] #1 -- #2 ;
}

\def\xcor{1}  % default unit in x direction. Can be overwritten locally within tikzpicture environment
\def\ycor{.7} % default unit in y direction. Can be overwritten locally within tikzpicture environment
\newcommand{\drawPropagationShort}{
  \begin{scope}[shift={(currentLocation)}]
    \coordinate (inA) at (currentLocation);
    \coordinate (outA) at ($(currentLocation)+(.6*\xcor,0)$); %(.6,0)
    \coordinate (inB) at ($(currentLocation)+(0*\xcor,-1*\ycor)$);
    \coordinate (outB) at ($(currentLocation)+(.6*\xcor,-1*\ycor)$);
    \drawShockwave{(inA)}{(outA)}
    \drawGravitonLine{(inB)}{(outB)}
  \end{scope}
  \coordinate (currentLocation) at (outA);
}
\newcommand{\drawPropagationShortDirected}{
  \begin{scope}[shift={(currentLocation)}]
    \coordinate (inA) at (currentLocation);
    \coordinate (outA) at ($(currentLocation)+(.6*\xcor,0)$);
    \coordinate (inB) at ($(currentLocation)+(0*\xcor,-1*\ycor)$);
    \coordinate (outB) at ($(currentLocation)+(.6*\xcor,-1*\ycor)$);
    \drawShockwave{(inA)}{(outA)}
    \drawGravitonLineDirected{(inB)}{(outB)}
  \end{scope}
  \coordinate (currentLocation) at (outA);
}
\newcommand{\drawPropagationMedium}{
  \begin{scope}[shift={(currentLocation)}]
    \coordinate (inA) at (currentLocation);
    \coordinate (outA) at ($(currentLocation)+(.8*\xcor,0)$);
    \coordinate (inB) at ($(currentLocation)+(0*\xcor,-1*\ycor)$);
    \coordinate (outB) at ($(currentLocation)+(.8*\xcor,-1*\ycor)$);
    \drawShockwave{(inA)}{(outA)}
    \drawGravitonLine{(inB)}{(outB)}
  \end{scope}
  \coordinate (currentLocation) at (outA);
}
\newcommand{\drawPropagationMediumDirected}{
  \begin{scope}[shift={(currentLocation)}]
    \coordinate (inA) at (currentLocation);
    \coordinate (outA) at ($(currentLocation)+(.8*\xcor,0)$);
    \coordinate (inB) at ($(currentLocation)+(0*\xcor,-1*\ycor)$);
    \coordinate (outB) at ($(currentLocation)+(.8*\xcor,-1*\ycor)$);
    \drawShockwave{(inA)}{(outA)}
    \drawGravitonLineDirected{(inB)}{(outB)}
  \end{scope}
  \coordinate (currentLocation) at (outA);
}
\newcommand{\drawPropagationLong}{
  \begin{scope}[shift={(currentLocation)}]
    \coordinate (inA) at (currentLocation);
    \coordinate (outA) at ($(currentLocation)+(1.2*\xcor,0)$);
    \coordinate (inB) at ($(currentLocation)+(0*\xcor,-1*\ycor)$);
    \coordinate (outB) at ($(currentLocation)+(1.2*\xcor,-1*\ycor)$);
    \drawShockwave{(inA)}{(outA)}
    \drawGravitonLine{(inB)}{(outB)}
  \end{scope}
  \coordinate (currentLocation) at (outA);
}
\newcommand{\drawAbsorb}{
  \begin{scope}[shift={(currentLocation)}]
    \coordinate (inA) at (currentLocation);
    \coordinate (outA) at ($(currentLocation)+(.8*\xcor,0)$);
    \coordinate (inB) at ($(currentLocation)+(0*\xcor,-1*\ycor)$);
    \drawShockwave{(inA)}{(outA)}
    \draw [photonTest] (outA) to[in=0,out=180] (inB);
  \end{scope}
  \coordinate (currentLocation) at (outA);
}

\newcommand{\drawEmit}{
  \begin{scope}[shift={(currentLocation)}]
    \coordinate (inA) at (currentLocation);
    \coordinate (outA) at ($(currentLocation)+(.8*\xcor,0)$);
    \coordinate (outB) at ($(currentLocation)+(.8*\xcor,-1*\ycor)$);
    \drawShockwave{(inA)}{(outA)}
    \draw [photonTest] (inA) to[in=180,out=0] (outB);
  \end{scope}
  \coordinate (currentLocation) at (outA);
}
\newcommand{\drawEmitDirected}{
  \begin{scope}[shift={(currentLocation)}]
    \coordinate (inA) at (currentLocation);
    \coordinate (outA) at ($(currentLocation)+(.8*\xcor,0)$);
    \coordinate (outB) at ($(currentLocation)+(.8*\xcor,-1*\ycor)$);

    \drawShockwave{(inA)}{(outA)}

    \draw [photonTest] (inA) to[in=180,out=0] (outB);

    \draw[-{Latex[length=5.5pt,width=4pt]}] (outB) -- ++(4pt,0);

  \end{scope}
  \coordinate (currentLocation) at (outA);
}

\newcommand{\drawWL}{
  \begin{scope}[shift={(currentLocation)}]
    \coordinate (inA) at (currentLocation);
    \coordinate (outA) at ($(currentLocation)+(.6*\xcor,0)$);
    \drawSolidBHLine{(inA)}{(outA)}
    \drawPerturbativeVertex{(inA)}
    \drawPerturbativeVertex{(outA)}
  \end{scope}
  \coordinate (currentLocation) at (outA);
}
\newcommand{\drawGirlande}{
  \begin{scope}[shift={(currentLocation)}]
    \coordinate (inA) at (currentLocation);
    \coordinate (outA) at ($(currentLocation)+(.8*\xcor,0)$);
    \drawShockwave{(inA)}{(outA)}
    \drawCurvedGravitonLine{(inA)}{(outA)}{-90}{-90}
  \end{scope}
  \coordinate (currentLocation) at (outA);
}
\newcommand{\drawCubGirlande}{ % girlande connecting to a cubix vertex in the middle
  \begin{scope}[shift={(currentLocation)}]
    \coordinate (inA) at (currentLocation);
    \coordinate (outA) at ($(currentLocation)+(1.2*\xcor,0)$);
    \coordinate (worldlineMidway) at ($(inA)!.5!(outA)$);
    \coordinate (girlandeBottom) at ($(worldlineMidway)-(0,.84*\ycor)$);
    \drawPerturbativeVertex{(girlandeBottom)};
    \drawPerturbativeVertex{(worldlineMidway)};
    \drawGravitonLine{(worldlineMidway)}{(girlandeBottom)}
    \drawShockwave{(inA)}{(outA)}
    \drawCurvedGravitonLine{(inA)}{(outA)}{-90}{-90}
  \end{scope}
  \coordinate (currentLocation) at (outA);
}
\newcommand{\drawQuartic}{
  \begin{scope}[shift={(currentLocation)}]
    \coordinate (a1) at ($(currentLocation)+(-.3*\xcor,0)$);
    \coordinate (a2) at ($(currentLocation)+(.3*\xcor,0)$);
    \coordinate (b) at ($(currentLocation)+(0*\xcor,-1*\ycor)$);
    \drawGravitonLine{(a1)}{(b)}
    \drawGravitonLine{(a2)}{(b)}
    \drawPerturbativeVertex{(a1)}
    \drawPerturbativeVertex{(a2)}
    \drawPerturbativeVertex{(b)}
  \end{scope}
}
\newcommand{\drawCubic}{
  \begin{scope}[shift={(currentLocation)}]
    \coordinate (a) at (currentLocation);
    \coordinate (b) at ($(currentLocation)+(0*\xcor,-1*\ycor)$);
    \drawGravitonLine{(a)}{(b)}
    \drawPerturbativeVertex{(a)}
    \drawPerturbativeVertex{(b)}
  \end{scope}
}

\NewDocumentCommand{\drawArrowPoint}{O{.5} O{1.2} m m}{ % opt arg 1: position, opt arg2: scale, arg 1 = point 1, arg2 = point 2
\draw [swallow tail={#1}{#2}] #3 -- #4
}

\usepackage[compat=1.1.0]{tikz-feynman}

\usepackage{tikz}
\usetikzlibrary{math}
\usepackage{adjustbox}
\usetikzlibrary{decorations.pathmorphing}
\usetikzlibrary{decorations.markings}
\usetikzlibrary{positioning, shapes, snakes, arrows}
\usetikzlibrary{patterns.meta}
\usetikzlibrary{patterns}
\usetikzlibrary{backgrounds}

\pgfdeclarelayer{foreground}
\pgfsetlayers{background,main,foreground}

\colorlet{midgray}{gray!50!darkgray}
\tikzset{
  arrowpoint/.style n args={2}{
    draw=none,
    postaction={decorate},
    decoration={
      markings,
      mark=at position #1 with {
        \arrow[
          black,
          scale=#2,
          xshift=3.5pt
          ]{Latex[length=5pt,width=4pt]}
      }
    }
  },
swallow tail/.style n args={2}{
  draw=none,
    postaction={
      decorate,
      decoration={
        markings,
        mark=at position #1 with {
          \fill (-4pt,-2pt) -- (0,0) -- (-4pt,2pt) -- cycle;
        }
      }
    }
  },
  photonTest/.style={
    line width =.8pt,
    decorate,
    decoration={
      snake,
      segment length=5pt,
      amplitude=1.4pt,
      pre length=.0cm,
      post length=.0cm}},
  photonTestDirected/.style={
    line width =.8pt,
    decorate,
    decoration={
      snake,
      segment length=5pt,
      amplitude=1.4pt,
      pre length=.0cm,
      post length=.1cm},
      postaction={decorate,decoration={markings,mark=at position .999 with {\arrow[xshift=4.5pt]{Latex[length=5.5pt,width=4pt]}}}}},
  wl/.style={line width=1pt},
  graviton/.style={line width=.8pt, -latex,decorate, decoration={snake, segment length=4pt,amplitude=1.8pt, pre length=.15cm, post length=.25cm}},
  doublegraviton/.style={
    decorate,
    decoration={
      snake,
      segment length=4pt,
      amplitude=1.8pt,
      pre length=.15cm,
      post length=.25cm
    },
    double distance=1.2pt,
    line width=.6pt,
    preaction={draw, white, line width=2.2pt},
    -latex
  },
  fatgraviton/.style={line width=2pt,midgray,-{Latex[length=6pt,width=6pt]},decorate,decoration={snake,segment length=5pt,amplitude=1.8pt,pre length=.1cm,post length=.2cm,}}, % a fat graviton?
    fatscalar/.style={
    line width=2pt,
    draw=midgray,
    dashed,
    dash pattern=on 3pt off 2pt
  }, % fat scalar
  fatgravitonUndirected/.style={line width=2pt,midgray,decorate,decoration={snake,segment length=5pt,amplitude=1.8pt,pre length=.1cm,post length=.1cm,}},
  worldlineStatic/.style={dotted, line width=1pt},
	worldline/.style={gray, line width=1pt},
	worldlineBold/.style={black, line width=.6pt},
	background/.style={black,dotted,line width=1pt},
	zUndirected/.style={line width=1pt},
	zParticle/.style={line width=1pt,postaction={decorate},decoration={markings,mark=at position .5 with {\arrow[xshift=3.5pt]{Latex[length=5pt,width=4pt]}}}},
	doublezParticle/.style={
    line width=.8pt,
    double distance=1.4pt,
    preaction={draw, white, line width=2.4pt},
    postaction={decorate},
    decoration={
    markings,
      mark=at position .6 with {\arrow[scale=0.6]{latex}}
    }
  },
  doublezParticle/.style={
    line width=.8pt,
    double distance=1.4pt,
    postaction={decorate},
    decoration={
      markings,
      mark=at position .6 with {\arrow[scale=0.6]{latex}}
    }
  },
	zParticleF/.style={line width=1pt,postaction={decorate}},
  doublezParticleF/.style={
    line width=.8pt,
    double distance=1.4pt,
    preaction={draw, white, line width=2.4pt},
    postaction={decorate}},
  fatzParticleDirected/.style={line width=2pt,midgray,postaction={decorate},decoration={markings,mark=at position .5 with {\arrow[xshift=7.5pt]{Latex[length=10pt,width=6pt]}}}},
  fatzParticle/.style={line width=2pt,midgray},
	cscalar/.style={line width=1pt,postaction={decorate},decoration={markings,mark=at position .6 with {\arrow[#1]{latex}}}},
	cscalar2/.style={line width=1pt,postaction={decorate},decoration={markings,mark=at position .8 with {\arrow[#1]{latex}}}},
	photon/.style={line width =.8pt, decorate, decoration={snake, segment length=4pt, amplitude=1.8pt,  pre length=.1cm, post length=.1cm}},
	doublephoton/.style={
    decorate,
    decoration={
      snake,
      segment length=4pt,
      amplitude=1.8pt,
      pre length=.1cm,
      post length=.1cm
    },
    double distance=1.2pt,
    line width=.6pt,
    preaction={draw, white, line width=2.2pt}
  },
	photonRed/.style={red, line width =.8pt, decorate, decoration={snake, segment length=4pt, amplitude=1.8pt,  pre length=.1cm, post length=.1cm}},
	cross/.style={cross out, line width =.8pt, draw=black, minimum size=2*(#1-\pgflinewidth), inner sep=0pt, outer sep=0pt},
cross/.default={4pt},
perp/.style={ % style for x_perp propagator
    postaction={
      decorate,
      decoration={
        markings,
        mark=at position 0.45 with {\draw (0,2pt)--(0,-2pt);},
        mark=at position 0.55 with {\draw (0,2pt)--(0,-2pt);},
      }
    }
  }
}
\newcommand{\drawdots}[1]{\foreach \p in #1{
  \drawPerturbativeVertex{(\p)}
}}

\newcommand{\I}{\mathrm{i}}
\newcommand{\euler}{\mathrm{e}}

\DeclareFontFamily{OT1}{pzc}{} 
\DeclareFontShape{OT1}{pzc}{m}{it}{<-> s * [1.350] pzcmi7t}{}
\DeclareMathAlphabet{\mathpzc}{OT1}{pzc}{m}{it}

\def\eps{\epsilon}

\def\k{\kappa}

\def\d{\mathrm{d}}

\def\mn{{\mu\nu}}
\def\ab{{\alpha\beta}}
\def\rs{{\rho\sigma}}

\def\i\math

\def\bH{\hat{b}}

\def\dd{\delta\!\!\!{}^-\!}

\def\d{\mathrm{d}}
\def\eps{\epsilon}

\def\abs#1{\vert #1 \vert}

\renewcommand{\i}{\ensuremath{\mathrm{i}}}
\renewcommand{\d}{\ensuremath{\mathrm{d}}}

\def\nn{\nonumber}

\def\eqn#1{eq.~\eqref{#1}}

\def\app#1{appendix~{\ref{#1}}}

\newcommand{\vev}[1]{\langle #1\rangle}

\newcommand{\Xd}{{\dot X}}

\newcommand{\be}{\begin{equation}}
\newcommand{\ee}{\end{equation}}
\newcommand{\ba}{\begin{align}}
\newcommand{\ea}{\end{align}}
\ifx\genfrac\sdflkaj\else\fi

\DeclareMathAlphabet{\mathsfit}{OT1}{cmss}{m}{sl}
\newcommand{\xd}{\dot x}

\newcommand{\fh}{h}
\newcommand{\ch}{\mathsfit{h}}

\newcommand{\cz}{\mathsfit{z}}

\def\centerarc[#1](#2)(#3:#4:#5){ \draw[#1] ($(#2)+({#5*cos(#3)},{#5*sin(#3)})$) arc (#3:#4:#5); }
\usepackage{chngcntr}
\counterwithin*{equation}{section}
\renewcommand{\theequation}{\arabic{section}.\arabic{equation}}

\newcommand{\bperp}[1]{\vct{#1}_\perp}
\newcommand{\bperpn}[2]{{\vct{#1}_{#2}}_{\perp}} %Perpendicular vector with index
\newcommand{\bell}{\vct\ell_\perp} % perpendicular loop momentum
\newcommand{\belln}[1]{\vct{\ell}_{#1\perp}} % perpendicular loop momentum with index
\newcommand{\bra}[1]{\langle #1 \rvert} % bra and ket
\newcommand{\ket}[1]{\lvert #1 \rangle}
\newcommand{\footnoteref}[1]{\hyperref[#1]{\footnotemark[\getrefnumber{#1}]}} % references provided label as footnotemark, including hyperref-link
\begin{document}

\begin{flushright}
\begingroup\footnotesize\ttfamily
	HU-EP-26/16-RTG
\endgroup
\end{flushright}

\vspace{15mm}

\begin{center}
{\LARGE\bfseries 
Black Hole Response Theory \\ and its Exact Shockwave Limit
\par}

\vspace{15mm}

\begingroup\scshape\large 
      Lara Bohnenblust${}^{1,2}$,
        Carl Jordan Eriksen${}^{1}$,
     Jitze Hoogeveen${}^{1}$,
    	Gustav Uhre Jakobsen${}^{1}$,
	and Jan~Plefka${}^{1}$
\endgroup
\vspace{5mm}
					
\textit{${}^{1}$Institut f\"ur Physik, Humboldt-Universit\"at zu Berlin, 10099 Berlin, Germany} \\[0.25cm]
\textit{${}^{2}$Max-Planck-Institut f\"ur Gravitationsphysik
(Albert-Einstein-Institut), 14476 Potsdam, Germany } \\[0.25cm]

\bigskip
  
\texttt{\small\{lara.bohnenblust, carl.jordan.eriksen, jitze.hoogeveen, gustav.uhre.jakobsen, 
jan.plefka\}@hu-berlin.de}

\vspace{5mm}

\textbf{Abstract}\vspace{5mm}\par

\begin{minipage}{14.7cm}
{
We present a black hole response formalism formulated within the worldline approach to the classical gravitational two-body problem. The central objects are response functions: a hierarchy of correlators that encode, successively, the black hole's own gravitational field, the scattering of a gravitational wave off of the black-hole including recoil, and the nonlinear response to multiple gravitational perturbations. These functions serve as the natural building blocks for a systematic diagrammatic expansion in the mass ratio of a binary, the gravitational self-force expansion (SF), employing the worldline quantum field theory (WQFT) formalism.
As a first application we treat an ultra-relativistic black hole, whose field is the Aichelburg–Sexl shockwave. We show that our framework reproduces the exact shockwave geometry and the trajectories of probes crossing it. Our main result is the scattering of a gravitational wave off the shockwave, computed exactly in Newton's constant by resumming the full post-Minkowskian (PM) perturbative series, which we compute for off-shell
gravitons enabling later use in the SF expansion.
The exact on-shell answer takes a strikingly compact form: the leading-order result is dressed by an overall phase that captures the expected infrared (Weinberg) behaviour together with a Coulomb-like scattering phase.}
Our results provide the basic WQFT ingredients
for future $1$SF computations of observables such as the impulse and waveform in
the ultra high-energy regime.
\end{minipage}\par

\end{center}
\setcounter{page}{0}
\thispagestyle{empty}
\newpage
 
\tableofcontents
\clearpage\noindent
\newpage

\section{Introduction}

The increasing precision of gravitational-wave observations \cite{LIGOScientific:2016aoc,LIGOScientific:2017vwq,KAGRA:2021vkt,LIGOScientific:2025slb},
in particular in view of the 
projected third generation detectors \cite{LISA:2017pwj, Punturo:2010zz, Ballmer:2022uxx,ET:2025xjr}, has turned the
relativistic two-body problem into a high-accuracy frontier problem  in theoretical
physics akin to the predictions from the Standard Model of elementary particle physics required for the LHC. On the analytical side, due to the complexity of Einstein’s equations one exploits a perturbative expansion 
in a small parameter around a regime in
which the dynamics is under control. In the post-Newtonian (PN) expansion~\cite{Blanchet:2013haa, Porto:2016pyg, Levi:2018nxp}, one
assumes weak fields and slow motion, while in the post-Minkowskian (PM)
expansion~\cite{Kosower:2022yvp, Bjerrum-Bohr:2022blt, Buonanno:2022pgc, DiVecchia:2023frv, Jakobsen:2023oow}, one instead expands in Newton's constant $G$ at arbitrary velocity. The PM
regime is particularly natural for scattering processes and highly eccentric motion
\cite{Kovacs:1978eu, Westpfahl:1979gu, Bel:1981be, Damour:2017zjx, Hopper:2022rwo},
and over the last few years, quantum-field-theory based methods {--- applied to the
classical realm ---} have become the central 
tool for pushing this expansion to high orders -- both in their worldline~\cite{Kalin:2020mvi, Kalin:2020fhe, Kalin:2020lmz, Mogull:2020sak, Jakobsen:2021smu, Dlapa:2021npj, Dlapa:2021vgp, Mougiakakos:2021ckm, Riva:2021vnj, Dlapa:2022lmu, Dlapa:2023hsl, Liu:2021zxr, Mougiakakos:2022sic, Riva:2022fru, Jakobsen:2021lvp, Jakobsen:2021zvh, Jakobsen:2022fcj, Jakobsen:2022zsx, Jakobsen:2022psy, Shi:2021qsb,Bastianelli:2021nbs, Comberiati:2022cpm,Wang:2022ntx,Ben-Shahar:2023djm, Bhattacharyya:2024aeq, Jakobsen:2023ndj, Jakobsen:2023hig, Jakobsen:2023pvx} and  scattering amplitude \cite{Neill:2013wsa, Luna:2017dtq, Kosower:2018adc, Cristofoli:2021vyo, Bjerrum-Bohr:2013bxa, Bjerrum-Bohr:2018xdl, Bern:2019nnu, Bern:2019crd, Bjerrum-Bohr:2021wwt, Cheung:2020gyp, Bjerrum-Bohr:2021din, DiVecchia:2020ymx, DiVecchia:2021bdo, DiVecchia:2021ndb, DiVecchia:2022piu, Heissenberg:2022tsn, Damour:2020tta, Herrmann:2021tct, Damgaard:2019lfh, Damgaard:2021ipf, Damgaard:2023vnx, Aoude:2020onz, AccettulliHuber:2020dal, Brandhuber:2021eyq, Bern:2021dqo, Bern:2021yeh, Bern:2022kto, Bern:2023ity, Damgaard:2023ttc, Brandhuber:2023hhy, Brandhuber:2023hhl, DeAngelis:2023lvf, Herderschee:2023fxh, Caron-Huot:2023vxl, FebresCordero:2022jts, Bohnenblust:2023qmy,Bern:2025zno,Bern:2025wyd}
formulations. 

The quantum field theory (QFT) based worldline approach enjoys a particularly direct
classical limit: compact objects are modelled as point particles coupled to the
gravitational field, which is integrated out perturbatively in a $G$ expansion \cite{Goldberger:2004jt,Porto:2016pyg,Kalin:2020mvi}. 
In its worldline quantum field theory (WQFT) formulation~\cite{Mogull:2020sak, Jakobsen:2022psy, Jakobsen:2023oow, Haddad:2024ebn}, one quantises not only the
bulk graviton but also the fluctuations of the worldlines themselves, thereby
obtaining a compact diagrammatic framework for perturbatively solving the classical equations of
motion and computing observables such as the impulse \cite{Dlapa:2022lmu, Dlapa:2023hsl, Jakobsen:2023ndj, Jakobsen:2023hig, Jakobsen:2023pvx,Driesse:2024xad, Driesse:2024feo,Driesse:2026qiz}, spin kick \cite{Jakobsen:2023ndj, Jakobsen:2023hig, Jakobsen:2023pvx}, waveform \cite{Jakobsen:2021smu,Jakobsen:2021lvp,Bohnenblust:2025gir} or 
the Magnus operator \cite{Kim:2024svw,Kim:2025gis,Gonzo:2026yha}. 
{State-of-the-art results in the PN expansion have also made heavy use of QFT techniques
to capture classical physics
\cite{Foffa:2016rgu, Foffa:2019rdf, Foffa:2019yfl,Foffa:2019hrb,Blumlein:2019zku,Foffa:2020nqe,Blumlein:2020pyo,Blumlein:2021txe,Porto:2024cwd,Porto:2026fsd,Blumlein:2021txj,Almeida:2026clf,Brunello:2025gpf,Brunello:2026anu}. }

While quantum field theory approaches are structurally rooted in a perturbative $G$ expansion,
general relativity also provides exact-in-$G$ information for
isolated compact objects through the metric: The
Schwarzschild geometry \cite{Schwarzschild:1916uq} for a point-like source, the Kerr solution \cite{Kerr:1963ud} for a rotating source, and the Aichelburg--Sexl shockwave \cite{Aichelburg:1970dh} for an
ultra-boosted source. Treating a second body as a
probe in one of these backgrounds yields the geodesics,  the leading term in the gravitational
self-force (SF) expansion~\cite{Mino:1996nk,Poisson:2011nh,Barack:2018yvs,Gralla:2021qaf,Pound:2019lzj}, i.e.~an expansion in the mass ratio of the two objects. 
The semi-analytic SF expansion is a highly developed subject, {see e.g.~\cite{Kavanagh:2015lva,vandeMeent:2017bcc,Pound:2021qin,Wardell:2021fyy,Lynch:2023gpu,Kuchler:2024esj,Honet:2025lmk,Kuchler:2025hwx}} and allows to access the strong-gravity regime, where the PM/PN expansions break down.
Closely related to the SF expansion is black hole perturbation theory (BHPT)~\cite{Pani:2013pma,Bini:2018qvd,Sasaki:2003xr,Berti:2025hly}, where one studies linear and non-linear
perturbations of an exact black hole background in order to extract its response to
external disturbances.

A natural question is therefore whether the exact-background logic of SF/BHPT can
be merged with the diagrammatic efficiency of perturbative quantum field theory methods, in particular in the
eminent worldline framework. A first step in this direction
is to ask whether the exact metrics themselves can be recovered by resumming the
PM-expanded worldline theory~\cite{Duff:1973zz}. For massive sources this question has recently been
answered in the affirmative: the one-point function sourced by a single worldline can
be resummed to the full Schwarzschild geometry \cite{Damgaard:2024fqj,Mougiakakos:2024nku}, as can the geodesic
be recovered \cite{Mougiakakos:2024lif,Damgaard:2026kqg}. In the present work we show that
the same logic extends to a massless source, where the one-point function resums to
the Aichelburg--Sexl (AS) shockwave and we also recover the exact geodesics \cite{Dray:1984ha,Steinbauer:1997dw}.

The next step is to move beyond the one-point function. Connected graviton
correlators in the presence of a heavy worldline may be viewed as \emph{black hole
response functions}: while the one-point response encodes the classical background, the
two-point response describes graviton propagation on that background including
the recoil of the black hole (BH), and higher-point responses capture non-linear response data. From the
effective field theory  perspective, these objects are the natural building blocks for a
diagrammatic gravitational SF expansion. Indeed, once the heavy worldline fluctuations are
integrated out, the two-point response acts as an exact background propagator while
the higher-point responses act as effective interaction vertices. Coupling this
response theory to a lighter secondary worldline then yields a systematic expansion
in the small mass ratio, in which an $n$SF computation requires response functions
only up to $(n+1)$-point order. Similar steps of establishing a QFT framed
SF expansion were
undertaken in \cite{Cheung:2023lnj,Cheung:2024byb,Bjerrum-Bohr:2025bqg} for the worldline approach and in \cite{Kosmopoulos:2023bwc} for the scattering amplitude
framework. Yet, concrete results could only be achieved upon expanding in $G$, thereby
effectively reproducing the PM expansion simply in a different setup.

In this paper we develop such a black hole-response formalism within WQFT and use
it to define a response-based effective theory adapted to the SF expansion. As a
first concrete application -- providing exact in $G$ results --
we specialise to the shockwave limit, i.e.~to a massless
primary source. Although the
usual mass-ratio parameter is then ill-defined, the expansion reorganises naturally in
powers of $m/E$, where $m$ is the mass of the secondary object and $E$ is the
shockwave energy. This setting is not only technically simpler than the generic
massive case, but is also physically interesting because it is tied to the ultra high-energy
limit of black hole scattering. 

The shockwave metric  results from an infinite boost of the Schwarzschild 
geometry, holding the energy fixed. It is flat everywhere except for curvature localised  on a null hypersurface
\cite{Aichelburg:1970dh} of distributional nature. Geodesics crossing the shockwave
experience a discontinuous shift in their straight line trajectories \cite{Dray:1984ha,Steinbauer:1997dw}. 
The shockwave  background has served as an arena for studying
ultra high-energy scattering in perturbative quantum gravity  \cite{tHooft:1987vrq,Muzinich:1987in,Verlinde:1991iu,Kabat:1992tb,Amati:1992zb,Camanho:2014apa} 
and string theory \cite{Amati:1987wq,Amati:1987uf}. 
The surprising feature here is that at energies above the Planck scale the
problem classicalises. Assuming eikonal kinematics and a dominance of ladder graphs
a resummation may be performed \cite{Muzinich:1987in,Kabat:1992tb,Amati:1992zb} and it is argued that for small enough impact parameter black hole formation should set in \cite{Eardley:2002re}. 
 Recent work in the amplitudes program
\cite{Cristofoli:2020hnk,Adamo:2022qci,Aoki:2026eos} have recovered the AS metric, as well as extensions to the spinning (Kerr) case \cite{Adamo:2022rob} using the KMOC formalism
\cite{Kosower:2018adc}, although one does not expect that spinning shockwaves are physical, as they exceed the extremality bound $a\leq GM$.
Very recently the propagator in the AS shockwave background was derived
for eikonal kinematics in \cite{Raj:2023iqn,Raj:2024xsi} by solving its defining Green’s function equation, observing a double copy structure.

Here, we show that the shockwave one-point response is exact already at leading PM
order and reproduces the full Aichelburg--Sexl (AS) metric. Upon coupling to a probe
worldline and resumming the resulting 0SF diagrams, we recover
the exact-in-$G$ geodesics crossing the shockwave first obtained by Dray and ’t Hooft \cite{Dray:1984ha,Steinbauer:1997dw}. These results provide a
useful check of the formalism and show that the response-theory point of view
correctly captures the exact background and its probe dynamics.

Our central new result is the exact-in-$G$ computation of the graviton two-point
response of the shockwave. Diagrammatically, this amounts to resumming the full
PM series for graviton propagation in the non-trivial AS background generated by the
massless source. Owing to the null nature of the shockwave momentum, large classes
of diagrams vanish and only a restricted set of contributions survives at high PM
orders, allowing the series to be resummed explicitly. 
The resulting off-shell
two-point response is the basic building block for future $1$SF observables. For
on-shell external gravitons it yields the exact transfer matrix $T$ -- equivalently, the
exact gravitational-wave scattering amplitude off the shockwave, including recoil.
The final exact answer has a remarkably compact structure, and in four dimensions its
infrared behaviour is controlled by the expected Weinberg phase $W/\epsilon$ \cite{Weinberg:1965nx}
along with a Coulomb  or ’t Hooft  scattering phase \cite{tHooft:1987vrq} of the form
\be
T(P,k,q) =  \frac{\Gamma(1-\I W)}{\Gamma(1+\I W)}\frac{\euler^{-\I W/\eps}}{(\bperp{q}^2L^2)^{-\I W}}\, T_{\text{Born}}(P,k,q)\, , \qquad W = 2G(P\cdot k)\,
\ee
with $P^\mu$ the null-momentum of the shockwave, $k^\mu$ the incoming graviton momentum and
$q^\mu$ the momentum transfer, $\epsilon=2-D/2$ is the dimensional regulator and $L$ the associated 
length-scale. The finite part matches the historic ultra high-energy
quantum scattering results in the eikonal regime \cite{tHooft:1987vrq,Muzinich:1987in,Verlinde:1991iu,Kabat:1992tb,Amati:1992zb}.
For the Schwarzschild background the analogue two-point response has been computed
in up to 3PM orders in \cite{Bjerrum-Bohr:2025bqg,Bjerrum-Bohr:2026fhs,Bautista:2026qse,Ivanov:2026icp}, see also \cite{Ivanov:2024sds,Correia:2024jgr,Caron-Huot:2025tlq,Correia:2025enx} for recent
amplitude connections to BHPT.

More broadly, the present work should be viewed as a first step toward a genuinely
response-based formulation of the gravitational self-force expansion within WQFT.
The shockwave problem provides a setting in which exact resummations are possible
and the hierarchy of response functions can be exhibited explicitly. The resulting
formalism supplies the ingredients needed for future computations of $1$SF
observables in the high-energy regime, such as the impulse and waveform, and at the
same time clarifies how exact-background information can be encoded in a
diagrammatic PM language.

Our results also revive the question of the convergence of the PM expansion \cite{Christodoulou:1979eu,Damour:1990rm} which, given our results for the response functions here, receives support in the affirmative.

Our paper is organised as follows. In section~\ref{Section2} we formulate black hole response
theory in WQFT and show how it generates a systematic SF expansion. In
section~\ref{Section3} we specialise to massless WQFT and recover the Aichelburg--Sexl metric,
together with the corresponding probe geodesics and resummed probe vertices. In
section~\ref{sec:2pt_response} we compute the exact two-point shockwave response function by resumming
the PM series. In section~\ref{sec:response_discussion} we discuss its main properties, including its position-space
interpretation, four-dimensional infrared structure and relation to the Magnus
operator. We conclude in section~\ref{sec:conclusions} with an outlook on future applications.

\section{Black Hole Response Theory}\label{Section2}

We begin our quest to establish a systematic gravitational self-force (SF) expansion of binary dynamics in the WQFT formalism by first establishing the framework of black hole
response theory, leveraging elementary objects in quantum field theory. With it we 
provide a diagrammatic framework to directly perform the SF expansion by employing  
the black hole response functions as effective vertices that directly yield 
the observables as is familiar from WQFT in the PM expansion.

\subsection{Black hole response expansion of WQFT}\label{sec:BH_response_expansion}
Our analysis will initially concern a single (primary) compact object such as a black hole (BH), neutron star (NS), or star of mass $M$ propagating through spacetime. Following the principles of effective field theory (EFT), we stay agnostic about the object's internal degrees of freedom by describing it as a massive point particle, with potential finite-size effects being incorporated through a tower of higher-derivative operators. In principle, this restricts the validity of the theory to distance scales much larger than the body's intrinsic size of order $2GM$, where $G$ is the four-dimensional Newton's constant. We work with $c = 1$, equating time and length, but refrain from setting $\hbar = 1$ as we are doing classical physics, where one most naturally keeps the dimensions of length $\ell$ and mass $m$ distinct. This choice implies that $[G] = \ell/m$, such that $2GM$ has units of length.

The worldline action of a single massive particle moving in curved spacetime 
is proportional to the invariant length of its worldline,
\be\label{mds}
  S[g,X]
  =
  -M \int \d s
  =
  -M\int\d\tau\,\sqrt{g_\mn\Xd^\mu\Xd^\nu}
  \,,
\ee
with a given parametrisation $X^{\mu}(\tau)$ of the classical trajectory. Importantly, the action is independent of the chosen
parametrisation, i.e. it is invariant under $\tau\to\tilde\tau(\tau)$. To circumvent the difficulties associated with the square root form of eq.~\eqref{mds}, we introduce an einbein $e(\tau)$ and reexpress the dynamics in the linearised Brink--Di Vecchia--Howe (BdVH) form \cite{Brink:1976uf}
\be\label{S1}
  S[g,X]
  =
  -\frac M2\int\d\tau\,\big(e^{-1}\,g_\mn\Xd^\mu\Xd^\nu + e\big)
  \,.
\ee
The solution to the algebraic equation of motion for $e$,
\be\label{EOMe}
\Xd^{2} - e^{2}=0\,,
\ee
yields \eqn{mds} upon being plugged into \eqn{S1}. In the BdVH form, the reparametrization invariance may be formulated infinitesimally as the invariance of the action under
\be
\delta X^{\mu} = \xi \Xd^{\mu}\,, \qquad \delta e = \frac{\d}{\d\tau} (\xi e)\,,
\ee
with $\xi(\tau)$ an infinitesimal gauge parameter. As the action must have units of $\ell m$, the einbein must have $[e] = 1$ if we also choose $[\tau]=\ell$. This and the reparametrization invariance allow us to gauge-fix $e=1$, leaving us with the on-shell constraint
$\Xd^{2} = 1$ from \eqn{EOMe}, which is preserved under time evolution.
The equation of motion for $X^{\mu}$ immediately follows:
\be\label{eq:eom-for-X}
\ddot X^{\rho} + \Gamma^{\rho}{}_{\mu\nu} \Xd^{\mu}\Xd^{\nu}=0\,,
\ee
and its flat-space solution reads
\begin{equation}
  X^{\mu}(\tau) = B^{\mu}+ V^{\mu}\tau\,, \qquad V^{2}=1\,.
\end{equation}

With these preliminaries out of the way, we now include the dynamics of spacetime itself and thereby arrive at the action for our primary black hole:
\begin{align} \label{monoBHS}
  S_{\rm BH}[X,g]
  =
  &-
  \frac{M}{2}\int\d\tau\,g_\mn(X) \dot X^\mu \dot X^\nu 
  -
  \int \d^D x\,
  \bigg[
    \frac{2}{\kappa_{D}^{2}}\sqrt{|g|} R
    -\eta^{\mu\nu}G_\mu G_\nu
  \bigg]
  \,,
\end{align}
where we dropped the non-dynamical constant term from eq.~\eqref{S1}. The second term is the Einstein--Hilbert action, and the third, which uses
\begin{equation}
  G_\mu
  =
  \partial_\nu h_\mu{}^\nu
  -
  \frac12\partial_\mu h^\nu{}_\nu\,,
\end{equation}
serves to put the graviton field, defined by
\begin{align}\label{eq:graviton}
  g_{\mu\nu}=\eta_{\mu\nu}+\kappa_{D} h_{\mu\nu}\,,
\end{align}
in de~Donder gauge. To regulate our computations, we generalised to $D=4-2\epsilon$ spacetime dimensions in \eqn{monoBHS} and introduced the $D$-dimensional gravitational coupling
\begin{equation}\label{eq:kD_def}
  \kappa_{D}^{2}
  =
  32\pi G \tilde L^{-2\epsilon}
  \,,
  \qquad
  \tilde L^2
  =
  4\pi\euler^{\gamma_\mathrm{E}}L^2
  \,,
\end{equation}
where $L$ is a fiducial length scale associated with the dimensional regularisation. A consequence of this regularisation which merits immediate discussion is that the metric becomes flat when evaluated at the position of the BH \cite{Cheung:2024byb},\footnote{This identity is equivalent to the scalelessness of the integrals of eq.~\eqref{eq:generic-bubble} to be discussed in section \ref{sec:diag_simplicity_shockwave}.}
\begin{equation}\label{eq:crucial-dim-reg-id}
  g_{\mu\nu}(X) = \eta_{\mu\nu}\,.
\end{equation}
This central identity is crucial in the context of the SF expansion, as it ensures that the equation of motion for the BH trajectory $X^\mu$ trivialises to $\ddot X^\mu = 0$, so that we may employ the background field expansion
\begin{equation}\label{eq:deflection}
  X^\mu(\tau) = B^\mu + V^\mu\tau + Z^\mu(\tau).
\end{equation}

{In the Worldline Quantum Field Theory (WQFT) approach to the gravitational two-body problem~\cite{Mogull:2020sak, Jakobsen:2022psy, Jakobsen:2023oow, Haddad:2024ebn} one couples a secondary black-hole to the primary BH and
bulk gravity action of \eqn{monoBHS}
\begin{align}
  &S_{\rm sec}[x]
  =
  -
  \frac m2
  \int \d\tau\,
  \dot {x}^\alpha(\tau)
  \dot {x}^\beta(\tau)
  g_\ab(x(\tau))\, ,
\end{align}
and performs a parallel background field expansion about straight line trajectories for
the secondary BH: $x^\mu(\tau) = b^\mu + v^\mu\tau + z^\mu(\tau)$. Then the graviton field $h_{\mn}$ and the deflections $z^{\alpha}$ and $Z^{\mu}$ are perturbatively 
integrated out in a PM-expansion using a
path integral formulation for the tree-level one-point functions $\vev{h_{\mn}}$, $\vev{X^{\mu}}$ and $\vev{x^{\alpha}}$, in turn giving rise to the two-body scattering observables of the wave-form and impulse. 
THis WQFT paradigm makes the efficient perturbative methods of QFT based on a Feynman diagrammatic expansion immediately available for \emph{classical} gravitational physics
and has proven to be very efficient \cite{Jakobsen:2023ndj,Jakobsen:2023hig,Driesse:2024xad,Driesse:2024feo,Driesse:2026qiz}. 
}
{
Here, we extend the WQFT paradigm to the Gravitational Self Force (SF) expansion
by first considering only our primary BH and bulk gravity theory  \eqn{monoBHS} in the path-integral coupling it to an \emph{external} stress-energy tensor $\mathcal{T}^{\mu\nu}$ via the graviton field.}
Keeping $\mT^\mn(x)$ generic, the tree-level partition function for the primary BH in the presence of the source is 
\begin{align}\label{Znaïve}
  Z[\mT]
  =
  \int \mathcal{D}[h,Z]\,\euler^{\frac\I\hbar(S_{\rm BH}[X,g]-\kappa_D\int\d^Dx\,\mT^{\mu\nu}(x)h_{\mu\nu}(x))}
  \Big|_\text{tree}
  =
  \euler^{\frac\I\hbar W[\mT]}
  \,,
\end{align}
where we introduced the Wilsonian effective action $W[\mathcal T]$ for the primary black hole in the second equality. We call this object the {\it black hole response effective action}, as it depends only on the source $\mathcal T^{\mu\nu}$ and thus characterises the response of the primary BH to external perturbations. It is also the generating functional of connected $n$-point functions in the primary BH theory.

In fact, the path integral in \eqn{Znaïve} is only half of the story,  as it pertains to an in-out formulation where both the initial and final configurations of our fluctuating fields $Z^\mu$ and $h_{\mu\nu}$ are specified as boundary conditions.  As we only specify initial conditions for our fields, one must instead apply the in-in or Schwinger--Keldysh formalism which was worked out for worldline effective (quantum) field theory in \cite{Kalin:2022hph,Jakobsen:2022psy,Galley:2009px}.
Here, one doubles the fluctuating fields and their sources as $Z\to Z^{(1,2)}$, $h\to h^{(1,2)}$, and $\mT\to\mT^{(1,2)}$, where the first (second) copy propagates forwards (backwards) in time, thereby specifying the initial boundary conditions at past infinity
and matching them at future infinity \cite{Galley:2009px}. One then has
\be\label{ininPI}
 \euler^{\frac{\I}{\hbar} W[\mT^{(1,2)}]}
 =
 \int \mathcal{D}[h^{(1,2)},Z^{(1,2)}]
  \,\euler^{\frac{\I}{\hbar} S[Z^{(1)},h^{(1)},\mT^{(1)}]
  -\frac{\I}{\hbar} S[Z^{(2)},h^{(2)},\mT^{(2)}]}\Bigr |_{\text{tree}}.
\ee
However, as was shown in \cite{Kalin:2022hph,Jakobsen:2022psy}, computing {classical}
correlation functions in the \emph{tree-level} theory simply reduces to the standard
in-out diagrammatic expansion, with the important difference of using \emph{retarded}
propagators throughout, respecting the causality flow towards the outgoing state.
This is manifested in the Keldysh basis $\mathcal{F}^{(\pm)}\sim \mathcal{F}^{(1)}\pm \mathcal{F}^{(2)}$ 
with $\mathcal{F}=\{h,Z,\mT\}$, where causality in one-point functions flows through retarded propagators into $m$-point vertices having $m-1$ incoming $(+)$ legs and a single outgoing $(-)$ leg \cite{Jakobsen:2022psy}\footnote{{This single-outgoing-many-incoming structure is
specific for \emph{classical} observables. More outgoing lines inevitably lead to
quantum contributions due to closed loops. }}. Importantly, the in-in
vertex rules are identical to the in-out vertices.

In conclusion, the black hole response (BHR) effective action admits the expansion
\begin{align}\label{BHresponsethy}
  \i W[\mT^{(\pm)}]
  &=
  -\i
  \kappa_D
  \int_{k_1}
  \mT^{(-)\ab}(k_1)
  \mR_\ab(k_1)
  -
  \kappa_D^2
  \int_{k_1,k_2}
  \mT^{(+)\ab}(k_1)
  \mT^{(-)\mn}(k_2)
  \mR_{\ab\mn}(k_1,k_2)
  \nn
  \\
  &\hspace{-.6cm}
  +
  \frac{\i}2
  \kappa_D^3
  \int_{k_{1},k_2,k_3}
  \mT^{(+)\ab}(k_1)
  \mT^{(+)\mn}(k_2)
  \mT^{(-)\rs}(k_3)
  \mR_{\ab\mn\rs}(k_1,k_2,k_3)
  +
  \dots,
\end{align}
where $\int_k = \int \frac{\d^D k}{(2\pi)^D}$. The expansion coefficients of the BHR effective action are the \textit{response functions}:
\begin{align}\label{eq:BHR_def_diagram}
  \mR_{\abn{1}\abn{2}\cdots\abn{n}}(k_1,k_2,\dots,k_n)
  &=
  \frac{\I^n}{\kappa_D^n}\frac{\delta^{n}\I W[\mT]}{\delta\mT^{(+)\abn{1}}(k_1)\delta\mT^{(+)\abn{2}}(k_2)\cdots\delta\mT^{(-)\abn{n}}(k_n)}
  \bigg|_{\mT=0} \notag\\[.3cm]
  &=\;
  \begin{tikzpicture}[baseline=(cAnchor)]
    \coordinate (cAnchor) at (0,-1.6) ;
    \coordinate (ca) at (-1,0) ;
    \coordinate (cb) at (-.2,0) ;
    \coordinate (cc) at (+1,0) ;
    \coordinate (cin)  at (-1.8,0) ;
    \coordinate (cout) at (+1.8,0) ;
    \coordinate (cmiddlein)  at (-.8,-1.) ;
    \coordinate (cmiddleout) at (+.8,-1.) ;
    \coordinate (cmiddle) at (0,-1.) ;
    \coordinate (caBelow) at (-1.4,-2.) ;
    \coordinate (cbBelow) at (-.2,-2.) ;
    \coordinate (ccBelow) at (+1.4,-2.) ;
    \drawDottedBHLine{(cmiddlein)}{(cmiddleout)}
    \drawResponseBlob{(cmiddle)}
    \drawGravitonLine{(cmiddle)}{(caBelow)}
    \drawGravitonLine{(cmiddle)}{(cbBelow)}
    \drawGravitonLineDirected{(cmiddle)}{(ccBelow)}
    \drawPerturbativeVertex{(caBelow)}
    \drawPerturbativeVertex{(cbBelow)}
    \drawPerturbativeVertex{(ccBelow)}
    \node [below left=0cm and -.3cm of caBelow] {$h^{(-)}_\abn{1}(k_1)$};
    \node at (cbBelow) [below] {$h^{(-)}_\abn{2}(k_2)$};
    \node [below right=0cm and -.2cm of ccBelow] {$h^{(+)}_\abn{n}(k_n)$};
    \node at ([shift={(22pt,0pt)}]cbBelow) {\ldots};
  \end{tikzpicture}.
\end{align}
{The response functions are nothing but the connected $n$-graviton correlation functions
  $\langle h_\abn{1}(k_1)\cdots h_\abn{n}(k_n)\rangle_{\rm con}$
generated by the effective action $W[\mathcal T]$. 
They are given by the sum of \emph{all} tree-level diagrams with any number of bulk-graviton and
heavy BH worldline vertices.}
They encode the response of a black hole to any number of gravitational perturbations, and are the central objects of interest in this work.

As they are not amputated in \eqn{eq:BHR_def_diagram}, we include a dot on their external legs.
In each term of eq.~\eqref{BHresponsethy}, only a single outgoing source $\mathcal T^{(-)\mu\nu}$ appears, whereas all remaining sources have $(+)$ labels. Consequently, an $n$-point response function
connects exactly one field $h^{(+)}_{\mu\nu}$ and $n-1$ fields $h^{(-)}_{\mu\nu}$, where we identify $h^{(+)}_{\mu\nu}$ with an outgoing field and $h^{(-)}_{\mu\nu}$ with an incoming one. We indicate the unique causality flow from the $n-1$ incoming fields to the $n$'th outgoing field with an arrow in \eqn{eq:BHR_def_diagram}.
As the diagrammatics, apart from this causality flow, is equivalent to the in-out theory, we may mostly ignore the in-in formalism in the following and simply use standard in-out Feynman rules, keeping in mind that all propagators are retarded and that there is a unique causality flow towards the outgoing leg.

\subsection{Perturbative structure of response functions}\label{sec:response_diagrammatics}

We now take up the diagrammatic representation of the response functions introduced above. In this context, it is natural to exploit the relationship that inherently exists between the response functions and the one-particle irreducible (1PI) effective action $\Gamma[h,Z]$, defined through the Legendre transform of the BHR effective action. 
{The effective action will serve two purposes, both rooted in its ``1PI Feynman rules'' to be introduced below.
First, as we will see, the response functions are given by a finite number of diagrams in terms the 1PI Feynman rules.
Second, the 1PI Feynman rules are quite easily related to (infinite sums of) the perturbative post-Minkowskian vertex rules.
In this way, the 1PI Feynman rules represent a useful middle point between the response functions and perturbative PM vertices.}
%Contrary to the usual PM expansion, this provides a concise pictorial representation of the PM-resummed responses.

Below, we first briefly review the usual PM Feynman rules, after which we in some detail describe how the 1PI effective action may be leveraged to derive PM resummed Feynman rules from \eqn{Znaïve}. Here we uncover that the two-point response acts as the propagator in the PM-resummed theory, providing a building block for higher-point response functions.

\subsubsection{Post-Minkowskian Feynman rules}\label{sec:feynman_rules_flat}
The PM  Feynman rules are well known.
One expands the action of \eqn{monoBHS} in the perturbative fields $h_\mn(x)$ and $Z^\mu(\tau)$ (eqs.~\eqref{eq:graviton} and~\eqref{eq:deflection}). This results in the propagators \cite{Mogull:2020sak},\footnote{We ignore the dependence of the Feynman rules on Planck's constant, as this dependence is trivial and invariably drops out in tree-level connected correlation functions.}
\begin{subequations}
\begin{align}
  \label{eq:deflection-propagator}
  \begin{tikzpicture}[baseline=(anchor)]
    \coordinate (anchor) at (0,-.1);
    \coordinate (in) at (-0.6,0);
    \coordinate (out) at (1.4,0);
    \coordinate (x) at (-.2,0);
    \coordinate (y) at (1.0,0);
    \drawPerturbativeVertex{(x)}
    \drawPerturbativeVertex{(y)}
    \drawSolidBHLineDirected{(x)}{(y)}
    \drawDottedBHLine{(in)}{(out)}
    \node at (x) [above left=0cm and -.7cm] {$Z^\mu(-\omega)$};
    \node at (y) [above right=0cm and -.4cm] {$Z^\nu(\omega)$};
  \end{tikzpicture} 
  &=
  \frac{-\i\eta^{\mu\nu}}{M(\omega+\iO)^{2}}
  \,, 
  \\
  \label{eq:gravProp}
  \begin{tikzpicture}[baseline=(anchor)]
    \coordinate (anchor) at (0,-.1);
    \coordinate (x) at (-.2,0);
    \coordinate (y) at (1.0,0);
    \drawPerturbativeVertex{(x)}
    \drawPerturbativeVertex{(y)}
    \drawGravitonLineDirected{(x)}{(y)}
    \node at (x) [above left=0cm and -.7cm] {$h_\mn(-k)$};
    \node at (y) [above right=0cm and -.4cm] {$h_\rs(k)$};
  \end{tikzpicture} 
  &= 
  \frac{\i \mathcal{P}_{\mu\nu\rho\sigma}
          }{k^{2}+ \iO(W\cdot k)}
  \,,
  \quad
  \mathcal{P}_{\mu\nu\rho\sigma} = \eta_{\mu(\rho}\eta_{\sigma)\nu}-\frac1{D-2}\eta_\mn\eta_\rs
  \,,
\end{align}
\end{subequations}
with arrows  indicating causality flow. The vector $W^{\mu}$ denotes a choice of frame in which to measure the time component of the graviton momentum.

The graviton interacts with the worldline through an infinite tower of vertices connecting any number of worldline deflections with a single graviton.
The first two are \cite{Mogull:2020sak,Jakobsen:2022psy,Driesse:2024feo}
\begin{equation}\label{eq:FeynmanRulesWL}
\begin{aligned}
  \begin{tikzpicture}[baseline=(anchor)]
    \coordinate (anchor) at (0,-.6);
    \coordinate (v) at (0,0);
    \coordinate (in) at (-.6,0);
    \coordinate (out) at (.6,0);
    \coordinate (gOut) at (0,-1);
    \drawDottedBHLine{(in)}{(out)}
    \drawPerturbativeVertex{(v)}
    \drawGravitonLine{(v)}{(gOut)}
    \node at (gOut) [below] {$h_{\mu\nu}(k)$};
  \end{tikzpicture}\;
  &=
  -\frac{\I M\kappa_{D}}{2}
  e^{\I k\cdot B}
  \dd(k\cdot V) 
  V^\mu V^\nu
  \ , 
  \\
  \begin{tikzpicture}[baseline=(anchor)]
    \coordinate (anchor) at (0,-.6);
    \coordinate (v) at (0,0);
    \coordinate (in) at (-1,0);
    \coordinate (out) at (.6,0);
    \coordinate (gOut) at (0,-1);
    \drawDottedBHLine{(v)}{(out)}
    \drawSolidBHLine{(in)}{(v)}
    \drawPerturbativeVertex{(v)}
    \drawGravitonLine{(v)}{(gOut)}
    \node at (gOut) [below] {$h_{\mu\nu}
    (k)$};
    \node at (in) [left] {$Z^\rho(\omega)$};
  \end{tikzpicture}\;
  &=
  \frac{M\kappa_{D}}{2}
  e^{\I k\cdot B}
  \dd(k\cdot V+\omega)
  (
    V^\mu V^\nu k_\rho
    +2\omega V^{(\mu}\delta^{\nu)}_{\rho}
  )
  \, ,
\end{aligned}
\end{equation}
{ using $\dd(x)=2\pi \delta(x)$}.
In addition, we have the $n$-point bulk vertices coming from the non-linearity of the Einstein--Hilbert term,
\be\label{gravitonvertexscaling}
\begin{tikzpicture}[baseline={(anchor)}]
\begin{feynman}
  \coordinate (a1) at (-0.87,0.5);
  \coordinate (a2) at (0.87,0.5);
  \coordinate (a3) at (0,-1);
  \coordinate (o) at (0,0);
  \coordinate (anchor) at (0,-.2);
\diagram*{ (o) -- [photonTest] (a1), (o) -- [photonTest] (a2), (o) -- [photonTest] (a3) };
\draw [fill] (a1) circle (.0);
\draw [fill] (a2) circle (.0);
\draw [fill] (a3) circle (.0);
\draw [fill] (o) circle (.06);
\end{feynman} 
\end{tikzpicture}\; ,
\qquad
\begin{tikzpicture}[baseline={(anchor)}]
\begin{feynman}
	\coordinate (a1) at (-.707,.707);
	\coordinate (a2) at (.707,.707);
  \coordinate (a3) at (.707,-.707);
  \coordinate (a4) at (-.707,-.707);
  \coordinate (o) at (0,0);
  \coordinate (anchor) at (0,-.2);
\diagram*{ (o) -- [photonTest] (a1), (o) -- [photonTest] (a2), (o) -- [photonTest] (a3), (o) -- [photonTest] (a4)};
\draw [fill] (o) circle (.06);
\end{feynman} 
\end{tikzpicture}  \; , \qquad
\begin{tikzpicture}[baseline={(anchor)}]
\begin{feynman}
  \coordinate (a1) at (0.951, -0.309);
  \coordinate (a2) at (0.588, 0.809);
  \coordinate (a3) at (-0.588, 0.809);
  \coordinate (a4) at (-0.951, -0.309);
  \coordinate (a5) at (0,-1);
  \coordinate (o) at (0,0);
  \coordinate (anchor) at (0,-.2);
\diagram*{ (o) -- [photonTest] (a1), (o) -- [photonTest] (a2), (o) -- [photonTest] (a3), (o) -- [photonTest] (a4), (o) -- [photonTest] (a5)};
\draw [fill] (o) circle (.06);
\end{feynman} 
\end{tikzpicture}   \;, \qquad
\begin{tikzpicture}[baseline={(anchor)}]
\begin{feynman}
  \coordinate (a1) at (0.5,0.866);
	\coordinate (a2) at (-0.5,0.866);
  \coordinate (a3) at (-1,0);
  \coordinate (a4) at (-0.5,-0.866);
  \coordinate (a5) at (0.5,-0.866);
  \coordinate (a6) at (1,0);
  \coordinate (o) at (0,0);
  \coordinate (anchor) at (0,-.2);
\diagram*{ (o) -- [photonTest] (a1), (o) -- [photonTest] (a2), (o) -- [photonTest] (a3), (o) -- [photonTest] (a4), (o) -- [photonTest] (a5), (o) -- [photonTest] (a6)};
\draw [fill] (o) circle (.06);
\end{feynman} 
\end{tikzpicture}  \;, \qquad \ldots
\ee
which we do not spell out in detail (see e.g.~\cite{Driesse:2024feo} in a non-linear de Donder gauge for explicit expressions up to six legs).

\subsubsection{Black hole response Feynman rules}
Having reviewed the perturbative Feynman rules, we will now detail the construction of PM-resummed Feynman rules from \eqn{Znaïve}.
{To orient the reader, we first provide a brief roadmap of the construction.
First, the 1PI effective action is introduced as the Legendre transform of the BHR effective action $W[\mT]$.
Then, demanding that the one-point 1PI vertex vanish defines the background solutions of the equations of motion. These, along with the higher-point response functions are, as mentioned above, depicted by grey blobs. The higher-point 1PI vertices, which we depict by white blobs, are then obtained by expanding around these solutions.
With the vertices in hand, we show that the two-point response function plays the role of the propagator in this background-expanded theory, so that the higher-point response functions can be assembled from the 1PI vertices and the two-point response.
The result is a compact diagrammatic language in which one easily derives equations for the higher-point response functions.}

We begin by generalising \eqn{Znaïve} to include a source $f_\mu(\tau)$ for the deflection $Z^\mu(\tau)$, which one can think of as a force:
\begin{align}\label{Znaïve-with-f}
  \euler^{\frac\I\hbar W[\mT,f]}
  =
  \int \mathcal{D}[h,Z]\,\euler^{\frac\I\hbar(S_{\rm BH}[X,g]-\kappa_D\int\d^Dx\,\mT^{\mu\nu}(x)h_{\mu\nu}(x)-\int\d\tau\,f_\mu(\tau)Z^\mu(\tau))}
  \Big|_\text{tree}.
\end{align}
It is useful for brevity's sake to introduce a DeWitt notation for the expectation values and the sources,
\begin{equation}\label{eq:dewitt}
  \vev{\phi_A}_{J}
  =
  \big\{
    \,\vev{h_{\mu\nu}(x)}_{\{\mT,f\}}\,,
    \,\vev{Z^\mu(\tau)}_{\{\mT,f\}}\,
  \big\},
  \qquad
  J_A
  =
  \big\{
    \,\mT^{\mu\nu}(x)\,,
    \,f_\mu(\tau)\,
  \big\},
\end{equation}
and to use a summation convention on the collective index, such that, e.g.,
\begin{equation}
  \vev{\phi_A }_{J}J_A
  =
  \kappa_D\int\d^Dx\,\mT^{\mu\nu}(x)\vev{h_{\mu\nu}(x)}_{\{\mT,f\}}
  +
  \int\d\tau\,f_\mu(\tau)\vev{Z^\mu(\tau)}_{\{\mT,f\}}.
\end{equation}
Using this notation, we introduce the tree-level 1PI effective action through the Legendre transform of the BHR effective action 
\begin{equation}\label{eq:legendre}
  \Gamma_\text{BH}[\vev{\phi}_J] = W[J] + \vev{\phi_A}_{J} J_A.
\end{equation}
In our context, where we work only at tree-level, $\Gamma_\text{BH}[\vev{\phi}_J]$ coincides with the classical action $S_\text{BH}[\phi]$ as a functional; its argument $\vev{\phi}_J$, however, is the expectation value of the field in the presence of sources. That is, we classically have that
\begin{equation}
  \Gamma_\text{BH}[\vev{\phi}_J]
  =
  S_\text{BH}[\vev{\phi}_J]
  \, .
\end{equation}
It is the generating functional of 1PI $n$-point vertices
\begin{equation}
  \begin{tikzpicture}[baseline=(anchor)]
    \coordinate (anchor) at (0,-.1);
    \coordinate (v) at (0,0);
    \coordinate (phi1) at (1,1);
    \coordinate (in) at (-1,0);
    \coordinate (phi2) at (1,.5);
    \coordinate (phi3) at (1,0);
    \coordinate (phin) at (1,-1);
    \drawPerturbativeVertex{(v)}
    \drawSolidBHLine{(v)}{(phi1)}
    \drawSolidBHLine{(v)}{(phi2)}
    \drawSolidBHLine{(v)}{(phi3)}
    \drawSolidBHLine{(v)}{(phin)}
    \drawDottedBHLine{(in)}{(v)}
    \node at (phi1) [right] {\footnotesize $\vev{\phi_{A_1}}_J$};
    \node at (phi2) [right] {\footnotesize $\vev{\phi_{A_2}}_J$};
    \node at (phi3) [right] {\footnotesize $\vev{\phi_{A_3}}_J$};
    \node at (phin) [right] {\footnotesize $\vev{\phi_{A_n}}_J$};
    \node at ([shift={(-1pt,-10pt)}]phi3) {\vdots};
  \end{tikzpicture}\;
  =
  \frac{
    \delta^n
    \i
    \Gamma_\text{BH}[\vev{\phi}_J]
    }{
      \delta\vev{\phi_{A_1}}_J\delta\vev{\phi_{A_2}}_J\delta\vev{\phi_{A_3}}_J\cdots\delta\vev{\phi_{A_n}}_J}\bigg|_{\vev{\phi}_J=0},
\end{equation}
where, in this graph, solid lines denote both gravitons and deflections. As indicated, when evaluated at $\vev{\phi}_J=0$ they are equivalent to the perturbative vertices of eqs.~\eqref{eq:FeynmanRulesWL} and \eqref{gravitonvertexscaling} at tree-level. As such, we will in our application not set $\vev{\phi}_J=0$, but rather $J=0$, such that we obtain vertices expanded around the exact solutions of the equations of motion, which are
\begin{subequations}
\begin{align}
  \label{eq:h-one-point}
  \vev{h_{\mu\nu}(x)}_{J=0}
  &=
  \;\begin{tikzpicture}[baseline=(anchor)]
    \coordinate (anchor) at (0,-.5);
    \coordinate (v) at (0,0);
    \coordinate (in) at (-.8,0);
    \coordinate (out) at (.8,0);
    \coordinate (gOut) at (0,-1);
    \drawDottedBHLine{(in)}{(out)}
    \drawResponseBlob{(v)}
    \drawGravitonLineDirected{(v)}{(gOut)}
    \drawPerturbativeVertex{(gOut)}
  \end{tikzpicture}\;
  =
  -\kappa_D^{-1}
  \frac{\delta W[J]}{\delta\mT^{\mu\nu}(x)}\bigg|_{J=0}
  \,,
  \\
  \label{eq:Z-one-point}
  \qquad
  \vev{Z^\mu(\tau)}_{J=0}
  &=
  \;\begin{tikzpicture}[baseline=(anchor)]
    \coordinate (anchor) at (0,-.1);
    \coordinate (v) at (0,0);
    \coordinate (in) at (-.8,0);
    \coordinate (out) at (.8,0);
    \drawDottedBHLine{(in)}{(v)}
    \drawResponseBlob{(v)}
    \drawSolidBHLineDirected{($(v)!.3!(out)$)}{(out)}
    \drawPerturbativeVertex{(out)}
  \end{tikzpicture}\;
  =
  -
  \frac{\delta W[J]}{\delta f_\mu(\tau)}\bigg|_{J=0}
  \,,
\end{align}
\end{subequations}
defined implicitly through
\begin{equation}
  \frac{\delta\Gamma_\text{BH}[\vev{\phi}_J]}{\delta\vev{\phi_A}_J}\bigg|_{J=0} = 0.
\end{equation}
Notice that the momentum space one-point function of the graviton is equivalent to the one-point (static) response, $\vev{h_\mn(k)}_{J=0} = \mR_\mn(k)$. Diagrammatically, the defining equations become Berends--Giele recursion relations:
\begin{subequations}
\begin{align}
  \label{eq:graviton-one-point}
  \begin{tikzpicture}[baseline=(anchor)]
    \coordinate (anchor) at (0,-.8);
    \coordinate (v) at (0,0);
    \coordinate (in) at (-.7,0);
    \coordinate (out) at (.7,0);
    \coordinate (gOut) at (0,-1.4);
    \drawDottedBHLine{(in)}{(out)}
    \drawResponseBlob{(v)}
    \drawGravitonLineDirected{(v)}{(gOut)}
    \drawPerturbativeVertex{(gOut)}
  \end{tikzpicture}\;
  &=
  \;\begin{tikzpicture}[baseline=(anchor)]
    \coordinate (anchor) at (0,-.8);
    \coordinate (v) at (0,0);
    \coordinate (in) at (-.5,0);
    \coordinate (out) at (.5,0);
    \coordinate (gOut) at (0,-1.4);
    \drawDottedBHLine{(in)}{(out)}
    \drawPerturbativeVertex{(v)}
    \drawGravitonLineDirected{(v)}{(gOut)}
    \drawPerturbativeVertex{(gOut)}
  \end{tikzpicture}\;
  +
  \frac12
  \;\begin{tikzpicture}[baseline=(anchor)]
    \coordinate (anchor) at (0,-.8);
    \coordinate (v1) at (-.8,0);
    \coordinate (v2) at (.8,0);
    \coordinate (c) at (0,-.8);
    \coordinate (in) at (-1.3,0);
    \coordinate (out) at (1.3,0);
    \coordinate (gOut) at (0,-1.4);
    \drawDottedBHLine{(in)}{(out)}
    \drawResponseBlob{(v1)}
    \drawResponseBlob{(v2)}
    \drawPerturbativeVertex{(c)}
    \drawGravitonLine{(v1)}{(c)}
    \drawGravitonLine{(v2)}{(c)}
    \drawGravitonLineDirected{(c)}{(gOut)}
    \drawPerturbativeVertex{(gOut)}
  \end{tikzpicture}\;
  +
  \frac1{3!}
  \;\begin{tikzpicture}[baseline=(anchor)]
    \coordinate (anchor) at (0,-.8);
    \coordinate (v1) at (-.8,0);
    \coordinate (v2) at (0,0);
    \coordinate (v3) at (.8,0);
    \coordinate (c) at (0,-.8);
    \coordinate (in) at (-1.3,0);
    \coordinate (out) at (1.3,0);
    \coordinate (gOut) at (0,-1.4);
    \drawDottedBHLine{(in)}{(out)}
    \drawResponseBlob{(v1)}
    \drawResponseBlob{(v2)}
    \drawResponseBlob{(v3)}
    \drawPerturbativeVertex{(c)}
    \drawGravitonLine{(v1)}{(c)}
    \drawGravitonLine{(v2)}{(c)}
    \drawGravitonLine{(v3)}{(c)}
    \drawGravitonLineDirected{(c)}{(gOut)}
    \drawPerturbativeVertex{(gOut)}
  \end{tikzpicture}\;
  +\cdots,
  \\[.5cm]
  \label{eq:deflection-one-point}
  \begin{tikzpicture}[baseline=(anchor)]
    \coordinate (anchor) at (0,-.1);
    \coordinate (v) at (0,0);
    \coordinate (in) at (-.7,0);
    \coordinate (out) at (.7,0);
    \drawDottedBHLine{(in)}{(v)}
    \drawResponseBlob{(v)}
    \drawSolidBHLineDirected{($(v)!.3!(out)$)}{(out)}
    \drawPerturbativeVertex{(out)}
  \end{tikzpicture}\;
  &=
  \;\begin{tikzpicture}[baseline=(anchor)]
    \coordinate (anchor) at (0,-.1);
    \coordinate (s1) at (-.7,0);
    \coordinate (v) at (0,0);
    \coordinate (in) at (-1.2,0);
    \coordinate (out) at (.6,0);
    \drawDottedBHLine{(in)}{(v)}
    \drawResponseBlob{(s1)}
    \drawCurvedGravitonLine{(v)}{(s1)}{-90}{-90}
    \drawPerturbativeVertex{(v)}
    \drawSolidBHLineDirected{(v)}{(out)}
    \drawPerturbativeVertex{(out)}
  \end{tikzpicture}\;
  +
  \;\begin{tikzpicture}[baseline=(anchor)]
    \coordinate (anchor) at (0,-.1);
    \coordinate (s1) at (-1.4,0);
    \coordinate (s2) at (-.6,0);
    \coordinate (v) at (0,0);
    \coordinate (in) at (-1.9,0);
    \coordinate (out) at (.6,0);
    \drawDottedBHLine{(in)}{(s2)}
    \drawResponseBlob{(s1)}
    \drawResponseBlob{(s2)}
    \drawSolidBHLine{(s2)}{(v)}
    \drawCurvedGravitonLine{(v)}{(s1)}{-90}{-90}
    \drawPerturbativeVertex{(v)}
    \drawSolidBHLineDirected{(v)}{(out)}
    \drawPerturbativeVertex{(out)}
  \end{tikzpicture}\;
  +
  \frac12
  \;\begin{tikzpicture}[baseline=(anchor)]
    \coordinate (anchor) at (0,-.1);
    \coordinate (s1) at (-2.2,0);
    \coordinate (s2) at (-1.4,0);
    \coordinate (s3) at (-.6,0);
    \coordinate (v) at (0,0);
    \coordinate (in) at (-2.7,0);
    \coordinate (out) at (.6,0);
    \drawDottedBHLine{(in)}{(s2)}
    \drawResponseBlob{(s1)}
    \drawResponseBlob{(s2)}
    \drawResponseBlob{(s3)}
    \drawSolidBHLine{(s3)}{(v)}
    \draw [solid,thick] (s2) to[out=70,in=180-70,looseness=1.2] (v);
    \drawCurvedGravitonLine{(v)}{(s1)}{-90}{-90}
    \drawPerturbativeVertex{(v)}
    \drawSolidBHLineDirected{(v)}{(out)}
    \drawPerturbativeVertex{(out)}
  \end{tikzpicture}\;
  +\cdots,
\end{align}
\end{subequations}
for the graviton and deflection fields, respectively. 
{On the right-hand-side of each equation, every possible post-Minkowskian vertex rule (small black dot) appears with an appropriate outgoing line and otherwise fully contracted with one-point black hole response functions (including $Z^\mu$ fluctuations).
The dotted lines refer to the background expansion of $X^\mu(\tau)$ in Eq.~\eqref{eq:deflection} and they do not represent any propagating degree of freedom.
We generally draw the dotted lines such that dotted and solid lines of the black hole always form a tree topology.}

Eq.~\eqref{eq:graviton-one-point} simply constitutes a recursive definition of the metric, which yields the usual PM expansion by repeatedly substituting the left-hand side into the right-hand side:
\begin{align}\label{eq:perturbative-metric}
  \langle h_\mn(x) \rangle \big|_{J=0}
  =
  \;
  \begin{tikzpicture}[baseline={(0,-.7)},yscale=-1]
    \begin{feynman}
      \draw[dotted,thick] (-.3,0)--(.3,0);
      \vertex at (0,0) (p1);
      \drawPerturbativeVertex{(0,0)}
      \drawGravitonLineDirected{(p1)}{(0,1.25)}
      \drawPerturbativeVertex{(0,1.25)}
    \end{feynman}
  \end{tikzpicture}
  \;
  +
  \frac12
  \;
  \begin{tikzpicture}[baseline={(0,-.7)},yscale=-1]
    \begin{feynman}
      \draw[dotted,thick] (-.9,0)--(.9,0);
      \vertex[black,dot] at (0,0.6) (inter) ;
      \draw[photonTest] (-0.6,0)--(inter);
      \draw[photonTest] (0.6,0)--(inter);
      \drawGravitonLineDirected{(inter)}{(0,1.25)}
      \drawPerturbativeVertex{(0,1.25)}
      \drawPerturbativeVertex{(inter)}
      \drawPerturbativeVertex{(.6,0)}
      \drawPerturbativeVertex{(-.6,0)}
    \end{feynman}
  \end{tikzpicture}
  \;
  % \\
  % &
  +
  \frac1{3!}
  \;
  \begin{tikzpicture}[baseline={(0,-.7)},yscale=-1]
    \begin{feynman}
      \draw[dotted,thick] (-.9,0)--(.9,0);
      \vertex at (0,0.6) (inter);
      \draw[photonTest] (-0.6,0)--(inter);
      \draw[photonTest] (0,0)--(inter);
      \draw[photonTest] (0.6,0)--(inter);
      \drawGravitonLineDirected{(inter)}{(0,1.25)}
      \drawPerturbativeVertex{(0,1.25)}
      \drawPerturbativeVertex{(inter)}
      \drawPerturbativeVertex{(.6,0)}
      \drawPerturbativeVertex{(0,0)}
      \drawPerturbativeVertex{(-.6,0)}
    \end{feynman}
  \end{tikzpicture}
  \;
  +
  \frac12
  \;
  \begin{tikzpicture}[baseline={(0,-.7)},yscale=-1]
    \begin{feynman}
      \draw[dotted,thick] (-.9,0)--(.9,0);
      \vertex at (0,0.6) (inter);
      \vertex at (-0.6,0) (p1);
      \vertex at (0.6,0) (p2);
      \vertex at (0,1.25) (top);
      \vertex at ($(inter)!0.5!(p1)$) (verthalf);
      \draw[photonTest] (p1)--(inter);
      \draw[photonTest] (p2)--(inter);
      \drawGravitonLineDirected{(inter)}{(top)}
      \drawPerturbativeVertex{(top)}
      \draw[photonTest] (0,0)--(verthalf);
      \drawPerturbativeVertex{(inter)}
      \drawPerturbativeVertex{(verthalf)}
      \drawPerturbativeVertex{(.6,0)}
      \drawPerturbativeVertex{(0,0)}
      \drawPerturbativeVertex{(-.6,0)}
    \end{feynman}
  \end{tikzpicture}
  \;
  +\cdots.
\end{align}
This is nothing but the background Schwarzschild metric~\cite{Duff:1973zz,Jakobsen:2020ksu,Mougiakakos:2024nku,Damgaard:2024fqj}.
To determine the deflection one-point function in \eqn{eq:deflection-one-point}, we first note that the dimensional regularisation identity of \eqn{eq:crucial-dim-reg-id} has the diagrammatic manifestation
\begin{equation}\label{eq:crucial-dim-reg-id-diagram}
  \langle h_\mn(X(\tau)) \rangle \big|_{J=0}
  =  
    \;\begin{tikzpicture}[baseline=(anchor)]
    \coordinate (anchor) at (0,-.1);
    \coordinate (s1) at (-1.4,0);
    \coordinate (v) at (0,0);
    \coordinate (in) at (-2,0);
    \coordinate (out) at (.4,0);
    \drawDottedBHLine{(in)}{(out)}
    \drawResponseBlob{(s1)}
    \draw [photonTestDirected] (s1) to[out=-90,in=-90,looseness=1.5] (v);
    \drawPerturbativeVertex{(v)}
  \end{tikzpicture}\;
  =
  0
  \ .
\end{equation}
As this subdiagram occurs in all terms of \eqn{eq:deflection-one-point}, we are able to conclude that
\begin{equation}
  \begin{tikzpicture}[baseline=(anchor)]
    \coordinate (anchor) at (0,-.1);
    \coordinate (v) at (0,0);
    \coordinate (in) at (-.7,0);
    \coordinate (out) at (.7,0);
    \drawDottedBHLine{(in)}{(v)}
    \drawResponseBlob{(v)}
    \drawSolidBHLineDirected{($(v)!.3!(out)$)}{(out)}
    \drawPerturbativeVertex{(out)}
  \end{tikzpicture}\; = 0
  \ .
\end{equation}
We now define the 1PI (white blob) vertices
\begin{equation}
  \begin{tikzpicture}[baseline=(anchor)]
    \coordinate (anchor) at (0,-.1);
    \coordinate (v) at (0,0);
    \coordinate (phi1) at (1,1);
    \coordinate (in) at (-1,0);
    \coordinate (phi2) at (1,.5);
    \coordinate (phi3) at (1,0);
    \coordinate (phin) at (1,-1);
    \drawResummedBlob{(v)}
    \drawSolidBHLine{(v)}{(phi1)}
    \drawSolidBHLine{(v)}{(phi2)}
    \drawSolidBHLine{(v)}{(phi3)}
    \drawSolidBHLine{(v)}{(phin)}
    \drawDottedBHLine{(in)}{(v)}
    \node at (phi1) [right] {\footnotesize $\vev{\phi_{A_1}}_J$};
    \node at (phi2) [right] {\footnotesize $\vev{\phi_{A_2}}_J$};
    \node at (phi3) [right] {\footnotesize $\vev{\phi_{A_3}}_J$};
    \node at (phin) [right] {\footnotesize $\vev{\phi_{A_n}}_J$};
    \node at ([shift={(-1pt,-10pt)}]phi3) {\vdots};
  \end{tikzpicture}\;
  =
  \frac{\delta^n
  \i\Gamma_\text{BH}[\vev{\phi}_J]
  }{
    \delta\vev{\phi_{A_1}}_J\delta\vev{\phi_{A_2}}_J\delta\vev{\phi_{A_3}}_J\cdots\delta\vev{\phi_{A_n}}_J}\bigg|_{J=0},
\end{equation}
where, again, the solid lines in this graph denote both deflections and gravitons. 
{{For the purpose of interpreting this as a vertex rule on curved spacetime, we restrict to $n\geq3$ and treat the cases $n\leq2$ separately.}}
{These vertices are not 1PI in a traditional sense, as they contain the one-point function of the graviton from \eqn{eq:graviton-one-point}. As such, cutting any line connecting a vertex to such a one-point function would split the diagram in two. In spite of this, we shall still refer to these vertices as 1PI, but with the caveat that our definition of one-particle irreducibility is with respect to external legs only.} Expressed in terms of the background metric of \eqn{eq:graviton-one-point}, we find
\begin{align}\label{eq:ResummedBlobs}
  \begin{tikzpicture}[baseline=(cAnchor)]
    \coordinate (cAnchor) at (0,-1.5) ;
    \coordinate (ca) at (-1,0) ;
    \coordinate (cb) at (-.2,0) ;
    \coordinate (cc) at (+1,0) ;
    \coordinate (cin)  at (-1.5,0) ;
    \coordinate (cout) at (+1.5,0) ;
    \coordinate (cmiddlein)  at (-.8,-1.) ;
    \coordinate (cmiddleout) at (+.8,-1.) ;
    \coordinate (cmiddle) at (0,-1.) ;
    \coordinate (caBelow) at (-1,-2.) ;
    \coordinate (cbBelow) at (-.2,-2.) ;
    \coordinate (ccBelow) at (+1,-2.) ;
    \drawDottedBHLine{(cmiddlein)}{(cmiddleout)}
    \drawResummedBlob{(cmiddle)}
    \drawGravitonLine{(caBelow)}{(cmiddle)}
    \drawGravitonLine{(cbBelow)}{(cmiddle)}
    \drawGravitonLine{(ccBelow)}{(cmiddle)}
    \node at ([shift={(17pt,0pt)}]cbBelow) {\ldots};
    \draw[decorate,thick,decoration={brace,mirror}]  
      ([shift={(-2pt,-5pt)}]caBelow) -- ([shift={(+2pt,-5pt)}]ccBelow)
    node[midway, below=4pt] {$n$} ;
  \end{tikzpicture}\;
  =
  \;\sum_{m=0}^\infty
  \frac1{m!}\;
  \begin{tikzpicture}[baseline=(cmiddle)]
    \drawDottedBHLine{(cin)}{(cout)}
    \drawResponseBlob{(ca)}
    \drawResponseBlob{(cb)}
    \drawResponseBlob{(cc)}
    \drawGravitonLine{(ca)}{(cmiddle)}
    \drawGravitonLine{(cb)}{(cmiddle)}
    \drawGravitonLine{(cc)}{(cmiddle)}
    \drawPerturbativeVertex{(cmiddle)}
    \drawGravitonLine{(caBelow)}{(cmiddle)}
    \drawGravitonLine{(cbBelow)}{(cmiddle)}
    \drawGravitonLine{(ccBelow)}{(cmiddle)}
    \node at ([shift={(18pt,5pt)}]cb) {\footnotesize{\ldots}};
    \node at ([shift={(17pt,0pt)}]cbBelow) {\ldots};
    \draw[decorate,thick,decoration={brace,mirror}]  
      ([shift={(-2pt,-5pt)}]caBelow) -- ([shift={(+2pt,-5pt)}]ccBelow)
      node[midway, below=4pt] {$n$} ;
    \draw[decorate,thick,decoration={brace}]  
      ([shift={(-8pt,12pt)}]ca) -- ([shift={(+8pt,12pt)}]cc)
      node[midway, above=2pt] {$m$} ;
  \end{tikzpicture}\;,
  \qquad
  \begin{tikzpicture}[baseline=(anchor)]
    \coordinate (anchor) at (0,-.1);
    \coordinate (v) at (0,0);
    \coordinate (gOut) at (0,-1);
    \coordinate (in) at (-1,0);
    \coordinate (out3) at (1,.8);
    \coordinate (out2) at (1,0);
    \coordinate (out1) at (1,-.5);
    \drawDottedBHLine{(in)}{(v)}
    \drawGravitonLine{(v)}{(gOut)}
    \drawResummedBlob{(v)}
    \drawCurvedSolidBHLine{(v)}{(out1)}{180}{-40}
    \drawSolidBHLine{(v)}{(out2)}
    \drawCurvedSolidBHLine{(v)}{(out3)}{180}{40}
    \node at ($(out2)+(-.15,0.48)$) {\vdots};
  \end{tikzpicture}\;
  =
  \;\begin{tikzpicture}[baseline=(anchor)]
    \coordinate (anchor) at (0,-.1);
    \coordinate (v) at (0,0);
    \coordinate (gOut) at (0,-1);
    \coordinate (in) at (-1,0);
    \coordinate (out3) at (1,.8);
    \coordinate (out2) at (1,0);
    \coordinate (out1) at (1,-.5);
    \drawDottedBHLine{(in)}{(v)}
    \drawGravitonLine{(v)}{(gOut)}
    \drawPerturbativeVertex{(v)}
    \drawCurvedSolidBHLine{(v)}{(out1)}{180}{-40}
    \drawSolidBHLine{(v)}{(out2)}
    \drawCurvedSolidBHLine{(v)}{(out3)}{180}{40}
    \node at ($(out2)+(-.15,0.48)$) {\vdots};
  \end{tikzpicture}\;.
\end{align}
{To derive these, we exploited \eqn{eq:h-one-point} along with the aforementioned fact that we classically have $\Gamma_\text{BH}[\vev{\phi}_J] = S_\text{BH}[\vev{\phi}_J]$.}
The equivalence between the 1PI and the PM worldline vertices in the second equality is again guaranteed by the dimensional regularisation identity from \eqn{eq:crucial-dim-reg-id-diagram}.

In the resummed framework defined by the 1PI vertices, the role of the propagator is played by the two-point response function. This follows from the relation
\begin{align}\label{eq:propagatorsEquation}
  \frac{
    \delta^2 
    \Gamma_\text{BH}[\vev{\phi}_J]
  }{\delta \vev{\phi_A}_J \delta \vev{\phi_B}_J}
  \frac{
    \delta^2 
    W[J]
    }{\delta J_B \delta J_C}
  \bigg|_{J=0}
  =
  -\delta_{AC}
  \ ,
\end{align}
which may be derived by taking two variational derivatives 
$\delta^{2}/\delta \vev{\phi_{A}}_{J} \delta J_{C}$ of \eqn{eq:legendre}
and using $\vev{\phi_{B}}_{J}= -\delta W[J]/\delta J_{B}$. 
The ``diagonal'' kinetic terms coming from the 1PI effective action are
\begin{subequations}\label{eq:kinetic-1pi}
\begin{align}
  \frac{
    \delta^2
    \i \Gamma_\text{BH}[\vev{\phi}_J]
    }{
      \delta\vev{h_{\mu\nu}(x_1)}_J\delta\vev{h_{\rho\sigma}(x_2)}_J}\bigg|_{J=0}
  &=
  -(\;\begin{tikzpicture}[baseline=(anchor)]
    \coordinate (anchor) at (0,-.1);
    \coordinate (gIn) at (-.66,0);
    \coordinate (gOut) at (.66,0);
    \drawGravitonLine{(gIn)}{(gOut)}
    \drawPerturbativeVertex{(gIn)}
    \drawPerturbativeVertex{(gOut)}
  \end{tikzpicture}\;)^{-1}
  +
  \;\begin{tikzpicture}[baseline=(cAnchor)]
    \coordinate (cAnchor) at (0,-.5);
    \coordinate (cmiddlein)  at (-.8,0);
    \coordinate (cmiddleout) at (+.8,0);
    \coordinate (cmiddle) at (0,0);
    \coordinate (caBelow) at (-.8,-.8);
    \coordinate (ccBelow) at (+.8,-.8);
    \drawDottedBHLine{(cmiddlein)}{(cmiddleout)}
    \drawResummedBlob{(cmiddle)}
    \drawGravitonLine{(caBelow)}{(cmiddle)}
    \drawGravitonLine{(ccBelow)}{(cmiddle)}
  \end{tikzpicture}
  \ , 
  \\[.5cm]
  \frac{
    \delta^2\i \Gamma_\text{BH}[\vev{\phi}_J]
    }{
      \delta\vev{Z^\mu(\tau_1)}_J\delta\vev{Z^\nu(\tau_2)}_J}\bigg|_{J=0}
  &=
  -(\;\begin{tikzpicture}[baseline=(anchor)]
    \coordinate (anchor) at (0,-.1);
    \coordinate (in) at (-.7,0);
    \coordinate (out) at (.7,0);
    \coordinate (gIn) at (-.4,0);
    \coordinate (gOut) at (.4,0);
    \drawDottedBHLine{(in)}{(gIn)}
    \drawDottedBHLine{(gOut)}{(out)}
    \drawSolidBHLine{(gIn)}{(gOut)}
    \drawPerturbativeVertex{(gIn)}
    \drawPerturbativeVertex{(gOut)}
  \end{tikzpicture}\;)^{-1}
  \ .
\end{align}
\end{subequations}
Here, the two-point vertex in the first equation is the self-energy of the graviton,
\begin{equation}
  \begin{tikzpicture}[baseline=(cAnchor)]
    \coordinate (cAnchor) at (0,-.5);
    \coordinate (cmiddlein)  at (-.8,0);
    \coordinate (cmiddleout) at (+.8,0);
    \coordinate (cmiddle) at (0,0);
    \coordinate (caBelow) at (-1,-1);
    \coordinate (ccBelow) at (+1,-1);
    \drawDottedBHLine{(cmiddlein)}{(cmiddleout)}
    \drawResummedBlob{(cmiddle)}
    \drawGravitonLine{(caBelow)}{(cmiddle)}
    \drawGravitonLine{(ccBelow)}{(cmiddle)}
  \end{tikzpicture}\;
  =
  \sum_{m=1}^\infty
  \frac1{m!}
  \;\begin{tikzpicture}[baseline=(cAnchor)]
    \coordinate (cAnchor) at (0,-.5) ;
    \coordinate (ca) at (-1,0) ;
    \coordinate (cb) at (-.2,0) ;
    \coordinate (cc) at (+1,0) ;
    \coordinate (cin)  at (-1.5,0) ;
    \coordinate (cout) at (+1.5,0) ;
    \coordinate (cmiddlein)  at (-.8,-1.) ;
    \coordinate (cmiddleout) at (+.8,-1.) ;
    \coordinate (cmiddle) at (0,-1.) ;
    \coordinate (caBelow) at (-1,-1.) ;
    \coordinate (cbBelow) at (+1,-1.) ;
    \drawDottedBHLine{(cin)}{(cout)}
    \drawResponseBlob{(ca)}
    \drawResponseBlob{(cb)}
    \drawResponseBlob{(cc)}
    \drawGravitonLine{(ca)}{(cmiddle)}
    \drawGravitonLine{(cb)}{(cmiddle)}
    \drawGravitonLine{(cc)}{(cmiddle)}
    \drawPerturbativeVertex{(cmiddle)}
    \drawGravitonLine{(caBelow)}{(cmiddle)}
    \drawGravitonLine{(cbBelow)}{(cmiddle)}
    \node at ([shift={(18pt,5pt)}]cb) {\footnotesize{\ldots}};
    \draw[decorate,thick,decoration={brace}]  
      ([shift={(-8pt,12pt)}]ca) -- ([shift={(+8pt,12pt)}]cc)
      node[midway, above=2pt] {$m$} ;
  \end{tikzpicture}
\end{equation}
That the self-energy of the deflection field vanishes is again a consequence of \eqn{eq:crucial-dim-reg-id-diagram}. The ``off-diagonal'' kinetic term is simply
\begin{equation}
  \frac{
    \delta^2\i\Gamma_\text{BH}[\vev{\phi}_J]
    }{
      \delta\vev{Z^\rho(\tau)}_J\delta\vev{h_\mn(x)}_J}\bigg|_{J=0}
  =
  \;\begin{tikzpicture}[baseline=(anchor)]
    \coordinate (anchor) at (0,-.5);
    \coordinate (v) at (0,0);
    \coordinate (in) at (-1,0);
    \coordinate (out) at (1,0);
    \coordinate (gOut) at (0,-1);
    \drawDottedBHLine{(v)}{(out)}
    \drawSolidBHLine{(in)}{(v)}
    \drawPerturbativeVertex{(v)}
    \drawGravitonLine{(v)}{(gOut)}
  \end{tikzpicture}\;.
\end{equation}
These equations tell us that non-trivial effects stem only from the gravitational self-energy.

Inversion of the system of equations \eqref{eq:propagatorsEquation} allow us to express the resummed propagators (i.e.~connected two-point functions) in terms of the two-point response function.
We find:
\begin{subequations}
\begin{align}
  \langle h_{\abn{1}}(k_1) h_{\abn{2}}(k_2) \rangle_{\rm con}
  &=
  \mR_{\abn{1}\abn{2}}(k_1,k_2)
  =
  \;\begin{tikzpicture}[baseline=(anchor)]
    \coordinate (anchor) at (0,-.4);
    \coordinate (in) at (-.6,0);
    \coordinate (out) at (.6,0);
    \coordinate (v) at (0,0);
    \coordinate (gIn) at (-.7,-.7);
    \coordinate (gOut) at (.7,-.7);
    \drawDottedBHLine{(in)}{(out)}
    \drawResponseBlob{(v)}
    \drawGravitonLine{(gIn)}{(v)}
    \drawGravitonLineDirected{(v)}{(gOut)}
    \drawPerturbativeVertex{(gIn)}
    \drawPerturbativeVertex{(gOut)}
  \end{tikzpicture}\;,
  \\
  \langle h_\ab(k) Z^{\sigma}(\omega) \rangle_{\rm con}
  &=
  \;\begin{tikzpicture}[baseline=(anchor)]
    \coordinate (anchor) at (0,-.4);
    \coordinate (in) at (-.6,0);
    \coordinate (out) at (1.6,0);
    \coordinate (v1) at (0,0);
    \coordinate (gIn) at (-.7,-.7);
    \coordinate (v2) at (1,0);
    \drawDottedBHLine{(in)}{(out)}
    \drawResponseBlob{(v)}
    \drawPerturbativeVertex{(gIn)}
    \drawGravitonLine{(gIn)}{(v)}
    \drawCurvedGravitonLine{(v1)}{(v2)}{270}{-90}
    \drawPerturbativeVertex{(v2)}
    \drawSolidBHLineDirected{(v2)}{(out)}
    \drawPerturbativeVertex{(out)}
  \end{tikzpicture}\;,
  \\
  \langle Z^{\sigma_1}(\omega_1) Z^{\sigma_2}(\omega_2) \rangle_{\rm con}
  &=
  \;\begin{tikzpicture}[baseline=(anchor)]
    \coordinate (anchor) at (0,-.1);
    \coordinate (in) at (-.7,0);
    \coordinate (out) at (.7,0);
    \coordinate (inZ) at (-.4,0);
    \coordinate (outZ) at (.4,0);
    \drawDottedBHLine{(in)}{(out)}
    \drawSolidBHLineDirected{(inZ)}{(outZ)}
    \drawPerturbativeVertex{(inZ)}
    \drawPerturbativeVertex{(outZ)}
  \end{tikzpicture}\;
  +
  \;\begin{tikzpicture}[baseline=(anchor)]
    \coordinate (anchor) at (0,-.1);
    \coordinate (in) at (-1.6,0);
    \coordinate (out) at (1.6,0);
    \coordinate (v0) at (-1,0);
    \coordinate (v1) at (0,0);
    \coordinate (gIn) at (-1,-1);
    \coordinate (v2) at (1,0);
    \drawSolidBHLine{(in)}{(v0)}
    \drawPerturbativeVertex{(in)}
    \drawPerturbativeVertex{(v0)}
    \drawDottedBHLine{(v0)}{(v2)}
    \drawResponseBlob{(v1)}
    \drawCurvedGravitonLine{(v0)}{(v1)}{270}{-90}
    \drawCurvedGravitonLine{(v1)}{(v2)}{270}{-90}
    \drawPerturbativeVertex{(v2)}
    \drawSolidBHLineDirected{(v2)}{(out)}
    \drawPerturbativeVertex{(out)}
  \end{tikzpicture}\;.
\end{align}
\end{subequations}
Further, still using eq.~\eqref{eq:propagatorsEquation}, we may derive the two-point response function from the gravitational self-energy through a Schwinger-Dyson like recursive equation \cite{Bautista:2026qse}:
\begin{align}\label{eq:SchwingerDyson}
  &\bigg[
  \;\begin{tikzpicture}[baseline=(cAnchor)]
    \coordinate (cAnchor) at (0,-.4);
    \coordinate (cmiddlein)  at (-.5,0);
    \coordinate (cmiddleout) at (+.5+.5,0);
    \coordinate (cmiddle1) at (0,0);
    \coordinate (cmiddle2) at (.5,0);
    \coordinate (caBelow) at (-.7,-.7);
    \coordinate (ccBelow) at (+.7+.5,-.7);
    \drawDottedBHLine{(cmiddlein)}{(cmiddleout)}
    \drawGravitonLine{(caBelow)}{(cmiddle1)}
    \drawPerturbativeVertex{(cmiddle1)}
    \drawSolidBHLine{(cmiddle1)}{(cmiddle2)}
    \drawPerturbativeVertex{(cmiddle2)}
    \drawGravitonLineDirected{(cmiddle2)}{(ccBelow)}
  \end{tikzpicture}\;
  +
  \;\begin{tikzpicture}[baseline=(cAnchor)]
    \coordinate (cAnchor) at (0,-.4);
    \coordinate (cmiddlein)  at (-.6,0);
    \coordinate (cmiddleout) at (+.6,0);
    \coordinate (cmiddle) at (0,0);
    \coordinate (caBelow) at (-.7,-.7);
    \coordinate (ccBelow) at (+.7,-.7);
    \drawDottedBHLine{(cmiddlein)}{(cmiddleout)}
    \drawResummedBlob{(cmiddle)}
    \drawGravitonLine{(caBelow)}{(cmiddle)}
    \drawGravitonLineDirected{(cmiddle)}{(ccBelow)}
  \end{tikzpicture}\;
  \bigg]
  \times
  \;\begin{tikzpicture}[baseline=(anchor)]
    \coordinate (anchor) at (0,-.4);
    \coordinate (in) at (-.6,0);
    \coordinate (out) at (.6,0);
    \coordinate (v) at (0,0);
    \coordinate (gIn) at (-.7,-.7);
    \coordinate (gOut) at (.7,-.7);
    \drawDottedBHLine{(in)}{(out)}
    \drawResponseBlob{(v)}
    \drawGravitonLine{(gIn)}{(v)}
    \drawGravitonLineDirected{(v)}{(gOut)}
    \drawPerturbativeVertex{(gIn)}
    \drawPerturbativeVertex{(gOut)}
  \end{tikzpicture}\;
  =
  \;\begin{tikzpicture}[baseline=(anchor)]
    \coordinate (anchor) at (0,-.4);
    \coordinate (in) at (-.6,0);
    \coordinate (out) at (.6,0);
    \coordinate (v) at (0,0);
    \coordinate (gIn) at (-.7,-.7);
    \coordinate (gOut) at (.7,-.7);
    \drawDottedBHLine{(in)}{(out)}
    \drawResponseBlob{(v)}
    \drawGravitonLine{(gIn)}{(v)}
    \drawGravitonLineDirected{(v)}{(gOut)}
    \drawPerturbativeVertex{(gIn)}
    \drawPerturbativeVertex{(gOut)}
  \end{tikzpicture}\;
  -
  \;\begin{tikzpicture}[baseline=(anchor)]
    \coordinate (anchor) at (0,-.1);
    \coordinate (in) at (-.5,0);
    \coordinate (out) at (.5,0);
    \drawGravitonLineDirected{(in)}{(out)}
    \drawPerturbativeVertex{(in)}
    \drawPerturbativeVertex{(out)}
  \end{tikzpicture}\,.
\end{align}
A formal inversion of the terms in the square brackets leads to a geometric series for the two-point black hole response function. 

The BH response functions $\mR_{\alpha_{1}\beta_{1}\ldots \alpha_{n}\beta_{n}}$ for $n\ge3$ are then given by connected tree-level diagrams of these resummed rules. We find, for instance, for $n=3$ that
\begin{align}\label{eq:response_3_diagrams_resum}
  \begin{tikzpicture}[baseline=(anchor)]
    \coordinate (anchor) at (0,-.5) ;
    \coordinate (a) at (0,0) ;
    \coordinate (in) at (-0.7,0) ;
    \coordinate (out) at(+0.7,0) ;    
    \coordinate (outA)at(-.8,-1) ;
    \coordinate (outB)at(0,-1) ;
    \coordinate (outC)at(+.8,-1) ;
    \drawResponseBlob{(a)}
    \drawGravitonLine{(a)}{(outA)}
    \drawGravitonLine{(a)}{(outB)}
    \drawGravitonLineDirected{(a)}{(outC)}
    \drawPerturbativeVertex{(outA)}
    \drawPerturbativeVertex{(outB)}
    \drawPerturbativeVertex{(outC)}
    \drawDottedBHLine{(in)}{(out)}
  \end{tikzpicture}\;
  &=
  \;\begin{tikzpicture}[baseline=(anchor)]
    \coordinate (anchor) at (0,-.5) ;
    \coordinate (a) at (-1,0) ;
    \coordinate (b) at (0,0) ;
    \coordinate (c) at (1,0) ;
    \coordinate (d) at (2,0) ;
    \coordinate (in) at (-1.7,0) ;
    \coordinate (out) at (2.7,0) ;
    \coordinate (aOut) at (-1.5,-1) ;
    \coordinate (cOut) at (+1.5,-1) ;
    \coordinate (dOut) at (+2.5,-1) ;
    \drawResponseBlob{(a)}
    \drawResummedBlob{(b)}
    \drawResponseBlob{(c)}
    \drawResponseBlob{(d)}
    \drawDottedBHLine{(in)}{(out)}
    \drawCurvedGravitonLine{(b)}{(a)}{-90}{-90}
    \drawCurvedGravitonLine{(b)}{(c)}{-90}{-90}
    \drawCurvedGravitonLine{(b)}{(d)}{-140}{-90}
    \drawGravitonLine{(a)}{(aOut)}
    \drawShadowLine{(c)}{(cOut)}
    \drawGravitonLine{(c)}{(cOut)}
    \drawGravitonLineDirected{(d)}{(dOut)}
    \drawPerturbativeVertex{(aOut)}
    \drawPerturbativeVertex{(cOut)}
    \drawPerturbativeVertex{(dOut)}
  \end{tikzpicture}
  \\&\qquad
  + \Bigg[
  \;\begin{tikzpicture}[baseline=(anchor)]
    \coordinate (anchor) at (0,-.5) ;
    \coordinate (aOut) at (-1.5,-1) ;
    \coordinate (eOut) at (+3,-1) ;
    \coordinate (fOut) at (+4,-1) ;
    \coordinate (a) at (-1,0) ;
    \coordinate (b) at (0,0) ;
    \coordinate (c) at (.75,0) ;
    \coordinate (d) at (1.5,0) ;
    \coordinate (e) at (2.5,0) ;
    \coordinate (f) at (3.5,0) ;
    \coordinate (in) at (-1.7,0) ;
    \coordinate (out) at (4.2,0) ;
    \coordinate (aOut) at (-1.5,-1) ;
    \coordinate (cOut) at (+1.5,-1) ;
    \coordinate (dOut) at (+2.5,-1) ;
    \drawResponseBlob{(a)}
    \drawResponseBlob{(e)}
    \drawResponseBlob{(f)}
    \drawDottedBHLine{(in)}{(b)}
    \drawDottedBHLine{(d)}{(out)}
    \drawSolidBHLine{(b)}{(d)}
    \drawPerturbativeVertex{(b)}
    \drawPerturbativeVertex{(c)}
    \drawPerturbativeVertex{(d)}
    \drawCurvedGravitonLine{(b)}{(a)}{-90}{-90}
    \drawCurvedGravitonLine{(d)}{(e)}{-90}{-90}
    \drawCurvedGravitonLine{(c)}{(f)}{-140}{-60}
    \drawGravitonLine{(a)}{(aOut)}
    \drawShadowLine{(e)}{(eOut)}
    \drawGravitonLine{(e)}{(eOut)}
    \drawGravitonLineDirected{(f)}{(fOut)}
    \drawPerturbativeVertex{(aOut)}
    \drawPerturbativeVertex{(eOut)}
    \drawPerturbativeVertex{(fOut)}
  \end{tikzpicture}\;
  +
  (\text{two permutations})
  \Bigg].
  \nn
\end{align}
In the square brackets, the two permutations refer to cyclic permutations of the three external gravitons.
We remind the reader that one leg should be singled out as outgoing and that the causality flow of all propagators should be directed toward that outgoing line.

We note in passing that if one removes the primary BH, the response functions do not vanish but reproduce the 
tree-level graviton scattering amplitudes, i.e.~the ``response of the vacuum’’.
In this case, the resummed 1PI vertices (white blobs) reduce to the perturbative $n$-point interaction vertices~\eqref{gravitonvertexscaling}.

\subsection{Gravitational self-force expansion from black hole response}
In order to consider a binary system of a heavy black hole and a secondary lighter one, we add the action of the lighter body
\begin{align}
  -\int\d^Dx\,\mT^\mn_\text{sec}(x)g_\mn(x)
  =
  -\int\d^Dx\,\eta_\mn \mT^\mn_{\rm sec}(x)
  -
  \kappa_D\int\d^Dx\,h_\mn(x)\mT^\mn_\text{sec}(x),
\end{align}
to the action of our primary from \eqn{monoBHS}, where the energy-momentum tensor of the secondary {is given by 
\begin{align}\label{eq:secTmn}
  \mT^\mn_{\rm sec}(x)
  =
  \frac m2\int\d\tau  \,\delta^{(D)}(x-x(\tau))\xd^\mu(\tau) \xd^\nu(\tau)
  \,.
\end{align}
}
This yields
\begin{align}
  S[\mT_\text{sec},X,g]
  =
  S_\text{BH}[X,g]
  -
  \int\d^Dx\,g_\mn(x)\mT^\mn_\text{sec}(x).
\end{align}
Then, following the steps of black hole response theory, we integrate out $h_\mn(x)$ and $X^\mu(\tau)$ and end up with an effective description of our secondary interacting with the BH.
We define,
\begin{align}\label{eq:SBHR_def}
  \euler^{
    \frac{\i}{\hbar} S_{\rm BHR}[\mT_\text{sec}]
  }
  =
  \int \mathcal{D}[h,Z]\,
  \euler^{
    \frac{\i}{\hbar}S[\mT_\text{sec},X,g]
  }
  \, ,
\end{align}
and find
\begin{align}
  S_{\rm BHR}[\mT_\text{sec}]
  =
  -\int\d^Dx\,\eta_\mn \mT^\mn_{\rm sec}(x)
  +
  W[\mT_{\rm sec}^\mn]
  \, .
\end{align}
This is our ``black hole response'' action for the secondary particle, with the first term being the kinetic term for deflections on the secondary worldline. Explicitly,
\begin{align}
  &S_{\rm BHR}[x]
  =
  -
  \frac m2
  \int \d\tau\,
  \dot x^\alpha(\tau)
  \dot x^\beta(\tau)
  g^{\rm BH}_\ab(x(\tau))
  \\
  &\qquad+
  \frac{\i}8m^2\kappa_D^2
  \int \d\tau_1\d\tau_2\,
  \dot x^\an{1}(\tau_1)
  \dot x^\bn{1}(\tau_1)
  \dot x^\an{2}(\tau_2)
  \dot x^\bn{2}(\tau_2)
  \mR_{\abn{1}\abn{2}}(x(\tau_1),x(\tau_2))
  \nn
  \\
  &\qquad+
  \frac{1}{48}m^3\kappa_D^3
  \int \d\tau_1\d\tau_2\d\tau_3\,
  \dot x^\an{1}_1
  \dot x^\bn{1}_1
  \dot x^\an{2}_2
  \dot x^\bn{2}_2
  \dot x^\an{3}_3
  \dot x^\bn{3}_3
  \mR_{\abn{1}\abn{2}\abn{3}}(x_1,x_2,x_3)
  +
  \cdots
  \nn
\end{align}
Here, $g^{\rm BH}_\mn(x)=\eta_\mn +\kappa_D \mR_\mn(x)$ is the background BH metric and in the third line we used the shorthand $x_i^\mu=x^\mu(\tau_i)$.

The BH response action is ideally suited for (diagrammatic) self-force expansions.
Its interactions scale with increasing powers of $m$ and for a fixed $n$SF calculation (i.e.~powers in $m/M$) one requires only the  BH response functions up to $(n+1)$-points.
To make this manifest, it is advantageous to integrate out the 0SF behaviour and expand the trajectory of the secondary around its $m/M\to0$ limit,
\begin{align}\label{eq:backgroundSFexpansion}
  x^\mu(\tau)
  =
  \bar x^\mu(\tau)
  +
  \cz^\mu(\tau)
  \, ,
\end{align}
with $\cz^\mu(\tau)$ denoting the deflection away from geodesic motion.
The trajectory $\bar x^\mu(\tau)$ solves the geodesic equation in the BH background $g^{\rm BH}_\mn(x)$ that follow from the action,
\begin{align}
  &S_{\rm probe}[x]
  =
  -
  \frac m2
  \int \d\tau\,
  \dot {x}^\alpha(\tau)
  \dot {x}^\beta(\tau)
  g^{\rm BH}_\ab(x(\tau))\, .
\end{align}
The geodesic trajectory may be known either analytically or perturbatively.
In particular, its diagrammatic expansion has been considered carefully in Ref.~\cite{Hoogeveen:2025tew}.

The secondary black hole always interacts with the black hole response functions through the dependence of $\mT_{\rm sec}^\mn(x)$ on $x^\mu(\tau)$.
Feynman vertices may thus be constructed through the expansion of $\mT^\mn(x)$ in $\cz^\mu(\tau)$. Going to momentum space, we define
\begin{align}\label{eq:crossbuildingblocks}
  \begin{tikzpicture}[baseline=(anchor)]
    \coordinate (anchor) at (0,-.1) ;
    \coordinate (in) at (-1,0) ;
    \coordinate (out1) at (1,0) ;
    \coordinate (out2) at (1,0.5) ;
    \coordinate (out3) at (1,-0.8) ;
    \coordinate (x) at (0,0) ;
    \coordinate (k) at (0,1) ;
    \draw (out1) node [right] {$\cz^{\alpha_2}(\omega_2)$};
    \draw (out2) node [right] {$\cz^{\alpha_1}(\omega_1)$};
    \draw (out3) node [right] {$\cz^{\alpha_n}(\omega_n)$};
    \drawDottedSecondaryLine{(in)}{(x)}
    \draw [zUndirected] (x) -- (out1);
    \draw [zUndirected] (x) to[out=40,in=180] (out2);
    \draw [zUndirected] (x) to[out=-40,in=180] (out3);
    \draw [photonTest] (x) -- (k);
    \node [above] at (k) {$h_\mn(k)$};
    \node at ([xshift=-4pt,yshift=-8pt]out1) {\vdots};
    \drawSecondaryVertex{(x)};
  \end{tikzpicture}\;
  =
  -\I\kappa_D\frac{
    \delta^n\mT^\mn_{\rm sec}(k)
    }{
    \delta x^\an{1}(\omega_1)
    \delta x^\an{2}(\omega_2)
    \dots
    \delta x^\an{n}(\omega_n)
  }
  \bigg|_{x^\mu\to\bar x^\mu}
  \,.
\end{align}
We note that $\bar x^\mu$ naturally solves the equation of motion given by 
\begin{align}
  \frac{\delta S_{\rm probe}}{
    \delta x^\mu
  }
  \bigg|_{x^\mu \to \bar x^\mu}
  =
  0
  \, ,
\end{align}
which implies that, diagrammatically, $\bar x^\mu(\omega)$ is given as, 
\begin{align}\label{eq:geodesic}
  \bar x^\rho(\omega)
  =
  \;\begin{tikzpicture}[baseline=(anchor)]
    \coordinate (anchor) at (0,.6);
    \coordinate (in) at (-.6,0);
    \coordinate (out) at (.6,0);
    \coordinate (x) at (0,0);
    \coordinate (k) at (0,1.3);
    \drawResponseBlob{(0,1.3)}
    \drawDottedBHLine{(-.6,1.3)}{(.6,1.3)}
    \draw (out) node [right] {$\cz^\rho(\omega)$};
    \draw [zParticle] ([xshift=3pt]x) -- (out);
    \drawPerturbativeVertex{(out)}
    \drawDottedSecondaryLine{(in)}{(x)}
    \drawGravitonLine{(x)}{(k)}
    \drawSecondaryVertex{(x)};
  \end{tikzpicture}
  \, .
\end{align}
This, in fact, implies a similar equation as \eqn{eq:ResummedBlobs} for the cross interactions:
\begin{align}\label{eq:cross-recursion}
  \begin{tikzpicture}[baseline={(anchor)}]
    \coordinate (in) at (-1,0) ;
    \coordinate (out1) at (1,0) ;
    \coordinate (out2) at (1,0.5) ;
    \coordinate (out3) at (1,-0.8) ;
    \coordinate (x) at (0,0) ;
    \coordinate (anchor) at (0,0) ;
    \coordinate (k) at (0,1.) ;
    \draw (out1) node [right] {$\cz^{\rho_2}(\omega_2)$};
    \draw (out2) node [right] {$\cz^{\rho_1}(\omega_1)$};
    \draw (out3) node [right] {$\cz^{\rho_n}(\omega_n)$};
    \drawDottedSecondaryLine{(in)}{(x)}
    \draw [zUndirected] (x) -- (out1);
    \draw [zUndirected] (x) to[out=40,in=180] (out2);
    \draw [zUndirected] (x) to[out=-40,in=180] (out3);
    \draw [photonTest] (x) -- (k);
    \node at ([xshift=-4pt,yshift=-8pt]out1) {\vdots};
    \drawSecondaryVertex{(x)};
  \end{tikzpicture}
  \ \ =\ 
  \sum_{m=0}^\infty
  \ \frac1{m!}\ \ 
  \begin{tikzpicture}[baseline={(anchor)}]
    \coordinate (in) at (-1,0) ;
    \coordinate (out1) at (1,0) ;
    \coordinate (out2) at (1,0.5) ;
    \coordinate (out3) at (1,-0.8) ;
    \coordinate (in2) at (-.7,0) ;
    \coordinate (in1) at (-1.4,0) ;
    \coordinate (in3) at (-2.5,0) ;
    \coordinate (up2) at (-.7,1.4) ;
    \coordinate (up1) at (-1.4,1.4) ;
    \coordinate (up3) at (-2.5,1.4) ;
    \coordinate (x) at (0,0) ;
    \coordinate (anchor) at (0,0) ;
    \coordinate (k) at (0,1.) ;
    \draw (out1) node [right] {$\cz^{\rho_2}(\omega_2)$};
    \draw (out2) node [right] {$\cz^{\rho_1}(\omega_1)$};
    \draw (out3) node [right] {$\cz^{\rho_n}(\omega_n)$};
    \drawSolidBHLine{(x)}{(out1)}
    \drawCurvedSolidBHLine{(x)}{(out2)}{180}{40}
    \drawCurvedSolidBHLine{(x)}{(out3)}{180}{-40}
    \drawSolidBHLine{(in2)}{(x)}
    \drawCurvedSolidBHLine{(in1)}{(x)}{-140-5}{-40+5}
    \drawCurvedSolidBHLine{(in3)}{(x)}{-140+5}{-40-5}
    \draw [photonTest] (x) -- (k);
    \drawGravitonLine{(in1)}{(up1)}
    \drawGravitonLine{(in2)}{(up2)}
    \drawGravitonLine{(in3)}{(up3)}
    \drawSecondaryVertex{(in1)}
    \drawSecondaryVertex{(in2)}
    \drawSecondaryVertex{(in3)}
    \drawResponseBlob{(up1)}
    \drawResponseBlob{(up2)}
    \drawResponseBlob{(up3)}
    \drawDottedSecondaryLine{([xshift=-17pt]in3)}{(in3)}
    \drawDottedBHLine{([xshift=-17pt]up3)}{([xshift=17pt]up2)}
    \node at ([xshift=-16pt,yshift=-20pt]up1) {\ldots};
    \node at ([xshift=-4pt,yshift=-8pt]out1) {\vdots};
    \drawPerturbativeVertex{(x)};
    \draw[decorate,thick,decoration={brace}]  
      ([shift={(-8pt,12pt)}]up3) -- ([shift={(+8pt,12pt)}]up2)
      node[midway, above=2pt] {$m$} ;
      %bg line
      \drawDottedSecondaryLine{(in3)}{(in2)}
  \end{tikzpicture}
  \,.
\end{align}
Note that for $n=1$ this equation is a recursive definition of the geodesic (which is the same recursive structure used in Ref.~\cite{Hoogeveen:2025tew}). 

Further, the two-point interaction, or the self-energy of the probe in the presence of the black hole background, appears in the kinetic term of the probe. It results in the following Schwinger--Dyson equation for the propagator of the probe deflection:
\begin{align}
  \label{eq:propagator}
  \langle \cz^{\mu}(\omega_1)\,  \cz^{\nu}(\omega_2)\rangle
  =
  \;\begin{tikzpicture}[baseline=(anchor)]
    \coordinate (anchor) at (0,-.1);
    \coordinate (in) at (-0.6,0);
    \coordinate (out) at (1.4,0);
    \coordinate (x) at (-.2,0);
    \coordinate (y) at (1.0,0);
    \drawSolidSecondaryLineDirected{(x)}{(y)}
    \draw [background] (in) -- (x);
    \draw [background] (y) -- (out);
    \drawSecondaryVertex{(x)};
    \drawSecondaryVertex{(y)};
  \end{tikzpicture}\;
  =
  \;\begin{tikzpicture}[baseline=(anchor)]
    \coordinate (anchor) at (0,-.1);
    \coordinate (in) at (-.7,0);
    \coordinate (out) at (.7,0);
    \coordinate (gIn) at (-.4,0);
    \coordinate (gOut) at (.4,0);
    \drawDottedBHLine{(in)}{(gIn)}
    \drawDottedBHLine{(gOut)}{(out)}
    \drawSolidBHLineDirected{(gIn)}{(gOut)}
    \drawPerturbativeVertex{(gIn)}
    \drawPerturbativeVertex{(gOut)}
  \end{tikzpicture}\;
  +
  \;\begin{tikzpicture}[baseline={(anchor)}]
    \coordinate (in) at (-.8,0);
    \coordinate (out) at (.8,0);
    \coordinate (x) at (0,0);
    \coordinate (anchor) at (0,.4);
    \drawResponseBlob{(0,1)}
    \drawDottedBHLine{(-.8,1)}{(.8,1)}
    \draw [zUndirected] (in) -- (x);
    \drawSolidSecondaryLineDirected{(x)}{(out)}
    \draw [photonTest] (x) -- (k);
    \drawSecondaryVertex{(x)}
    \drawSecondaryVertex{(out)}
    \drawPerturbativeVertex{(in)}
  \end{tikzpicture}
  \;.
\end{align}
Finally, the interaction vertices are any combinations of the vertices in eq.~\eqref{eq:crossbuildingblocks} with the black hole response functions (except the two combinations considered in eqs.~\eqref{eq:geodesic} and~\eqref{eq:propagator}).
That is, for the one-point black hole response function, we have for three worldline deflections and beyond
\be
\begin{tikzpicture}[baseline={(anchor)}]
  \def\length{.9};
  \coordinate (in) at (-\length,0);
  \coordinate (out1) at (\length,0);
  \coordinate (out2) at (\length,0.5);
  \coordinate (x) at (0,0);
  \coordinate (up)at (0,1.2);
  \coordinate (anchor) at (0,0);
  \draw (in) node [left] {$\cz^{\rho_1}(\omega_1)$};
  \draw (out1) node [right] {$\cz^{\rho_2}(\omega_2)$};
  \draw (out2) node [right] {$\cz^{\rho_3}(\omega_3)$};
  \draw [zUndirected] (in) -- (x);
  \draw [zUndirected] (x) -- (out1);
  \draw [zUndirected] (x) to[out=40,in=180] (out2);
  \drawSecondaryVertex{(x)}
  \drawGravitonLine{(x)}{(up)}
  \drawResponseBlob{(up)}
  %bg line
  \drawDottedSecondaryLine{($(up)-(.8,0)$)}{($(up)+(.8,0)$)}
\end{tikzpicture}\,,
\qquad
\begin{tikzpicture}[baseline={(anchor)}]
  \def\length{.9};
  \coordinate (in) at (-\length,0);
  \coordinate (out1) at (\length,0);
  \coordinate (out2) at (\length,.4);
  \coordinate (out3) at (\length,.8);
  \coordinate (x) at (0,0);
  \coordinate (anchor) at (0,0);
  \draw (in) node [left] {$\cz^{\rho_1}(\omega_1)$};
  \draw (out1) node [right] {$\cz^{\rho_2}(\omega_2)$};
  \draw (out2) node [right] {$\cz^{\rho_3}(\omega_3)$};
  \draw (out3) node [right] {$\cz^{\rho_4}(\omega_4)$};
  \draw [zUndirected] (in) -- (x);
  \draw [zUndirected] (x) -- (out1);
  \draw [zUndirected] (x) to[out=40,in=180] (out2);
  \draw [zUndirected] (x) to[out=40,in=180] (out3);
  \drawSecondaryVertex{(x)};
  \drawGravitonLine{(x)}{(up)}
  \drawResponseBlob{(up)}
  %bg line
  \drawDottedSecondaryLine{($(up)-(.8,0)$)}{($(up)+(.8,0)$)}
\end{tikzpicture}\,,
\qquad
\ldots.
\ee 
The two-point black hole response can connect to any number of worldline deflections
\begin{equation}
  \begin{tikzpicture}[baseline=(anchor)]
    \def\length{.75};
    \coordinate (anchor) at (0,-1.5) ;
    \coordinate (aAbove) at (0, 0) ;
    \coordinate (inAbove)at (-.8,0) ;
    \coordinate (outAbove)at(+.8,0) ;
    \coordinate (aBelow) at (-.5,-1.5) ;
    \coordinate (bBelow) at (+.5,-1.5) ;
    \coordinate (inBelow) at (-1,-1.5) ;
    \coordinate (outBelow) at ($(bBelow)+(\length,0)$);
    \drawSecondaryVertex{(aBelow)}
    \drawSecondaryVertex{(bBelow)}
    \drawDottedSecondaryLine{(inBelow)}{(bBelow)}
    \drawResponseBlob{(aAbove)}
    \drawDottedBHLine{(inAbove)}{(outAbove)}
    \drawGravitonLine{(aBelow)}{(aAbove)}
    \drawGravitonLine{(bBelow)}{(aAbove)}
    \drawSolidBHLine{(bBelow)}{(outBelow)}
    \node at (outBelow) [above] {$\cz^{\rho_1}(\omega_1)$};
  \end{tikzpicture}
  ,\quad
  \begin{tikzpicture}[baseline=(anchor)]
    \def\length{.75};
    \coordinate (anchor) at (0,-1.5) ;
    \coordinate (aAbove) at (0, 0) ;
    \coordinate (inAbove)at (-.8,0) ;
    \coordinate (outAbove) at (+.8,0) ;
    \coordinate (aBelow) at (-.5,-1.5) ;
    \coordinate (bBelow) at (+.5,-1.5) ;
    \coordinate (inBelow) at ($(aBelow)-(\length,0)$); 
    \coordinate (outBelow) at ($(bBelow)+(\length,0)$); 
    \drawSecondaryVertex{(aBelow)}
    \drawSecondaryVertex{(bBelow)}
    \drawSolidBHLine{(inBelow)}{(aBelow)}
    \drawDottedSecondaryLine{(aBelow)}{(bBelow)}
    \drawResponseBlob{(aAbove)}
    \drawDottedBHLine{(inAbove)}{(outAbove)}
    \drawGravitonLine{(aBelow)}{(aAbove)}
    \drawGravitonLine{(bBelow)}{(aAbove)}
    \drawSolidBHLine{(bBelow)}{(outBelow)}
    \node at (outBelow) [above] {$\cz^{\rho_2}(\omega_2)$};
    \node at (inBelow) [above] {$\cz^{\rho_1}(\omega_1)$};
  \end{tikzpicture}
  ,\quad
  \begin{tikzpicture}[baseline=(anchor)]
    \def\length{.75};
    \coordinate (anchor) at (0,-1.5) ;
    \coordinate (aAbove) at (0, 0) ;
    \coordinate (inAbove)at (-.8,0) ;
    \coordinate (outAbove) at (+.8,0) ;
    \coordinate (aBelow) at (-.5,-1.5) ;
    \coordinate (bBelow) at (+.5,-1.5) ;
    \coordinate (inBelow) at (-1,-1.5) ;
    \coordinate (outBelow) at ($(bBelow)+(\length,0)$); %(+1.25,-1.5)
    \coordinate (outBelow2) at ($(outBelow)+(0,.5)$);
    \drawSecondaryVertex{(aBelow)}
    \drawSecondaryVertex{(bBelow)}
    \drawDottedSecondaryLine{(inBelow)}{(bBelow)}
    \drawResponseBlob{(aAbove)}
    \drawDottedBHLine{(inAbove)}{(outAbove)}
    \drawGravitonLine{(aBelow)}{(aAbove)}
    \drawGravitonLine{(bBelow)}{(aAbove)}
    \drawSolidBHLine{(bBelow)}{(outBelow)}
    % draw secondary curved z line
    \draw [zUndirected] (bBelow) to[out=40,in=180] (outBelow2);
    %\drawSecondaryVertex{(outBelow)}
    \node at (outBelow) [right] {$\cz^{\rho_1}(\omega_1)$};
    \node at (outBelow2) [right] {$\cz^{\rho_2}(\omega_2)$};
  \end{tikzpicture}
  ,\quad\ldots
  \,,
\end{equation}
and similarly for higher-point responses. These vertices, along with the probe propagator, are the fundamental building blocks of the diagrammatic expansion of SF observables. Importantly all interactions of $\cz$ scale as $m$, while the propagator scales as $1/m$.

The two-body observables are now given as: 
\begin{subequations}\label{eq:SF_expansion}
\begin{align}\label{eq:1sfWaveform}
  \langle h_{\mu\nu}(k)\rangle
  \  = \ 
   \underbrace{ \begin{tikzpicture}[baseline=(anchor)]
    \coordinate (anchor) at (0,-.8) ;
    \coordinate (aAbove) at (0, 0) ;
    \coordinate (inAbove)at (+.8,0) ;
    \coordinate (outAbove)at(-.8,0) ;
    \coordinate (aBelow) at (0,-1.5) ;
    \coordinate (inBelow)at (+.5,-1.5) ;
    \coordinate (outBelow)at(-.5,-1.5) ;
    \coordinate (out) at (1.2,-.75) ;
    \drawSecondaryVertex{(aBelow)}
    \drawDottedSecondaryLine{(inBelow)}{(outBelow)}
    \drawResponseBlob{(aAbove)}
    \drawDottedBHLine{(inAbove)}{(outAbove)}
    \drawGravitonLine{(aBelow)}{(aAbove)}
    \draw [photonTestDirected] (aAbove) to[out=-45,in=-180,looseness=1.5] (out);
    \drawPerturbativeVertex{(out)}
  \end{tikzpicture}}_{{m^{1}}}
  \ +\ 
  \underbrace{ \frac12\  \begin{tikzpicture}[baseline=(anchor)]
    \coordinate (aAbove) at (0, 0) ;
    \coordinate (inAbove)at (+.8,0) ;
    \coordinate (outAbove)at(-.8,0) ;
    \coordinate (aBelow) at (-.5,-1.5) ;
    \coordinate (bBelow) at (+.5,-1.5) ;
    \coordinate (inBelow)at (+1,-1.5) ;
    \coordinate (outBelow)at(-1,-1.5) ;
    \coordinate (out) at (1.2,-.75) ;
    \drawSecondaryVertex{(aBelow)}
    \drawGravitonLine{(aBelow)}{(aAbove)}
    \drawSecondaryVertex{(bBelow)}
    \drawGravitonLine{(bBelow)}{(aAbove)}
    \drawDottedSecondaryLine{(inBelow)}{(outBelow)}
    \drawResponseBlob{(aAbove)}
    \drawDottedBHLine{(inAbove)}{(outAbove)}
    \draw [photonTestDirected] (aAbove) to[out=-45,in=-180,looseness=1.5] (out);
    \drawPerturbativeVertex{(out)}
  \end{tikzpicture}
  \ +\ \ 
    \begin{tikzpicture}[baseline=(anchor)]
    \coordinate (aAbove) at (0, 0) ;
    \coordinate (bAbove) at (1.5, 0) ;
    \coordinate (outAbove)at (2.3,0) ;
    \coordinate (inAbove)at(-.8,0) ;
    \coordinate (aBelow) at (-.5,-1.5) ;
    \coordinate (bBelow) at (+.5,-1.5) ;
    \coordinate (cBelow) at (1.5,-1.5) ;
    \coordinate (outBelow)at (+2,-1.5) ;
    \coordinate (inBelow)at(-1,-1.5) ;
    \coordinate (out) at (2.7,-.75) ;
    \drawSecondaryVertex{(cBelow)}
    \drawSecondaryVertex{(aBelow)}
    \drawGravitonLine{(aBelow)}{(aAbove)}
    \drawSecondaryVertex{(bBelow)}
    \drawGravitonLine{(bBelow)}{(aAbove)}
    \drawSolidSecondaryLine{(bBelow)}{(cBelow)}
    \drawDottedSecondaryLine{(inBelow)}{(bBelow)}
    \drawDottedSecondaryLine{(outBelow)}{(cBelow)}
    \drawResponseBlob{(aAbove)}
    \drawResponseBlob{(bAbove)}
    \drawGravitonLine{(cBelow)}{(bAbove)}
    \drawDottedBHLine{(inAbove)}{(outAbove)}
    \draw [photonTestDirected] (bAbove) to[out=-45,in=-180,looseness=1.5] (out);
    \drawPerturbativeVertex{(out)}
  \end{tikzpicture}
  }_{m^{2}}
  \ +\ \mathcal{O}\Big(\frac{m^3}{M^3}\Big)\,,
\end{align}
\begin{align}
  \langle \cz^\sigma(\omega)\rangle
  \  &= \ 
    \underbrace{\begin{tikzpicture}[baseline=(anchor)]
    \coordinate (anchor) at (0,-.8) ;
    \coordinate (aAbove) at (0, 0) ;
    \coordinate (inAbove)at (-.8,0) ;
    \coordinate (outAbove)at(+.8,0) ;
    \coordinate (aBelow) at (-.5,-1.5) ;
    \coordinate (bBelow) at (+.5,-1.5) ;
    \coordinate (inBelow)at (-1,-1.5) ;
    \coordinate (outBelow)at(+1.25,-1.5) ;
    \drawSecondaryVertex{(aBelow)}
    \drawSecondaryVertex{(bBelow)}
    \drawDottedSecondaryLine{(inBelow)}{(bBelow)}
    \drawResponseBlob{(aAbove)}
    \drawDottedBHLine{(inAbove)}{(outAbove)}
    \drawGravitonLine{(aBelow)}{(aAbove)}
    \drawGravitonLine{(bBelow)}{(aAbove)}
    \drawSolidSecondaryLineDirected{(bBelow)}{(outBelow)}
    \drawSecondaryVertex{(outBelow)}
  \end{tikzpicture}}_{m^{1}}
  \ +\ \underbrace{\frac12\ 
  \begin{tikzpicture}[baseline=(anchor)]
    \coordinate (aAbove) at (0, 0) ;
    \coordinate (inAbove)at (+.8,0) ;
    \coordinate (outAbove)at(-.8,0) ;
    \coordinate (aBelow) at (-.6,-1.5) ;
    \coordinate (bBelow) at (0,-1.5) ;
    \coordinate (cBelow) at (+.6,-1.5) ;
    \coordinate (inBelow)at (-1,-1.5) ;
    \coordinate (outBelow)at(+1.3,-1.5) ;
    \drawSecondaryVertex{(aBelow)}
    \drawGravitonLine{(aBelow)}{(aAbove)}
    \drawSecondaryVertex{(bBelow)}
    \drawGravitonLine{(bBelow)}{(aAbove)}
    \drawSecondaryVertex{(cBelow)}
    \drawGravitonLine{(cBelow)}{(aAbove)}
    \drawDottedSecondaryLine{(inBelow)}{(cBelow)}
    \drawResponseBlob{(aAbove)}
    \drawDottedBHLine{(inAbove)}{(outAbove)}
    \drawSolidSecondaryLineDirected{(cBelow)}{(outBelow)}
    \drawSecondaryVertex{(outBelow)}
  \end{tikzpicture}
  \ +\ \ 
    \begin{tikzpicture}[baseline=(anchor)]
    \coordinate (aAbove) at (0, 0) ;
    \coordinate (bAbove) at (2, 0) ;
    \coordinate (outAbove)at (2.8,0) ;
    \coordinate (inAbove)at(-.8,0) ;
    \coordinate (aBelow) at (-.5,-1.5) ;
    \coordinate (bBelow) at (+.5,-1.5) ;
    \coordinate (cBelow) at (1.5,-1.5) ;
    \coordinate (dBelow) at (2.5,-1.5) ;
    \coordinate (outBelow)at (+3.2,-1.5) ;
    \coordinate (inBelow)at(-1,-1.5) ;
    \drawSecondaryVertex{(cBelow)}
    \drawSecondaryVertex{(aBelow)}
    \drawGravitonLine{(aBelow)}{(aAbove)}
    \drawSecondaryVertex{(bBelow)}
    \drawGravitonLine{(bBelow)}{(aAbove)}
    \drawSolidSecondaryLine{(bBelow)}{(cBelow)}
    \drawDottedSecondaryLine{(inBelow)}{(bBelow)}
    \drawDottedSecondaryLine{(dBelow)}{(cBelow)}
    \drawResponseBlob{(aAbove)}
    \drawResponseBlob{(bAbove)}
    \drawGravitonLine{(cBelow)}{(bAbove)}
    \drawGravitonLine{(dBelow)}{(bAbove)}
    \drawDottedBHLine{(inAbove)}{(outAbove)}
    \drawSecondaryVertex{(dBelow)}
    \drawSolidSecondaryLineDirected{(dBelow)}{(outBelow)}
    \drawSecondaryVertex{(outBelow)}
  \end{tikzpicture}}_{m^2}
  \nn
  \\&
  \hspace{-.3cm}+\ \ \underbrace{
  \begin{tikzpicture}[baseline=(anchor)]
    \coordinate (aAbove) at (0, 0) ;
    \coordinate (bAbove) at (2, 0) ;
    \coordinate (outAbove)at (2.8,0) ;
    \coordinate (inAbove)at(-.8,0) ;
    \coordinate (aBelow) at (-.5,-1.5) ;
    \coordinate (bBelow) at (+.5,-1.5) ;
    \coordinate (cBelow) at (1.5,-1.5) ;
    \coordinate (dBelow) at (2.5,-1.5) ;
    \coordinate (dBelowBelow1) at (2.4,-1.8) ;
    \coordinate (dBelowBelow2) at (2.6,-1.8) ;
    \coordinate (outBelow)at (+3.5-.15,-1.5) ;
    \coordinate (inBelow)at(-1,-1.5) ;
    \drawSecondaryVertex{(cBelow)}
    \drawSecondaryVertex{(aBelow)}
    \drawGravitonLine{(aBelow)}{(aAbove)}
    \drawSecondaryVertex{(bBelow)}
    \drawGravitonLine{(bBelow)}{(aAbove)}
    \drawSolidSecondaryLine{(bBelow)}{(cBelow)}
    \drawDottedSecondaryLine{(inBelow)}{(bBelow)}
    \drawDottedSecondaryLine{(dBelow)}{(cBelow)}
    \drawResponseBlob{(aAbove)}
    \drawResponseBlob{(bAbove)}
    \drawGravitonLine{(cBelow)}{(bAbove)}
    \drawGravitonLine{(dBelow)}{(bAbove)}
    \drawDottedBHLine{(inAbove)}{(outAbove)}
    \drawSecondaryVertex{(dBelow)}
    \draw [fatzParticleDirected] ($(.15,0)+(cBelow)$) to[in=180,out=0] (dBelowBelow1) -- (dBelowBelow2) to[in=180,out=0] (outBelow) ;
    \drawSecondaryVertex{(outBelow)}
  \end{tikzpicture}
    \ +\ \frac12\ 
  \begin{tikzpicture}[baseline=(anchor)]
    \coordinate (aAbove) at (-0.2, 0) ;
    \coordinate (bAbove) at (2.2, 0) ;
    \coordinate (mAbove) at (1, 0) ;
    \coordinate (outAbove)at (3,0) ;
    \coordinate (inAbove)at(-1,0) ;
    \coordinate (aBelow) at (-.7,-1.5) ;
    \coordinate (bBelow) at (+.3,-1.5) ;
    \coordinate (mBelow) at (1,-1.5) ;
    \coordinate (cBelow) at (1.7,-1.5) ;
    \coordinate (dBelow) at (2.7,-1.5) ;
    \coordinate (dBelowBelow1) at (2-.1,-1.8) ;
    \coordinate (dBelowBelow2) at (2.4+.1+.1,-1.8) ;
    \coordinate (outBelow) at (+3.4+.1,-1.5) ;
    \coordinate (inBelow) at(-1.2,-1.5) ;
    \drawSecondaryVertex{(cBelow)}
    \drawSecondaryVertex{(aBelow)}
    \drawSecondaryVertex{(mBelow)}
    \drawGravitonLine{(aBelow)}{(aAbove)}
    \drawSecondaryVertex{(bBelow)}
    \drawGravitonLine{(bBelow)}{(aAbove)}
    \drawGravitonLine{(mBelow)}{(mAbove)}
    \drawSolidSecondaryLine{(bBelow)}{(cBelow)}
    \drawDottedSecondaryLine{(inBelow)}{(bBelow)}
    \drawDottedSecondaryLine{(dBelow)}{(cBelow)}
    \drawResponseBlob{(aAbove)}
    \drawResponseBlob{(bAbove)}
    \drawResponseBlob{(mAbove)}
    \drawGravitonLine{(cBelow)}{(bAbove)}
    \drawGravitonLine{(dBelow)}{(bAbove)}
    \drawDottedBHLine{(inAbove)}{(outAbove)}
    \drawSecondaryVertex{(dBelow)}
    \draw [fatzParticleDirected] ($(0,0)+(mBelow)$) to[in=180,out=0] (dBelowBelow1) -- (dBelowBelow2) to[in=180,out=0] (outBelow) ;
    \drawSecondaryVertex{(outBelow)}
  \end{tikzpicture}}_{m^2}
  \ +\ \mathcal{O}\Big(\frac{m^3}{M^3}\Big)\,.
%  \nn
  \label{eq:z_SF_expansion}
\end{align}
\end{subequations}
We remind the reader that the cross on the external worldline signals the use of a non-amputated resummed propagator.
We see that the SF order is captured by the number of times the secondary worldline is ``touched’’ -- mirroring the 
mass scalings known from the flat-space WQFT expansion. Particularly, the $n$SF contribution to observables requires at most the $(n-1)$-point response function.

The second term of each line with the three-point BHR function gives rise to so-called ``memory'' effects of the waveform or impulse~\cite{Almeida:2024lbv,Porto:2024cwd}.
Equivalently, these terms of the observables are dependent on non-linear black hole perturbation theory encapsulated, again, in the three-point BHR function.

\section{Massless WQFT and the Shockwave at 0SF}\label{Section3}
For the remainder of this paper, we will apply the formalism of black hole response theory to the gravitational shockwave.
Such a spacetime may be understood as an ultra-boosted Schwarzschild black hole, and in this sense it represents a ``massless'' black hole. 
At first glance, it may seem orthogonal to our stated aim of developing the SF expansion to study this object, as the perturbative parameter $m/M$ is ill-defined in this limit.
However, our setup extends naturally to this regime.
The relevant expansion parameter is now $m/E$, the ratio of the mass of the secondary BH to the energy of the shockwave, and a systematic perturbative expansion in this parameter is perfectly well-defined.
After describing how massless BHs are accommodated in WQFT, we demonstrate how the full shockwave metric, first derived by Aichelburg and Sexl \cite{Aichelburg:1970dh}, may be obtained by summing the perturbative WQFT diagrams in eq.~\eqref{eq:perturbative-metric}.
This requires the introduction of massless Feynman rules for the WQFT, which in turn require a \emph{finite-width regulator} in addition to the standard dimensional regulator, in order to control the subtle light-cone divergencies that occur in this setting.
We then proceed with deriving the geodesics of the secondary BH, from which we derive the form of the vertices~\eqref{eq:crossbuildingblocks} for the shockwave setup.

\subsection{Massless WQFT}
To accommodate massless particles or shockwaves in the WQFT, we must take the massless limit of our worldline action. In the square-root form of eq.~\eqref{mds}, this is not possible. The Brink--Di Vecchia--Howe form~\cite{Brink:1976uf} of eq.~\eqref{S1} is ideally suited for this limit, once one  rescales the einbein by a factor of $M^{-1}$, 
\begin{equation}
  S = -\frac12\int\d\tau\,\Big(e^{-1}\,g_{\mu\nu}(X)\Xd^\mu\Xd^\nu + eM^2\Big)\,,
\end{equation}
after which one easily takes $M\to0$ to wit
\begin{equation}\label{eq:massless-polyakov}
  S = -\frac12\int\d\tau\,e^{-1}\,g_{\mu\nu}(X)\Xd^\mu\Xd^\nu\,.
\end{equation}
Now, the algebraic equation of motion for the einbein of eq.~\eqref{EOMe} simply reads
\begin{equation}
  \Xd^2 = 0.
\end{equation}
Though the einbein now carries dimensions of inverse mass, as we absorbed a factor of $M^{-1}$ in it, we will see shortly that there is still a physically motivated way to gauge-fix it. One easily checks that the equation of motion for $X^\mu$ implied by eq.~\eqref{eq:massless-polyakov} coincides with the one in eq.~\eqref{eq:eom-for-X}, implying one has the flat-space solution
\begin{equation}
  X^\mu(\tau) = B^\mu + V^\mu\tau\,, \qquad V^2 = 0\,,
\end{equation}
where the second equality is the on-shell constraint. The canonical momentum is in this case
\begin{equation}
  P_\mu = \left.\frac{\delta S}{\delta\dot X^\mu}\right\vert_\text{flat} = e^{-1}V_\mu\,.
\end{equation}
From this relation, $e^{-1}$ naturally plays the role of the energy of the shockwave. Hence, introducing a frame defined by a time-like vector $v^\mu$ it is natural to gauge-fix the einbein to be the energy as measured in this frame,
\begin{equation}\label{eq:energy-gauge-fix}
  e^{-1} = P\cdot v = E\,.
\end{equation}
This implies that in the frame $v^\mu=(1,\vct{0})$, we have $V^\mu = (1,\hat{\vct{V}})$ for motion of the shockwave in the $\hat{\vct{V}}$-direction.
We conclude that for a shockwave, the gauge-fixed worldline action and the background field expansion are given by
\begin{equation}
  S = -\frac{E}{2}\int\d\tau\,g_{\mu\nu}(X)\Xd^\mu\Xd^\nu\,, \qquad X^\mu(\tau) = B^\mu + \frac{P^\mu}E\tau + Z^\mu(\tau)\,.
\end{equation}
Comparison with eq.~\eqref{monoBHS} reveals that, neglecting the non-dynamical constant term, the above action is identical to the massive one, with $E$ now simply playing the role of $M$. Hence, the massless PM Feynman rules are easily obtained directly from the massive ones given in eq.~\eqref{eq:FeynmanRulesWL} as
\begin{equation}\label{eq:massless-replacements}
  \begin{tikzpicture}[baseline=(anchor)]
    \coordinate (anchor) at (0,-.1);
    \coordinate (v) at (0,0);
    \coordinate (gOut) at (0,-1);
    \coordinate (in) at (-1,0);
    \coordinate (out3) at (1,.8);
    \coordinate (out2) at (1,0);
    \coordinate (out1) at (1,-.5);
    \drawShockwave{(in)}{(v)}
    \drawGravitonLine{(v)}{(gOut)}
    \drawPerturbativeVertex{(v)}
    \drawCurvedSolidBHLine{(v)}{(out1)}{180}{-40}
    \drawSolidBHLine{(v)}{(out2)}
    \drawCurvedSolidBHLine{(v)}{(out3)}{180}{40}
    \node at ($(out2)+(-.15,0.48)$) {\vdots};
    \node at (gOut) [below] {$h_{\mu\nu}(k)$};
    \node at (out1) [right] {$Z^{\rho_1}(\omega_1)$};
    \node at (out2) [right] {$Z^{\rho_2}(\omega_2)$};
    \node at (out3) [right] {$Z^{\rho_n}(\omega_n)$};
  \end{tikzpicture}\;=\;\left.\begin{tikzpicture}[baseline=(anchor)]
    \coordinate (anchor) at (0,-.1);
    \coordinate (v) at (0,0);
    \coordinate (gOut) at (0,-1);
    \coordinate (in) at (-1,0);
    \coordinate (out3) at (1,.8);
    \coordinate (out2) at (1,0);
    \coordinate (out1) at (1,-.5);
    \drawDottedBHLine{(in)}{(v)}
    \drawGravitonLine{(v)}{(gOut)}
    \drawPerturbativeVertex{(v)}
    \drawCurvedSolidBHLine{(v)}{(out1)}{180}{-40}
    \drawSolidBHLine{(v)}{(out2)}
    \drawCurvedSolidBHLine{(v)}{(out3)}{180}{40}
    \node at ($(out2)+(-.15,0.48)$) {\vdots};
    \node at (gOut) [below] {$h_{\mu\nu}(k)$};
    \node at (out1) [right] {$Z^{\rho_1}(\omega_1)$};
    \node at (out2) [right] {$Z^{\rho_2}(\omega_2)$};
    \node at (out3) [right] {$Z^{\rho_n}(\omega_n)$};
  \end{tikzpicture}\right|_{M\to E}.
\end{equation}
The propagator of the massless worldline fluctuation is obtained in the same way. We distinguish the massless Feynman rules from the massive ones by drawing the background worldline to evoke the image of a wave-train.

Finally, time-ordering on the shockwave can essentially be measured by $P^\mu$ instead of a time-like vector.
In particular, when working with the shockwave, we will find it useful to use a ``Mandelstam-Leibbrandt''-like prescription \cite{Mandelstam:1982cb,Leibbrandt:1983pj} for the graviton propagator instead of the causal retarded propagator:
\begin{align}
  \begin{tikzpicture}[baseline=(anchor)]
    \coordinate (anchor) at (0,-.1);
    \coordinate (x) at (-.2,0);
    \coordinate (y) at (1.0,0);
    \drawPerturbativeVertex{(x)}
    \drawPerturbativeVertex{(y)}
    \drawGravitonLineDirected{(x)}{(y)}
    \node at (x) [above left=0cm and -.7cm] {$h_\mn(-k)$};
    \node at (y) [above right=0cm and -.4cm] {$h_\rs(k)$};
  \end{tikzpicture}\;
  &=
  \frac{\i\mathcal{P}_{\mn\rs}}{k^{2}+ \iO(P\cdot k)}
  \, .
\end{align}
The retarded prescription and the Mandelstam-Leibbrandt prescription are equivalent under the following conditions:
First, they play a role only when the momentum is on-shell: $k^2=0$.
For this case, one may show that $k^0$ and $k\cdot P$ have the same sign (and thus lead to the same pole displacments) as long as $k^\mu$ is not proportional to $P^\mu$. To be precise, for this case and an on-shell momentum, we have $k^0 = \frac{k\cdot P + \bperp{k}^2 E^2/(k\cdot P)}{2E}$, from which we can directly read of that $k^0$ and $k\cdot P$ have the same sign, given that the energy $E$ of the shockwave is larger than zero.
In the following, the integration domain of our loop momentum obey these restrictions. To be prescise, we will see that the only momenta contributing to the pole prescription of active propagators are the external momenta of the probe mass, for which we assume $k\cdot P \neq 0$. We may hence safely use the Mandelstam-Leibbrandt prescription as equivalent to the causal retarded prescription.

\subsection{Light-cone coordinates and associated notation}
Before applying black hole response theory to the shockwave background, we introduce a system of light-cone coordinates and establish some associated notation that will be used throughout the rest of this paper. We begin by introducing a ``conjugate'' null momentum $\bar P^\mu$, defined through 
\begin{equation}\label{eq:conjugate-normalization}
  P\cdot\bar P = 2E^2\, ,
\end{equation}
where $E$ is the energy of the shockwave from eq.~\eqref{eq:energy-gauge-fix}.
The frame-dependence of the energy is reflected in the non-uniqueness of the choice of $\bar P^\mu$ (similarly to the choice of frame $v^\mu$ above).
Next, we introduce a pair of light-cone basis vectors $e^\mu$ and $\bar e^\mu$, chosen such that
\begin{equation}\label{eq:light-cone-basis}
  P^\mu = Ee^\mu\, , \qquad \bar P^\mu = E\bar e^\mu\, ,
\end{equation}
which implies that $\bar e\cdot e = 2$ and $\bar e^2 = e^2 = 0$. We define the corresponding light-cone coordinates
\begin{equation}
    x^- = e\cdot x =  (x^0 - x^1)\, , \qquad
    x^+ = \bar e\cdot x =  (x^0 + x^1)\, .
\end{equation}
Similarly, for a general momentum $k^\mu$ we define
\begin{equation}
  k^- = k\cdot e\, , \qquad k^+ = k\cdot \bar e\, ,
\end{equation}
such that the light-cone decomposition of $k^\mu$ and its associated integral measure read
\be
k^{\mu} = \frac{k^+e^{\mu} + k^-{\bar e}^{\mu}}{2} + k_{\perp}^{\mu}\,, \qquad \d^{D}k = \frac{\d k^{+} \d k^{-}}{2}\d^{{D-2}}\bperp{k}\,.
\ee
Here, $k_{\perp}^\mu = (0,0,\bperp{k})$ where $\bperp{k}$ is the part of $k^\mu$ lying in the $(D-2)$-dimensional subspace orthogonal to $P^\mu$ and $\bar P^\mu$.
We generally keep the light-cone indices $\pm$ raised and thus avoid the non-diagonal metric associated with this basis.

As this transverse space is defined through $\bar P^\mu$, it inherits the non-uniqueness of this vector. This is akin to the necessity for an auxiliary vector in the definition of massless polarization vectors. To be concrete, let us introduce the transverse projector
\begin{equation}
   \pi_{\mu\nu} = \eta_{\mu\nu} - \frac{2P_{(\mu}\bar P_{\nu)}}{P\cdot\bar P}\, ,
\end{equation}
with which the transverse momentum can be written $k_{\perp}^{\mu} = \pi^\mu{}_\nu k^\nu$. In the coming sections, it will be convenient to be able to choose $\bar P^\mu$ such that $\pi^{\mu\nu}$ is orthogonal to a given vector $k_i^\mu$, as this makes doing certain integrals easier and allows us to express certain tensor structures in a compact manner. This is achieved by taking,
\be
  \bar P^\mu = {2(P\cdot k_i)k_i^\mu - k_i^2P^\mu}\, ,
\ee
which leads to the ``gauge-fixed'' transverse projector
\be\label{eq:gauge-fixed-projector}
  \pi_i^{\mn}=\pi^{\mu\nu}(k_i) = \eta^{\mu\nu} + \frac{k_i^2 P^\mu P^\nu - 2(P\cdot k_i)k_i^{(\mu}P^{\nu)}}{(P\cdot k_i)^2}\, .
\ee
With this choice one finds $P\cdot\bar P = 2(P\cdot k_i)^2$, so that the $E$ appearing in eq.~\eqref{eq:conjugate-normalization} corresponds to the energy of the shockwave in the frame defined by $k_i^\mu$.

\subsection{Aichelburg--Sexl from WQFT}
\label{Vienna}
In our exploration of the shockwave limit of black hole response theory, it is a natural first step to sum the perturbative WQFT diagrams in eq.~\eqref{eq:perturbative-metric} and thereby obtain the Aichelburg--Sexl (AS) metric, which one may also think of as the shockwave's zeroth (static) response function. Despite eq.~\eqref{eq:perturbative-metric} containing an infinite number of diagrams, we will see that the sum truncates at 1PM.

To compute the full metric, we first need to compute the PM vertex describing the emission of a graviton from the shockwave. Using the massless replacement rules of \eqref{eq:massless-replacements}, we find it to be 
\begin{align}\label{eq:vertexH}
  \begin{tikzpicture}[baseline=(anchor)]
    \coordinate (anchor) at (0,-.6);
    \coordinate (in) at (-.7,0);
    \coordinate (out) at (.7,0);
    \coordinate (a) at (0,0);
    \coordinate (b) at (0,-1);
    \drawShockwave{(in)}{(out)};
    \drawPerturbativeVertex{(a)};
    \drawGravitonLineDirected{(a)}{(b)};
    \node at (b) [below] {$h_{\mu\nu}(k)$};
  \end{tikzpicture}\;= -\frac{\I\kappa_{D}}{2}\euler^{\I k\cdot B} \dd(P\cdot k)P^\mu P^\nu\, ,
\end{align}
with $k^\mu$ outgoing. From now on, without loss of generality, we set $B^\mu = 0$ implying that the shockwave crosses the origin at $\tau = 0$. 

Using this vertex, the leading-order (1PM) contribution to the metric is then given by the diagram
\begin{align}
  \vev{\fh_{\mu\nu}(x)}_{\mathrm{1PM}}&= \int_{k} e^{-\I k\cdot x}\;
  \begin{tikzpicture}[baseline=(anchor)]
    \coordinate (in) at (-.7,0);
    \coordinate (out) at (.7,0);
    \coordinate (a) at (0,0);
    \coordinate (b) at (0,-1);
    \drawShockwave{(in)}{(out)}
    \drawPerturbativeVertex{(a)}
    \drawGravitonLineDirected{(a)}{(b)}
    \drawPerturbativeVertex{(b)}
    \node at (b) [below] {$h_{\mu\nu}(k)$};
  \end{tikzpicture}\;= \int_{k} e^{-\I k\cdot x}\bigg(
  \frac{-\I \kappa_{D}}{2} \dd(k\cdot P) P^{\alpha}P^{\beta}\bigg)\frac{\I \mP_{\alpha\beta \mu\nu}}{k^{2}} \nn\\
  &= \frac{\kappa_{D}}{2} P_{\mu} P_{\nu} \int_{k} e^{-\I k\cdot x} \frac{\dd(k\cdot P)}{k^{2}}\, ,
\end{align}
where we used $P^2=0$ in the last step. To resolve the delta function and perform the remaining integral, we plug in the light-cone decomposition of the momentum and coordinates and obtain
\begin{align}
  \int_{k} \euler^{-\I k\cdot x} \frac{\dd(P\cdot k)}{k^{2}} &= 
  \frac1{2E}\int_{\bperp{k},k^+,k^-}
  \euler^{-\I\big(\frac{k^+x^-+k^-x^+}2 - \bperp{k}\cdot \bperp{x}\big)}
  \frac{\dd(k^-)}{k^+k^--\bperp{k}^2} \nn\\
  &= \delta(P\cdot x) \,  \int_{k_\perp} \frac{e^{\I \bperp{k}\cdot \bperp{x}}}
  {-\bperp{k}^2}
  = -\delta(P\cdot x) \, \frac{\Gamma(-\epsilon)(-x_\perp^2)^\eps}{4\pi^{1-\eps}}\,,
\end{align}
where we performed the $D-2=2-2\epsilon$ dimensional Fourier-transform in the last step. Recalling the definition of the $D$-dimensional gravitational coupling $\kappa_D$ in \eqn{eq:kD_def}, we can express the result covariantly in terms of the four-dimensional Newton's constant $G$ as
\begin{align}\label{ASfinal}
  \kappa_D\fh_\mn(x) = 4Gf(\bperp{x}^{2})\delta(P\cdot x)P_\mu P_\nu\, ,
\end{align}
where we have chosen to exhibit the dependence on $G$ through the dimensionless function 
\be\label{eq:Gf_def}
  G f(\bperp{x}^{2}) = -\Gamma(-\eps)\frac{\kappa_D^2 (\bperp{x}^2)^\eps}{32\pi^{1-\eps}} = -G\Gamma(-\eps)\bigg(\frac{\bperp{x}^2}{4\euler^\mascheroni L^2}\bigg)^\eps\, .
\ee

All higher-order contributions to the static background vanish. To explain why this is the case, we find it useful first to consider the next-to-leading order diagram,\footnote{{The vertex rules from interacting with the shockwave technically contributes an extra phase $e^{\I \ell_1 \cdot B}e^{\I \ell_2 \cdot B}=e^{\I k \cdot B}$, which can be factored out of the integral. }For simplicity, we haveomitted the Fourier transform and amputated the diagram, as these elements are immaterial for the argument.}
\begin{align}\label{ASG2}
  \begin{tikzpicture}[baseline=(anchor)]
    \coordinate (y) at (0,-.75);
    \coordinate (anchor) at (0,-.2);
    \coordinate (in2) at (-.9,.7);
    \coordinate (out2) at (.9,.7);
    \coordinate (v2) at (-0.5,.7);
    \coordinate (v1) at (0,0);
    \coordinate (v4) at (.5,.7);
    \drawShockwave{(in2)}{(v2)};
    \drawShockwave{(v2)}{(v4)};
    \drawShockwave{(v4)}{(out2)};
    \drawGravitonLineDirected{(v1)}{(y)};
    \drawGravitonLine{(v2)}{(v1)};
    \drawGravitonLine{(v4)}{(v1)};
    \drawPerturbativeVertex{(v1)};
    \drawPerturbativeVertex{(v2)};
    \drawPerturbativeVertex{(v4)};
    \node at (y) [below] {$h_{\mu\nu}(k)$};
  \end{tikzpicture}\;= \int_{\ell_1}\frac{\dd(P\cdot\ell_1)\dd(P\cdot\ell_2)P^{\mu_{1}}P^{\nu_{1}}P^{\mu_{2}}P^{\nu_{2}}\Omega_{\mu_1\nu_1\mu_2\nu_2;\mu\nu}(\ell_1,\ell_2)}{\ell_1^2\ell_2^2}\, ,
\end{align}
where $\ell_2 = k - \ell_1$. Here, one quickly realises that the numerator,
\begin{equation}
  \mathcal{N}^{(2)}_{\mu\nu}(\ell_1,\ell_2) = P^{\mu_{1}}P^{\nu_{1}}P^{\mu_{2}}P^{\nu_{2}}\Omega_{\mu_1\nu_1\mu_2\nu_2;\mu\nu}(\ell_1,\ell_2)\, ,
\end{equation}
is forced to vanish: The expression has two open indices, but is quartic in $P^\mu$. Hence, every term invariably picks up at least one of the scalar products $P^2$, $P\cdot \ell_{1}$, or $P\cdot\ell_2$. As these all vanish by virtue of the delta functions or the null property of $P^\mu$, we immediately have $\mathcal{N}^{(2)}_{\mu\nu}(\ell_1,\ell_2) = 0$. To generalise this argument to higher orders, consider that the generic diagrams we would be dealing with take the form
\begin{align}\label{sect:quickdeath}
  \begin{tikzpicture}[baseline=(anchor)]
    \coordinate (anchor) at (0,-.7);
    \coordinate (in) at (-1.4,0);
    \coordinate (out) at (1.4,0);
    \coordinate (a1) at (-1,0);
    \coordinate (a2) at (-.4,0);
    \coordinate (a3) at (1,0);
    \coordinate (b1) at ($(a1)-(0,.4)$);
    \coordinate (b2) at ($(a2)-(0,.4)$);
    \coordinate (b3) at ($(a3)-(0,.4)$);
    \coordinate (corner) at ($(b3)-(0,.6)$);
    \coordinate (c) at (0,-1);
    \coordinate (d) at (0,-1.6);
    \drawShockwave{(in)}{(out)}
    \node at (a1) [above] {\footnotesize $\ell_1$};
    \node at (a2) [above] {\footnotesize $\ell_2$};
    \node at (a3) [above] {\footnotesize $\ell_n$};
    \drawPerturbativeVertex{(a1)}
    \drawPerturbativeVertex{(a2)}
    \drawPerturbativeVertex{(a3)}
    \drawPerturbativeVertex{(b1)}
    \drawPerturbativeVertex{(b2)}
    \drawPerturbativeVertex{(b3)}
    \drawPerturbativeVertex{(c)}
    \drawGravitonLine{(a1)}{(b1)}
    \drawGravitonLine{(a2)}{(b2)}
    \drawGravitonLine{(a3)}{(b3)}
    \drawGravitonLineDirected{(c)}{(d)}
    \node at ($(a2)+(0.75,.24)$){$\cdots$};
    \draw [pattern=north east lines, thick] (b1) rectangle (corner);
    \node at ($(b1)!0.5!(corner)$) [circle, fill=white, inner sep=0pt] {\footnotesize $\Omega$};
    \node at (d) [below] {$h_{\mu\nu}(k)$};
  \end{tikzpicture}\; = \int_{\ell_1,\ldots,\ell_{n-1}}\Bigg(\prod_{i=1}^n\frac{\dd(P\cdot\ell_i)P^{\mu_i}P^{\nu_i}}{\ell_i^2}\Bigg)\Omega_{\mu_1\nu_1\cdots\mu_n\nu_n;\mu\nu}(\ell_1,\ldots,\ell_n)\, ,
\end{align}
where $k^\mu= \sum_{i=1}^{n}\ell^\mu_i$ by momentum conservation. 
The shaded rectangle represents all possible (tree-level) ways of connecting the graviton lines and thus the possible numerator structures $\Omega_{\mu_1\nu_1\cdots\mu_n\nu_n;\mu\nu}(\ell_1,\ldots,\ell_n)$.
Extending the argument from above, the numerator,
\begin{equation}
 \mathcal{N}^{(n)}_{\mu\nu}(\ell_i,\ldots,\ell_n) = \Bigg(\prod_{i=1}^n P^{\mu_i}P^{\nu_i}\Bigg)\Omega_{\mu_1\nu_1\cdots\mu_n\nu_n;\mu\nu}(\ell_1,\ldots,\ell_n)\, ,
\end{equation}
has two free indices, but contains $2n$ factors of the shockwave momentum. From this alone, it is clear that each term picks up at least $n-1$ factors of the scalar products $P\cdot P$ or $P\cdot \ell_i$. Therefore, the numerator vanishes for $n>1$, and with it, all contributions beyond 1PM to the static metric in eq.~\eqref{eq:perturbative-metric}. The shockwave metric thus truncates at 1PM order,
\begin{equation}\label{eq:metric-truncation}
  \begin{tikzpicture}[baseline=(anchor)]
    \coordinate (anchor) at (0,-.6);
    \coordinate (in) at (-.7,0);
    \coordinate (out) at (.7,0);
    \coordinate (a) at (0,0);
    \coordinate (b) at (0,-1);
    \drawShockwave{(in)}{(a)}
    \drawShockwave{(a)}{(out)}
    \drawResponseBlob{(a)}
    \drawGravitonLineDirected{(a)}{(b)}
    \drawPerturbativeVertex{(b)}
    \node at (b) [below] {$h_{\mu\nu}(k)$};
  \end{tikzpicture}\;=\;\begin{tikzpicture}[baseline=(anchor)]
    \coordinate (in) at (-.7,0);
    \coordinate (out) at (.7,0);
    \coordinate (a) at (0,0);
    \coordinate (b) at (0,-1);
    \drawShockwave{(in)}{(a)}
    \drawShockwave{(a)}{(out)}
    \drawPerturbativeVertex{(a)}
    \drawGravitonLineDirected{(a)}{(b)}
    \drawPerturbativeVertex{(b)}
    \node at (b) [below] {$h_{\mu\nu}(k)$};
  \end{tikzpicture}\,,
\end{equation}
leaving the result of \eqn{ASfinal} uncorrected by all higher-order diagrams and implying that the 1PM shockwave metric is an \emph{exact} solution of Einstein's equations.
Taking $\epsilon\to 0$ recovers the 4-dimensional shockwave metric obtained by Aichelburg and Sexl \cite{Aichelburg:1970dh} \footnote{Note that we dropped an epsilon pole independent of $x_\perp$ here, which can be removed by a gauge transformation.}
\be \label{eq:shockwave_light-conecoords}
\d s^{2} = \d x^-\, \d x^+ - \d \bperp{x}^{2}
+ 4G\delta(P\cdot x)\log\!\bigg(\frac{\bperp{x}^{2}}{4L^{2}}\bigg)(P\cdot \d x)^{2}\, .
\ee
A similar yet more involved derivation of this result using the amplitudes based KMOC formalism \cite{Kosower:2018adc} was presented in~\cite{Cristofoli:2020hnk}.

\subsection{Finite-width regularised massless Feynman rules}\label{sec:metric_regularisation}
While our argument for the vanishing of the higher-order diagrams appears solid, the alert reader might already have noticed a subtle flaw: While the numerator in \eqn{ASG2} vanishes, the integral over $\ell_{1}$ is actually divergent in a manner that is not dimensionally regulated. The issue becomes apparent by examining the scalar integral emerging from \eqn{ASG2}:
\begin{equation}\label{G2example}
  \int_{\ell_1}\frac{\dd(P\cdot \ell_{1}) \dd(P\cdot \ell_2)}{\ell_{1}^{2}\ell_2^{2}} = \frac1{2E^2}\int_{\ell_1}\frac{\dd(\ell_1^{-}) \dd(\ell_2^{-})}{\belln{1}^2\belln{2}^2} = \frac1{2E^2}\Bigg(\!\int\frac{\d\ell_1^+}{2\pi}\Bigg)\int_{\belln{1}}\frac1{\belln{1}^2\belln{2}^2}\, .
\end{equation}
In the final expression, we have a completely unregulated integral over the light-cone direction $\ell_{1}^{+}=\bar e\cdot \ell_{1}$, related to the distributional nature of the shockwave. This is similar in structure to the rapidity divergences occuring in
soft-collinear effective theory \cite{Chiu:2011qc}.
To regain control of this, and other problematic integrals, we will now introduce a finite-width regulator.

Tempering the distributional nature of the metric can be done by replacing the delta function $\delta(P\cdot x)$ with a \emph{nascent} delta function $\rho_\Lambda(P\cdot x)$ characterised by a width parameter $\Lambda^{-1}$ and satisfying 
\begin{equation}\label{eq:regulator-pos-space-cond}
   \lim_{\Lambda\to\infty}\rho_\Lambda(P\cdot x) = \delta(P\cdot x)\, .
\end{equation}
This is well-founded due to the following physical picture: Instead of considering the single shockwave localised on the null hyperplane $P\cdot x = 0$, we now imagine an arbitrary number of such shockwaves moving in parallel. These parallely moving shockwaves have identical momenta $P^\mu$, but are described by a superposition over shifted impact parameters $B_\sigma^\mu = \sigma\bar P^\mu/E$ distributed according to $\rho_\Lambda(\sigma)$ where $\sigma = P\cdot B/2$. Since they do not interact, superposing them yields a consistent solution of Einstein's equations, given by
\begin{equation}
  h_\mn^{\rm reg}(x) = \int\d\sigma\,\rho_\Lambda(\sigma)\Big(h_{\mn}(x)\big|_{P\cdot x\to P\cdot x-\sigma}\Big) = h_{\mu\nu}(x)\big|_{\delta(P\cdot x)\to\rho_\Lambda(P\cdot x)}\, .
\end{equation}

At the level of the perturbative Feynman rules for the massless particle, the implementation of this regulator has two elements. The first is the replacement
\begin{equation}
  \euler^{\I k\cdot B} \to \euler^{\I k\cdot B}\euler^{\I k\cdot\bar P\sigma/E}\,,
\end{equation}
from which follows that the $(n+1)$-point interaction of a graviton with $n$ massless fluctuations takes the schematic form
\begin{align}\label{eq:zzH}
  \begin{tikzpicture}[baseline=(anchor)]
    \coordinate (anchor) at (0,-.1);
    \coordinate (v) at (0,0);
    \coordinate (gOut) at (0,-1);
    \coordinate (in) at (-.7,0);
    \coordinate (out3) at (.9,.8);
    \coordinate (out2) at (.9,0);
    \coordinate (out1) at (.9,-.5);
    \drawShockwave{(in)}{(v)}
    \drawGravitonLine{(v)}{(gOut)}
    \drawPerturbativeVertex{(v)}
    \drawCurvedSolidBHLine{(v)}{(out1)}{180}{-40}
    \drawSolidBHLine{(v)}{(out2)}
    \drawCurvedSolidBHLine{(v)}{(out3)}{180}{40}
    \node at ($(out2)+(-.15,0.48)$) {\vdots};
    \node at (gOut) [below] {$h_{\mu\nu}(k)$};
    \node at (out1) [right] {\!$Z^{\rho_1}(\omega_1)$};
    \node at (out2) [right] {\!$Z^{\rho_2}(\omega_2)$};
    \node at (out3) [right] {\!$Z^{\rho_n}(\omega_n)$};
    \node at (in) [left] {$\sigma$};
  \end{tikzpicture}\!\! = \kappa_D\euler^{\I k\cdot B}\euler^{\I\bar P\cdot k\sigma/E}\dd\bigg(P\cdot k+E\sum_{i=1}^n\omega_i\bigg)F^{\mu\nu}_{\rho_1\cdots\rho_n}(P,k,E\omega_i)\,.
\end{align}
The second is the instruction that one must integrate on each $\sigma$ with the density $\rho_\Lambda(\sigma)$, resulting in the regulator
\begin{equation}\label{eq:Regulator}
  \int\d\sigma\,\euler^{\I K\cdot\bar P\sigma/E}\rho_\Lambda(\sigma) = \tilde\rho_\Lambda(K\cdot\bar P/E)\, .
\end{equation}
Here, $K^\mu$ is the sum of all momenta flowing out of the vertices connected by one worldline fluctuation, as one may only integrate $\sigma$ out after having contracted all solid lines. The reason for this simple structure is that all vertices connected by one worldline fluctuation carry the same $\sigma$-dependence since the propagator of the deflection $Z^\mu$ is diagonal in $\sigma$-space. On physical grounds, this follows from the statement that parallely moving shockwaves do not interact.
Due to eq.~\eqref{eq:regulator-pos-space-cond}, the Fourier transform $\tilde\rho_\Lambda(\omega)$ has the property
\begin{align}
  \lim_{\Lambda\to\infty}\tilde\rho_\Lambda(\omega) = 1\, .
\end{align}

Using the replacement rules from eq.~\eqref{eq:massless-replacements}, we now give explicit expressions for the perturbative, regulated, massless Feynman rules necessary for the computations in this paper. These are the deflection propagator,
\begin{equation}
  \begin{tikzpicture}[baseline=(anchor)]
    \coordinate (anchor) at (0,-.1);
    \coordinate (in) at (-0.6,0);
    \coordinate (out) at (1.4,0);
    \coordinate (x) at (-.2,0);
    \coordinate (y) at (1.0,0);
    \drawShockwave{(in)}{(x)}
    \drawShockwave{(y)}{(out)}
    \drawSolidBHLineDirected{(x)}{(y)}
    \drawPerturbativeVertex{(x)}
    \drawPerturbativeVertex{(y)}
    \node at (in) [left] {$\sigma$};
    \node at (out) [right] {$\sigma'$};
    \node at (x) [above left=.1cm and -.7cm] {$Z^\mu(-\omega)$};
    \node at (y) [above right=.1cm and -.4cm] {$Z^\nu(\omega)$};
  \end{tikzpicture}
  =
  \frac{-\I\eta^\mn}{E(\omega+\iO)^2}
  \delta(\sigma-\sigma')
  \, .
\end{equation}
and the one- and two-point vertices,
\begin{align}\label{eq:vertexHregularised}
  &\begin{tikzpicture}[baseline=(anchor)]
    \coordinate (anchor) at (0,-.6);
    \coordinate (in) at (-.7,0);
    \coordinate (out) at (.7,0);
    \coordinate (a) at (0,0);
    \coordinate (b) at (0,-1);
    \drawShockwave{(in)}{(out)};
    \drawPerturbativeVertex{(a)};
    \drawGravitonLine{(a)}{(b)};
    \node at (b) [below] {$h_{\mu\nu}(k)$};
    \node at (in) [left] {$\sigma$};
  \end{tikzpicture}\;= -\frac{\I\kappa_{D}}{2}\euler^{\I k\cdot B}\euler^{\I\bar P\cdot k\sigma/E}\dd(P\cdot k)P^\mu P^\nu\, , \\
  \label{eq:zH}
  &\begin{tikzpicture}[baseline=(anchor)]
    \coordinate (anchor) at (0,-.6);
    \coordinate (in) at (-.7,0);
    \coordinate (out) at (.7,0);
    \coordinate (a) at (0,0);
    \coordinate (b) at (0,-1);
    \drawShockwave{(in)}{(a)};
    \drawSolidBHLine{(a)}{(out)}
    \drawPerturbativeVertex{(a)};
    \drawGravitonLine{(a)}{(b)};
    \node at (b) [below] {$h_{\mu\nu}(k)$};
    \node at (in) [left] {$\sigma$};
    \node at (out) [right] {$Z^\rho(\omega)$};
  \end{tikzpicture}\;=
\frac{\kappa_{D}}{2}
\euler^{\I k\cdot B}\euler^{\I\bar P\cdot k\sigma/E}
\dd(P\cdot k+E \omega)
\big(
  2E \omega P^{(\mu}\delta^{\nu)}_\rho
  +
  P^\mu P^\nu k_\rho\big)\, .
\end{align}
The delta function in the deflection propagator implements the point made above, namely, that one should integrate on each $\sigma_i$ against $\rho_\Lambda(\sigma_i)$ for every time one touches the worldline. As an example, consider the emission vertex which is always isolated, so one may immediately integrate $\sigma$ out:
\begin{align}
  \int\d\sigma\,\rho_\Lambda(\sigma)\;
  \begin{tikzpicture}[baseline=(anchor)]
    \coordinate (anchor) at (0,-.6);
    \coordinate (in) at (-.7,0);
    \coordinate (out) at (.7,0);
    \coordinate (a) at (0,0);
    \coordinate (b) at (0,-1);
    \drawShockwave{(in)}{(out)};
    \drawPerturbativeVertex{(a)};
    \drawGravitonLine{(a)}{(b)};
    \node at (b) [below] {$h_{\mu\nu}(k)$};
    \node at (in) [left] {$\sigma$};
  \end{tikzpicture}\;=
-\frac{\I\kappa_{D}}{2}
\euler^{\I k\cdot B}
\tilde\rho_{\Lambda}(k\cdot \bar P/E)
\dd(P\cdot k)
P^\mu P^\nu.
\end{align}
The last point to address is the choice of density. We find a convenient choice to be
\begin{align}
  \rho_\Lambda(\sigma)
  =
  \frac{\Lambda}{2}
  \exp(-\Lambda |\sigma|)\, ,
\end{align}
as the Fourier transform is reminiscent of a propagator,
\begin{align}\label{eq:regulator_fourier}
  \tilde\rho_{\Lambda}(\omega)
  =
  \int \d \sigma\, \euler^{\I \omega\sigma}
  \frac{\Lambda}{2}
  \exp(-\Lambda |\sigma|)
  =
  \frac{\Lambda^2}{\omega^2+\Lambda^2}\, .
\end{align}
Going back to our initial example from \eqn{G2example}, our new $\Lambda$-regulator turns the divergence into  a well-defined integral
\be
  \int\frac{\d\ell^+_1}{2\pi}\;\rightarrow\;\int\frac{\d\ell^+_1}{2\pi}\frac{\Lambda^{2}}{\Lambda^{2}+(\ell_{1}^{+})^{2}}\frac{\Lambda^{2}}{\Lambda^{2}+(k^{+}-\ell_{1}^{+})^{2}} = \frac{\Lambda^{3}}{(k^{+})^{2}+4\Lambda^{2}}\, ,
\ee
which \textit{a posteriori} validates our truncation argument for the higher-order diagrams.
The use of this regulator will be crucial for the computation of the response function in section \ref{sec:2pt_response}.
One may also interpret its structure as a Pauli-Villars regularisation with imaginary mass.

Inclusion of such a regulator can conceivably introduce a scale for naïvely scaleless integrals. Here, we want to stress that the self-energy graphs on the massless worldline,
\begin{equation}\label{eq:generic-bubble}
  \begin{tikzpicture}[baseline=(anchor)]
    \coordinate (anchor) at (0,0);
    \coordinate (v1) at (-1.3,0);
    \coordinate (v2) at (0,0);
    \coordinate (in) at (-1.8,0);
    \coordinate (out3) at (.9,.8);
    \coordinate (out2) at (.9,0);
    \coordinate (out1) at (.9,-.5);
    \drawShockwave{(in)}{(v2)}
    \drawCurvedGravitonLine{(v1)}{(v2)}{-90}{-90}
    \drawPerturbativeVertex{(v2)}
    \drawPerturbativeVertex{(v1)}
    \drawCurvedSolidBHLine{(v2)}{(out1)}{180}{-40}
    \drawSolidBHLine{(v2)}{(out2)}
    \drawCurvedSolidBHLine{(v2)}{(out3)}{180}{40}
    \node at ($(out2)+(-.15,0.48)$) {\vdots};
  \end{tikzpicture}
  \;\;=\;\;
  0\;,
\end{equation}
remain scaleless and can be neglected, just as in the massive case.

\subsection{Dray--'t Hooft geodesics via WQFT}
\label{Utrecht}
Having computed the exact AS metric and introduced our finite-width regulator, the next step is naturally to compute the shockwave geodesics that the secondary BH will follow, which in WQFT are obtained directly as the Fourier transform of the one-point function of the trajectory \cite{Mogull:2025cfn}. These geodesics, which were first studied by Dray and 't Hooft \cite{Dray:1984ha}, will be used in the next section to determine the resummed vertices of the secondary worldline, cf. eq.~\eqref{eq:crossbuildingblocks}.

As described in the discussion surrounding eqs.~\eqref{eq:geodesic} and \eqref{eq:cross-recursion}, the trajectory is determined recursively from
\begin{align}\label{eq:geodesic-recursion}
  \begin{tikzpicture}[baseline={(anchor)}]
    \coordinate (inA) at (-.8,1.3);
    \coordinate (outA) at (.8,1.3);
    \coordinate (inB) at (-.8,0);
    \coordinate (outB) at (.8,0);
    \coordinate (x) at (0,0);
    \coordinate (y) at (0,1.3);
    \coordinate (anchor) at (0,.5);
    \drawResponseBlob{(y)}
    \drawShockwave{(inA)}{(outA)}
    \drawDottedSecondaryLine{(inB)}{(x)}
    \drawSolidBHLineDirected{(x)}{(outB)}
    \drawGravitonLine{(x)}{(y)}
    \drawSecondaryVertex{(x)}
    \drawPerturbativeVertex{(outB)}
    \node at (outB) [above=.1cm] {$\cz^\mu(\omega)$};
  \end{tikzpicture}\;
  =
  \;\begin{tikzpicture}[baseline={(anchor)}]
    \coordinate (inA) at (-.6,1.3);
    \coordinate (outA) at (.6,1.3);
    \coordinate (inB) at (-.6,0);
    \coordinate (outB) at (.6,0);
    \coordinate (x) at (0,0);
    \coordinate (y) at (0,1.3);
    \drawResponseBlob{(y)}
    \drawShockwave{(inA)}{(outA)}
    \draw (outB);
    \drawDottedSecondaryLine{(inB)}{(x)}
    \draw [zParticle] (x) -- (outB);
    \drawGravitonLine{(x)}{(y)}
    \drawPerturbativeVertex{(x)}
    \drawPerturbativeVertex{(outB)}
  \end{tikzpicture}\;
  +
  \;\begin{tikzpicture}[baseline={(anchor)}]
    \coordinate (inA) at (-1.4,1.3);
    \coordinate (outA) at (.6,1.3);
    \coordinate (inB) at (-1.4,0);
    \coordinate (outB) at (.6,0);
    \coordinate (x1) at (0,0);
    \coordinate (y1) at (0,1.3);
    \coordinate (x2) at (-.8,0);
    \coordinate (y2) at (-.8,1.3);
    \drawResponseBlob{(y1)}
    \drawResponseBlob{(y2)}
    \drawShockwave{(inA)}{(outA)}
    \draw (outB);
    \drawDottedSecondaryLine{(inB)}{(x2)}
    \draw [zUndirected] (x2) -- (x1);
    \draw [zParticle] (x1) -- (outB);
    \drawGravitonLine{(x1)}{(y1)}
    \drawGravitonLine{(x2)}{(y2)}
    \drawPerturbativeVertex{(x1)}
    \drawSecondaryVertex{(x2)}
    \drawPerturbativeVertex{(outB)}
  \end{tikzpicture}\;
  +
  \frac12
  \;\begin{tikzpicture}[baseline={(anchor)}]
    \coordinate (inA) at (-2.2,1.3);
    \coordinate (outA) at (.6,1.3);
    \coordinate (inB) at (-2.2,0);
    \coordinate (outB) at (.6,0);
    \coordinate (x1) at (0,0);
    \coordinate (y1) at (0,1.3);
    \coordinate (x2) at (-.8,0);
    \coordinate (y2) at (-.8,1.3);
    \coordinate (x3) at (-1.6,0);
    \coordinate (y3) at (-1.6,1.3);
    \drawResponseBlob{(y1)}
    \drawResponseBlob{(y2)}
    \drawResponseBlob{(y3)}
    \drawShockwave{(inA)}{(outA)}
    \drawDottedSecondaryLine{(inB)}{(x3)}
    \draw [zUndirected] (x2) -- (x1);
    \draw [zUndirected] (x3) to[out=-60,in=180+60] (x1);
    \draw [zParticle] (x1) -- (outB);
    \drawGravitonLine{(x1)}{(y1)}
    \drawGravitonLine{(x2)}{(y2)}
    \drawGravitonLine{(x3)}{(y3)}
    \drawPerturbativeVertex{(x1)}
    \drawSecondaryVertex{(x2)}
    \drawSecondaryVertex{(x3)}
    \drawPerturbativeVertex{(outB)}
  \end{tikzpicture}\;
  +
  \cdots\,
  ,
\end{align}
which is structurally identical to the relation provided in \cite{Hoogeveen:2025tew}. As we will prove below, the recursion truncates at 2PM in our present context, paralleling the 1PM truncation of the AS metric itself. Using eq.~\eqref{eq:geodesic-recursion}, the tree-level exactness of the metric, and the massless Feynman rules established in the previous section, we determine the 1PM contribution to be
\begin{align}
  \int_\omega\euler^{-\I\omega\tau}\;\begin{tikzpicture}[baseline=(anchor)]
    \coordinate (inA) at (-.8,0);
    \coordinate (outA) at (.8,0);
    \coordinate (inB) at (-.8,-1.2);
    \coordinate (outB) at (.8,-1.2);
    \coordinate (xA) at (0,0);
    \coordinate (xB) at (0,-1.2);
    \coordinate (anchor) at (0,-.7);
    \drawShockwave{(inA)}{(outA)}
    \drawDottedSecondaryLine{(inB)}{(xB)}
    \drawGravitonLine{(xA)}{(xB)}
    \draw [zParticle] (xB) -- (outB) node [above] {$\cz^\mu(\omega)$};
    \drawResponseBlob{(xA)}
    \drawPerturbativeVertex{(xB)}
    \drawPerturbativeVertex{(outB)}
  \end{tikzpicture}\! &= \frac{\I\kappa_D^2E}{4}\int_{\omega,q}\frac{\euler^{-\I(q\cdot b+\omega\tau)}\dd(\omega-q\cdot v)\dd(P\cdot q)(Eq^\mu-2\omega P^\mu)\Lambda^2}{q^2(\omega+\iO)^2(\Lambda^2+(\bar e\cdot q)^2)} \notag\\
  &= \Delta\bar v^{(1)\mu}\theta(\tau-\tau_0)(\tau-\tau_0) + \Delta\bar b^{\mu}\theta(\tau-\tau_0)\, ,
\end{align}
where $\tau_0 = -P\cdot b/E$ is the instant at which the geodesic crosses the shockwave, and
\be \label{eq:1PM_geodesic}
  \Delta\bar v^{(1)\mu}
  =
  -4GE\eps f(\bperp{b}^{2})
  \frac{b_\perp^\mu}{\abs{b_\perp}^2}
  \, ,
  \qquad
  \Delta\bar b^\mu
  =
  -2GP^\mu f(\bperp{b}^{2})
  \, .
\ee
{The names $\Delta\bar v^{(1)\mu}$ and $\Delta\bar b^\mu$ are chosen due to the dependence of the respective term on $\tau$.} The shift in the impact parameter $\Delta\bar b^\mu$ contains a pole in the dimensional regulator, but we will see in sec.~\ref{sec:graviton-light} that it does not contribute to observables. To obtain the stated result, one must compute the Fourier integral,
\begin{equation}\label{eq:fourier-trajectory}
  J_{\alpha,\beta}^{\mu_1\cdots\mu_m} = \int_q\frac{\euler^{-\I q\cdot(b+v\tau)}\dd(P\cdot q)q^{\mu_1}\cdots q^{\mu_m}}{q^{2\alpha}(q\cdot v+\iO)^\beta}\frac{\Lambda^2}{\Lambda^2 + (\bar e\cdot q)^2}\, ,
\end{equation}
in the limit as $\Lambda\to\infty$, which, at the level of the Fourier integral, can be taken immediately. We may confine our attention to the scalar case, as the tensor integrals follow by taking derivatives
\begin{equation}
  J_{\alpha,\beta}^{\mu_1\cdots\mu_m} = \I^m\frac{\partial}{\partial b_{(\mu_1}}\cdots\frac{\partial}{\partial b_{\mu_m)}}J_{\alpha,\beta}\, .
\end{equation}
We may further concentrate on the case where $\beta = 0$, as other $\beta\in \mathbb{N}$ cases are obtained using 
\begin{equation}
  J_{\alpha,\beta}(\tau) = (-\I)^\beta\int_{-\infty}^\tau\d\tau_1\cdots\int_{-\infty}^{\tau_{\beta-2}}\d\tau_{\beta-1}\int_{-\infty}^{\tau_{\beta-1}}\d\tau_{\beta}\,J_{\alpha,0}(\tau_\beta)\, .
\end{equation}
Introducing light-cone coordinates,
\begin{equation}
  q^\mu = \frac{q^+e^\mu + q^-\bar e^\mu}{2} + q_\perp^\mu\, , \qquad q_\perp^\mu = \pi^\mu{}_\nu(v)q^\nu\, ,
\end{equation}
where $\pi^{\mu\nu}(v)$ was defined in eq.~\eqref{eq:gauge-fixed-projector}, we obtain
\begin{equation}
  J_{\alpha,0} = \frac{1}{2E}\int_{q_\perp}\frac{\euler^{-\I q_\perp\cdot b_\perp}}{q_\perp^{2\alpha}}\int_{q^+}\euler^{-\I q^+(\tau-\tau_0)/2} = -\frac{(-1)^\alpha\Gamma(1-\alpha-\eps)}{4^\alpha\pi^{1-\eps}\Gamma(\alpha)(-b_\perp^2)^{1-\alpha-\eps}}\delta(\tau-\tau_0)\, .
\end{equation}

Proceeding to the next order using the recursive expansion~\eqref{eq:geodesic-recursion}, we determine the 2PM contribution to be
\begin{align}\label{eq:2pm-geodesic-result}
  &\int_\omega\euler^{-\I\omega\tau}\;\begin{tikzpicture}[baseline=(anchor)]
    \coordinate (inA) at (-1.3,0);
    \coordinate (outA) at (1.3,0);
    \coordinate (inB) at (-1.3,-1.2);
    \coordinate (outB) at (1.3,-1.2);
    \coordinate (xA) at (-.5,0);
    \coordinate (xB) at (-.5,-1.2);
    \coordinate (yA) at (.5,0);
    \coordinate (yB) at (.5,-1.2);
    \coordinate (anchor) at (0,-.7);
    \drawShockwave{(inA)}{(outA)}
    \drawGravitonLine{(xA)}{(xB)}
    \drawGravitonLine{(yA)}{(yB)}
    \drawDottedSecondaryLine{(inB)}{(xB)}
    \draw [zUndirected] (xB) -- (yB);
    \draw [zParticle] (yB) -- (outB) node [above] {$\cz^\mu(\omega)$};
    \drawResponseBlob{(xA)}
    \drawResponseBlob{(yA)}
    \drawPerturbativeVertex{(xB)}
    \drawPerturbativeVertex{(yB)}
    \drawPerturbativeVertex{(outB)}
  \end{tikzpicture} \notag\\[.2cm]
  &= \frac{\I\kappa_D^4E^3}{64}\int_{\omega,q}\frac{\euler^{-\I(q\cdot b+\omega\tau)}\dd(\omega-q\cdot v)\dd(P\cdot q)q^2}{(\omega+\iO)^2}\big((E q^\mu-3\omega P^\mu)\bar{I}_{1,1,2}-2P^\mu \bar{I}_{1,1,1}\big) \notag\\
  &= \Delta\bar v^{(2)\mu}\theta(\tau-\tau_0)(\tau-\tau_0)\, , \qquad \Delta\bar v^{(2)\mu} = \frac{8G^2E\eps^2f^2(b_\perp)P^\mu}{\abs{b_\perp}^2}\, .
\end{align}
The one-loop integrals required to obtain the above result are
\begin{align}\label{eq:geodesic-one-loop}
  \bar{ I}_{\nu_1,\nu_2,\nu_3}
  &=
  \int_\ell\frac{\dd(P\cdot\ell)}{\ell^{2\nu_1}(q-\ell)^{2\nu_2}(v\cdot\ell+\iO)^{\nu_3}}\frac{\Lambda^2}{\Lambda^2 + (\bar e\cdot\ell)^2}\frac{\Lambda^2}{\Lambda^2 + (\bar e\cdot(q-\ell))^2}
  \\
  &=
  \frac1{2E}\int_{\ell_\perp}\frac1{\ell_\perp^{2\nu_1}(q_\perp-\ell_\perp)^{2\nu_2}}\int_{\ell^+}\frac1{(\ell^+/2+\iO)^{\nu_3}}\frac{\Lambda^2}{\Lambda^2 + (\ell^+)^2}\frac{\Lambda^2}{\Lambda^2 + (2q\cdot v-\ell^+)^2}\, , \notag
\end{align}
where we introduced light-cone coordinates for the loop momentum and performed the integral over $\ell^-$ in the second step. We see here the necessity of our finite-width regularisation, as the contour integral over $\ell^+$ would in some cases be ill-defined without it. Integrals with $\nu_1<1$ or $\nu_2<1$ are scaleless and can thus be discarded in the first equality of eq.~\eqref{eq:2pm-geodesic-result}. Restricting to the relevant case where $\nu_1 = \nu_2 = 1$, we start by evaluating the $\ell^+$ integral by closing the contour in the upper half-plane. This yields
\begin{equation}
  \bar{I}_{1,1,\nu_3} = \frac1{2E}\int_{\ell_\perp}\frac1{\ell_\perp^2(q_\perp-\ell_\perp)^2}\frac{\Lambda^3}{2^{3-\nu_3}q\cdot v}\Bigg[\frac{(\I\Lambda)^{-\nu_3}}{q\cdot v-\I\Lambda} + \frac{(2q\cdot v+\I\Lambda)^{-\nu_3}}{q\cdot v+\I\Lambda}\Bigg]\,.
\end{equation}
Taking the limit as $\Lambda\to\infty$ leaves
\begin{equation}
  \bar{I}_{1,1,\nu_3} = \frac1{2E}\int_{\ell_\perp}\frac1{\ell_\perp^2(q_\perp-\ell_\perp)^2}\frac{(-\I)^{\nu_3}(1+\nu_3)}{2^{2-\nu_3}}\Lambda^{1-\nu_3}
\end{equation}
as the leading term. As expected, the integral is suppressed in the limit for $\nu_3 > 1$, leaving
\begin{equation}
  \bar{I}_{1,1,1} = \frac{\I(4\pi)^\eps\Gamma(-\eps)\csc\pi\eps}{4\Gamma(-2\eps)E}\frac{1}{(-q^2)^{1+\eps}}
\end{equation}
as the only requisite non-zero contribution. The result in eq.~\eqref{eq:2pm-geodesic-result} now follows immediately, since the remaining Fourier integral is a member of the family~\eqref{eq:fourier-trajectory}.

We now move to the truncation of the diagrammatic expansion. To explain the argument, it is useful to first introduce some terminology. At $n$PM order the recursion relation in \eqn{eq:geodesic-recursion} results in a sum of (amputated) diagrams
\begin{equation}
  \left.\begin{tikzpicture}[baseline={(anchor)}]
    \coordinate (inA) at (-.8,1.3);
    \coordinate (outA) at (.8,1.3);
    \coordinate (inB) at (-.8,0);
    \coordinate (outB) at (.8,0);
    \coordinate (x) at (0,0);
    \coordinate (y) at (0,1.3);
    \coordinate (anchor) at (0,.5);
    \drawResponseBlob{(y)};
    \drawShockwave{(inA)}{(outA)};
    \draw (outB) node [above] {$\cz^\mu(\omega)$};
    \drawDottedSecondaryLine{(inB)}{(x)};
    \drawSolidBHLineDirected{(x)}{(outB)};
    \drawGravitonLine{(x)}{(y)};
    \drawSecondaryVertex{(x)};
  \end{tikzpicture}\right\vert_\text{$n$PM} = \sum_{\Gamma_n}\frac{\Gamma_n^\mu}{S(\Gamma_n)}\, ,
\end{equation}
where the sum runs over $n$PM-0SF WQFT diagrams $\Gamma_n^\mu$ with $S(\Gamma_n)$ being the symmetry factor of each diagram.
Such a tree-level graph, $\Gamma_n = (\mathcal{V},\mathcal{E},n)$, is characterized by a vertex set $\mathcal{V}$, an edge set $\mathcal{E}$, and a root vertex $n\in \mathcal{V}$.
Each non-root vertex $i\in \mathcal{V}\setminus\{n\}$ has a unique parent $p(i)$, which is the first vertex on the path from $i$ to $n$. We associate to each vertex $i\in \mathcal{V}$ a loop momentum $\ell_i^\mu$, with the root momentum being given by
\begin{equation}
  \ell_n^\mu = q^\mu-\sum_{i\in \mathcal{V}\setminus\{n\}}\ell_i^\mu\, .
\end{equation}
Further, by defining a subtree $(\mathcal{V}_i,\mathcal{E}_i,i)=\Gamma_i\subseteq\Gamma$ as the rooted tree obtained by making $i\in \mathcal{V}$ the new root and discarding all ancestors of $i$, we can define a subtree momentum
\begin{equation}
  \ell_{\Gamma_i}^\mu = \sum_{j\in \mathcal{V}_i}\ell_j^\mu\, .
\end{equation}
See fig.~\ref{fig:rooted-tree-defs} for an illustration of these definitions.
Having defined these quantities, we find explicitly that\footnote{The 2PM expression in \eqn{eq:2pm-geodesic-result} also conforms to this pattern before one expands the scalar product and performs the tensor decomposition.}
\begin{align}\label{eq:geodesic-generic-expression}
  \Gamma_n^\mu = \frac{\I\kappa_D^{2n}E^{2n-1}}{4^n}&\int_{q,\ell_1,\cdots,\ell_n}\euler^{-\I q\cdot b}\dd(\omega-q\cdot v)\big(E\ell_n^\mu - 2(q\cdot v)P^\mu\big) \\
  &\times\Bigg[\prod_{i\in \mathcal{V}}\frac{\dd(P\cdot\ell_i)}{\ell_i^2}\frac{\Lambda^2}{\Lambda^2+(\bar e\cdot\ell_i)^2}\Bigg]\Bigg[\prod_{j\in \mathcal{V}\setminus\{n\}}\frac{\ell_j\cdot\ell_{p(j)}}{(v\cdot\ell_{\Gamma_j}+\iO)^2}\Bigg]\, . \notag
\end{align}
Recall that the one-loop integral $\bar{I}_{\nu_1,\nu_2,\nu_3}$ was suppressed in the regulator limit for $\nu_3 > 1$.
A completely analogous argument applies to any loop-order, implying that the integral in eq.~\eqref{eq:geodesic-generic-expression} is suppressed in the limit $\Lambda\to\infty$ as long as there is at least one squared denominator in the final bracket.
The expansion of $\ell_j\cdot\ell_{p(j)}$ in propagators will never contain the scalar products $v\cdot\ell_{\Gamma_j}$, implying that this factor never will be able to reduce any of the worldline propagators in the final bracket from a squared to a linear denominator.
Contrariwise, the scalar products $v\cdot\ell_{\Gamma_j}$ do occur in the vector decomposition of $\ell_n^\mu$.
However, the crux of the argument is that naturally there is at most one such scalar product per term.
These observations allow us to immediately conclude that eq.~\eqref{eq:geodesic-generic-expression} is suppressed at 3PM and above.
\begin{figure}[t!]
  \centering
  \begin{tikzpicture}[baseline=(anchor)]
    \coordinate (anchor) at (0,0);
    \coordinate (inA) at (-5.4,1.3);
    \coordinate (outA) at (.7,1.3);
    \coordinate (inB) at (-5.4,0);
    \coordinate (outB) at (.6,0);
    \coordinate (x1) at (0,0);
    \coordinate (y1) at (0,1.3);
    \coordinate (x2) at (-.8,0);
    \coordinate (y2) at (-.8,1.3);
    \coordinate (x3) at (-1.6,0);
    \coordinate (y3) at (-1.6,1.3);
    \coordinate (x4) at (-2.4,0);
    \coordinate (y4) at (-2.4,1.3);
    \coordinate (x5) at (-3.2,0);
    \coordinate (y5) at (-3.2,1.3);
    \coordinate (x6) at (-4,0);
    \coordinate (y6) at (-4,1.3);
    \coordinate (x7) at (-4.8,0);
    \coordinate (y7) at (-4.8,1.3);
    \drawResponseBlob{(y1)}
    \drawResponseBlob{(y2)}
    \drawResponseBlob{(y3)}
    \drawResponseBlob{(y4)}
    \drawResponseBlob{(y5)}
    \drawResponseBlob{(y6)}
    \drawResponseBlob{(y7)}
    \node at (y1) [above=.3cm] {\footnotesize 7};
    \node at (y2) [above=.3cm] {\footnotesize 6};
    \node at (y3) [above=.3cm] {\footnotesize 5};
    \node at (y4) [above=.3cm] {\footnotesize 4};
    \node at (y5) [above=.3cm] {\footnotesize 3};
    \node at (y6) [above=.3cm] {\footnotesize 2};
    \node at (y7) [above=.3cm] {\footnotesize 1};
    \drawShockwave{(inA)}{(outA)}
    \drawDottedSecondaryLine{(inB)}{(x7)}
    \draw [zParticle] (x1) -- (outB);
    \draw [zUndirected] (x2) -- (x1);
    \draw [zUndirected] (x3) -- (x2);
    \draw [zUndirected] (x4) to[out=-60,in=180+60] (x2);
    \draw [zUndirected] (x6) -- (x5) -- (x4);
    \draw [zUndirected] (x7) to[out=-60,in=180+60] (x5);
    \drawGravitonLine{(x1)}{(y1)}
    \drawGravitonLine{(x2)}{(y2)}
    \drawGravitonLine{(x3)}{(y3)}
    \drawGravitonLine{(x4)}{(y4)}
    \drawGravitonLine{(x5)}{(y5)}
    \drawGravitonLine{(x6)}{(y6)}
    \drawGravitonLine{(x7)}{(y7)}
    \drawPerturbativeVertex{(x1)}
    \drawPerturbativeVertex{(x2)}
    \drawPerturbativeVertex{(x3)}
    \drawPerturbativeVertex{(x4)}
    \drawPerturbativeVertex{(x5)}
    \drawPerturbativeVertex{(x6)}
    \drawPerturbativeVertex{(x7)}
    \node at (.6/2-5.4/2,-.8) [below] {(a)};
  \end{tikzpicture}
  \hspace{1.5cm}
  \begin{tikzpicture}[baseline=(anchor)]
    \coordinate (anchor) at (0,0);
    \coordinate (inA) at (-5.4,1.3);
    \coordinate (outA) at (-2.6,1.3);
    \coordinate (inB) at (-5.4,0);
    \coordinate (x4) at (-2.6,0);
    \coordinate (y4) at (-2.6,1.3);
    \coordinate (x5) at (-3.2,0);
    \coordinate (y5) at (-3.2,1.3);
    \coordinate (x6) at (-4,0);
    \coordinate (y6) at (-4,1.3);
    \coordinate (x7) at (-4.8,0);
    \coordinate (y7) at (-4.8,1.3);
    \drawResponseBlob{(y5)}
    \drawResponseBlob{(y6)}
    \drawResponseBlob{(y7)}
    \node at (y5) [above=.3cm] {\footnotesize 3};
    \node at (y6) [above=.3cm] {\footnotesize 2};
    \node at (y7) [above=.3cm] {\footnotesize 1};
    \drawShockwave{(inA)}{(outA)}
    \drawDottedSecondaryLine{(inB)}{(x7)}
    \draw [zParticle] (x5) -- (x4);
    \draw [zUndirected] (x6) -- (x5);
    \draw [zUndirected] (x7) to[out=-60,in=180+60] (x5);
    \drawGravitonLine{(x5)}{(y5)}
    \drawGravitonLine{(x6)}{(y6)}
    \drawGravitonLine{(x7)}{(y7)}
    \drawPerturbativeVertex{(x5)}
    \drawPerturbativeVertex{(x6)}
    \drawPerturbativeVertex{(x7)}
    \node at (-2.6/2-5.4/2,-.8) [below] {(b)};
  \end{tikzpicture}
  \caption{(a) An example of a 7PM-0SF WQFT graph $\Gamma_7$ where, e.g., $p(1) = 3$ and $p(5) = 6$. (b) The subtree $\Gamma_3 \subset \Gamma_7$, from which one sees that $\ell_{\Gamma_3} = \ell_1 + \ell_2 + \ell_3$.}
  \label{fig:rooted-tree-defs}
\end{figure}

Summarising our findings, we see that the geodesics in the AS background are piecewise straight-line trajectories, which enjoy an instantaneous shift in velocity,
\begin{equation}
  \Delta\bar v^\mu = \Delta\bar v^{(1)\mu}+\Delta\bar v^{(2)\mu}\,,
\end{equation}
and impact parameter $\Delta\bar b^\mu$ upon crossing the shockwave. Moreover, these shifts are 2PM-exact as we have shown with a WQFT diagrammatic argument. Explicitly,
\begin{equation}\label{eq:shockwave-geodesics-final}
\begin{aligned}
  \bar x^\mu(\tau)
  &=
  v^\mu\tau
  +
  b^\mu
  +
  \big(
    \Delta\bar v^\mu(\tau-\tau_0)
    +
    \Delta\bar b^\mu
  \big)\theta(\tau-\tau_0)
  \\
  &=
  (v^\mu\tau + b^\mu)\theta(\tau_0-\tau)
  +
  \big(
    v_\infty^\mu(\tau-\tau_0)
    +
    b_\infty^\mu
  \big)\theta(\tau-\tau_0)
  \, ,
\end{aligned}
\end{equation}
where
\begin{equation}
\begin{aligned}
  v_\infty^\mu
  &=
  v^\mu
  +
  \Delta\bar v^\mu
  =
  v^\mu
  -
  \frac{4GE\eps f(\bperp{b}^{2})}{\abs{b_\perp}^2}
  \Big(
    b_\perp^\mu
    -
    2G\eps f(\bperp{b}^{2})P^\mu
  \Big)
  \, ,
  \\
  b_\infty^\mu
  &=
  b^\mu
  +
  v^\mu\tau_0
  +
  \Delta\bar b^{\mu}
  =
  b^\mu
  +
  v^\mu\tau_0
  -
  2GP^\mu f(\bperp{b}^{2})
  \, .
\end{aligned}
\end{equation}
which explicitly exposes the piece-wise straight lines before and after scattering. 
%In fact, had we kept the finite-width regularisation $\Lambda$, we would have likely found a smooth trajectory instead of a kink at $\tau = \tau_0$, which would unambiguously fix $\theta(0)=1/2$.
As mentioned after \eqn{eq:1PM_geodesic}, the change in impact parameter, $\Delta b^\mu$, carries a pole in the dimensional regulator $\epsilon$ through its dependence on $f(\bperp b^2)$. This is a gauge artifact and may be absorbed by redefining the fiducial length scale $L\rightarrow L’=e^{1/2\epsilon}L$ in $f(\bperp{b}^2)$. We thus write
\begin{equation}
\begin{aligned}
  f(\bperp x^2)
  =\,&
  \frac1\epsilon+\log\frac{\bperp x^2}{4L^2}
  +
  \mathcal O(\eps)
  =
  \log\frac{\bperp x^2}{4L'^2}
  +
  \mathcal O(\eps)
  \\
  \Delta \hat b^\mu=\,&-2G(1/\epsilon+\log(\bperp x^2/4L^2))+\mathcal O(\epsilon)=\,-2G\log(\bperp x^2/4L'^2)+\mathcal O(\epsilon)
  \label{eq:b_infty_gauge_trans}
  \, .
\end{aligned}
\end{equation}
With this, the shockwave geodesic of \eqn{eq:shockwave-geodesics-final} agrees with that found in the literature \cite{Dray:1984ha,Steinbauer:1997dw}.
Another consistency check is afforded by the on-shell constraint,
\begin{equation}
  \bar g_{\mu\nu}(\bar x)\dot{\bar x}^\mu\dot{\bar x}^\nu
  =
  1
  \, ,
\end{equation}
which we find is satisfied everywhere except at $\tau = \tau_0$; we attribute this to the singular character of the spacetime at the crossing point. 
To separately verify eq.~\eqref{eq:shockwave-geodesics-final}, we also solved the geodesic equation~\eqref{eq:eom-for-X} directly and found perfect agreement.

\subsection{Resummed Feynman rules for the secondary black hole}
\label{sec:graviton-light}
With the geodesics in hand, we now turn to the derivation of the resummed vertices governing the interaction of
the secondary BH with the gravitational shockwave. 
This connection is done with the energy-momentum tensor of the secondary object $\mT^\mn_{\rm sec}(x)$ given explicitly in eq.~\eqref{eq:secTmn}.

As in Eq.~\eqref{eq:backgroundSFexpansion}, we expand the trajectory around geodesic motion as $x^\mu(\tau) = \bar x^\mu(\tau) + \cz^\mu(\tau)$ with $\bar x^\mu(\tau)$ given explicitly in eq.~\eqref{eq:shockwave-geodesics-final}.
To derive the vertices, we first transform the energy-momentum tensor to momentum space:
\begin{align}\label{eq:TmnOfK}
  \mT^\mn_{\rm sec}(k)
  =
  \int \d^D x
  \,
  \euler^{\i k\cdot x}
  \,
  \mT^\mn_{\rm sec}(x)
  =
  \frac{m}2
  \int \d\tau
  \,
  \euler^{\i k\cdot x(\tau)}
  \,
  \dot x^\mu(\tau)\dot x^\nu(\tau)
  \ .
\end{align}
Note, importantly, that the variable $x^\mu(\tau)$ in the final equality is the trajectory of the secondary which, naturally, is different from the space-time variable $x$.
Expanding $\mT^\mn_{\rm sec}(k)$ to zeroth and first order in the deflection $\cz^\mu$, we get
\begin{subequations}\label{eq:vertices_secondary}
\begin{align}
  \mT_{\rm sec}^\mn(k) \Big\vert_{\cz^0} 
  &= 
  \frac{m}{2}
  \int\d\tau
  \,
  \euler^{\I k\cdot\bar x(\tau)}\dot{\bar x}^\mu(\tau)\dot{\bar x}^\nu(\tau)
  \, , 
  \\
  \mT_{\rm sec}^\mn(k) \Big\vert_{\cz^1} 
  &= 
  \frac{\I m}{2}
  \int_{\omega}
  \cz^\rho(-\omega) 
  \int\d\tau
  \,
  \euler^{\I k\cdot\bar x(\tau)+\I\omega\tau}\Big(2\omega\dot{\bar x}^{(\mu}(\tau)\delta^{\nu)}_\rho + k_\rho\dot{\bar x}^\mu(\tau)\dot{\bar x}^\nu(\tau)\Big)\, .
\end{align}
\end{subequations}
From this, it is clear that one obtains the resummed interaction vertex for the secondary BH with $n$ deflections from the perturbative WQFT Feynman rules \eqn{eq:FeynmanRulesWL} by making the replacement
\begin{equation}\label{eq:resummed-vertices-replacement}
  \euler^{\I k\cdot b}\I\dd(k\cdot v + \Omega)\left\{\begin{array}{c}v^\mu v^\nu \\[.4em] v^\mu\end{array}\right\} \to \int\d\tau\,\euler^{\I k\cdot\bar x(\tau)+\I\Omega\tau}\left\{\begin{array}{c}\dot{\bar x}^{\mu}(\tau)\dot{\bar x}^{\nu}(\tau) \\[.4em] \dot{\bar x}^{\mu}(\tau)\end{array}\right\}\, ,
\end{equation}
where $\Omega$ is the sum of the deflection energies.
As the time derivative of the shockwave geodesic is
\begin{align}\label{eq:geodesic_velocity}
  \dot{\bar x}^\mu(\tau) &= v^\mu\theta(\tau_0-\tau) + v_\infty^\mu\theta(\tau-\tau_0) + \Delta\bar b^\mu\delta(\tau-\tau_0)\, ,
\end{align}
the integrals in eq.~\eqref{eq:resummed-vertices-replacement} evaluate to
\begin{align}
  &\int\d\tau\,\euler^{\I k\cdot\bar x(\tau)+\I\Omega\tau}\dot{\bar x}^{\mu}(\tau) \notag\\
  &= -\I\euler^{\I(k\cdot v+\Omega)\tau_0+\I k\cdot b}\bigg[\frac{v^\mu}{k\cdot v + \Omega - \iO} - \frac{\euler^{\I k\cdot\Delta\bar b}v_\infty^\mu}{k\cdot v_\infty + \Omega + \iO}\bigg] + \euler^{\I k\cdot\bar x(\tau_0)+\I\Omega\tau_0}\Delta\bar b^\mu\, ,
\end{align}
and
\begin{align}
  \int\d\tau\,\euler^{\I k\cdot\bar x(\tau)+\I\Omega\tau}&\dot{\bar x}^{\mu}(\tau)\dot{\bar x}^{\nu}(\tau) = -\I\euler^{\I(k\cdot v+\Omega)\tau_0+\I k\cdot b}\bigg[\frac{v^\mu v^\nu}{k\cdot v + \Omega - \iO} - \frac{\euler^{\I k\cdot\Delta\bar b}v_\infty^\mu v_\infty^\nu}{k\cdot v_\infty + \Omega + \iO}\bigg] \notag\\
  &+ \euler^{\I k\cdot\bar x(\tau_0)+\I\Omega\tau_0}\Big(2v^{(\mu}\Delta\bar b^{\nu)} + 2\theta(0)\Delta\bar v^{(\mu}\Delta\bar b^{\nu)} + \delta(0)\Delta\bar b^\mu \Delta\bar b^\nu\Big)\, .
\end{align}
We see the haunting presence of  $\delta(0)$ in the last term of the second integral. Furthermore, the ambiguous value $\theta(0)$ appears. Both are attributed to the $\delta$-distribution of the shockwave, particularly the impulsive kick $\Delta \bar b^\mu \delta(\tau-\tau_0)$ of the geodesic velocity in eq.~\eqref{eq:geodesic_velocity}.
The $\delta(0)$ term \emph{never} contributes to observables since, as we shall see, the response function of the shockwave is always transverse to $\Delta \bar b^\mu\Delta \bar b^\nu \propto P^\mu P^\nu$. For this reason, we completely drop this term going forward.
The ambiguous $\theta(0)$ term would be tempered by keeping the $\Lambda$ regulator finite throughout the computation, thereby smearing out the $\theta$-function. We hence expect the value of $\theta(0)$ to be uniquely fixed, as in refs.~\cite{Balasin:1996mq, Steinbauer:1997dw}. The full analysis is to be carried out when observables are computed, and we keep $\theta(0)$ symbolic for the time being.

Using the above, we find the resummed emission vertex and two-point vertex
\begin{align}
  \begin{tikzpicture}[baseline=(anchor)]
    \coordinate (anchor) at (0,.5);
    \coordinate (in) at (-1,0);
    \coordinate (out) at (1,0);
    \coordinate (a) at (0,0);
    \coordinate (b) at (0,1);
    \drawDottedSecondaryLine{(in)}{(out)};
    \drawGravitonLine{(a)}{(b)};
    \node at (b) [above] {$\ch_{\mu\nu}(k)$};
    \drawSecondaryVertex{(a)};
  \end{tikzpicture}
  &\;=\;
  \begin{aligned}[t]
     &-\frac{m\kappa_D}{2}\euler^{\I k\cdot(v\tau_0+b)}\bigg(\frac{v^\mu v^\nu}{v\cdot k-\iO}-\frac{\euler^{\I k\cdot\Delta\bar b}v_\infty^\mu v_\infty^\nu}{v_\infty\cdot k+\iO}\bigg)\\
  &- \I m\kappa_D\euler^{\I k\cdot\bar x(\tau_0)}\big(v^{(\mu}\Delta\bar b^{\nu)} + \theta(0)\Delta\bar v^{(\mu}\Delta\bar b^{\nu)}\big)\, ,
\end{aligned}\\[10pt]
\begin{tikzpicture}[baseline=(anchor)]
    \coordinate (anchor) at (0,.5);
    \coordinate (in) at (-1,0);
    \coordinate (out) at (1,0);
    \coordinate (a) at (0,0);
    \coordinate (b) at (0,1);
    \drawDottedSecondaryLine{(in)}{(a)};
    \drawSolidBHLine{(a)}{(out)};
    \drawGravitonLine{(a)}{(b)};
    \node at (b) [above] {$\ch_{\mu\nu}(k)$};
    \drawSecondaryVertex{(a)};
    \node at (out) [below] {$\cz^\rho(\omega)$};
  \end{tikzpicture}
  &\;=\;
  \begin{aligned}[t]
     &-\frac{\I m\kappa_D}{2}\euler^{\I k\cdot(v\tau_0+b)}\Bigg[2\omega\bigg(\frac{v^{(\mu}\delta^{\nu)}_\rho}{v\cdot k+\omega-\iO}-\frac{\euler^{\I k\cdot\Delta\bar b}v_\infty^{(\mu}\delta^{\nu)}_\rho}{v_\infty\cdot k+\omega+\iO}\bigg)\\
  &+ k_\rho\bigg(\frac{v^\mu v^\nu}{v\cdot k+\omega-\iO}-\frac{\euler^{\I k\cdot\Delta\bar b}v_\infty^\mu v_\infty^\nu}{v_\infty\cdot k+\omega+\iO}\bigg)\Bigg] \\ &- m\kappa_D\euler^{\I k\cdot\bar x(\tau_0)+\I\omega\tau_0}\big(\omega\Delta\bar b^{(\mu}\delta^{\nu)}_\rho + k_\rho v^{(\mu}\Delta\bar b^{\nu)} + \theta(0)k_\rho\Delta\bar v^{(\mu}\Delta\bar b^{\nu)}\big)\, .
\end{aligned}
\raisetag{4.1em}
\end{align}
When $\Delta\bar v^\mu = \Delta\bar b^\mu = 0$, these vertices reduce to the usual perturbative vertices, as
\begin{equation}
  \I\dd(x)
  =
  \frac{1}{x-\iO}
  -
  \frac{1}{x+\iO}
  \, .
\end{equation}

\section{Computation of Exact Two-Point Shockwave Response}\label{sec:2pt_response}

The metric and geodesics computed above represent the leading (0SF) contributions to dynamics in the shockwave background in the self-force expansion.
In this section, we provide the basis to extend calculations to the 1SF order by computing the full two-point shockwave response function appearing in \eqn{eq:SF_expansion}.
Diagrammatically it is given by
\begin{align}
  \mR_{\mu\nu\rho\sigma}(k_{1},k_{2}) 
  = 
  \vev{h_{\rho\sigma}(k_2)h_{\mu\nu}(k_1)}_{\text{con}}
  = 
  \;\begin{tikzpicture}[baseline=(anchor)]
    \coordinate (anchor) at (0,-.5);
    \coordinate (inA) at (-1,0);
    \coordinate (outA) at (1,0);
    \coordinate (v) at (0,0);
    \coordinate (c1) at (-1,-1);
    \coordinate (c2) at (1,-1);
    \drawShockwave{(inA)}{(outA)};
    \drawGravitonLine{(c1)}{(v)}
    \drawPerturbativeVertex{(c1)}
    \drawGravitonLineDirected{(v)}{(c2)}
    \drawPerturbativeVertex{(c2)}
    \drawResponseBlob{(v)}
    \node [below right=.1cm and -.4cm] at (c2) {$h_\rs(k_2)$};
    \node [below left=.1cm and -.5cm] at (c1) {$h_\mn(k_1)$};
  \end{tikzpicture}.
\end{align}
The response may be computed as the sum of all perturbative PM diagrams contributing to the geometric series arising from \eqn{eq:SchwingerDyson} (including recoil effects). It corresponds to the propagator of a graviton on the non-trivial shockwave background. This motivates the split of the response function $\mR_{\mn\ab}(k_1,k_2)$ into free propagation and an interaction labelled by $\Sigma^{\mn\ab}(k_1,k_2)$:
\begin{align}\label{eq:sigmasplit}
  \begin{tikzpicture}[baseline=(anchor)]
    \coordinate (anchor) at (0,-.5);
    \coordinate (inA) at (-1,0);
    \coordinate (outA) at (1,0);
    \coordinate (v) at (0,0);
    \coordinate (c1) at (-1,-1);
    \coordinate (c2) at (1,-1);
    \drawShockwave{(inA)}{(outA)};
    \drawGravitonLine{(c1)}{(v)}
    \drawPerturbativeVertex{(c1)}
    \drawGravitonLineDirected{(v)}{(c2)}
    \drawPerturbativeVertex{(c2)}
    \drawResponseBlob{(v)}
  \end{tikzpicture}
  \quad = \quad
  \begin{tikzpicture}[baseline=(anchor)]
    \coordinate (anchor) at (0,-.5);
    \coordinate (inA) at (-1,0);
    \coordinate (outA) at (1,0);
    \coordinate (c1) at (-1,-1);
    \coordinate (c2) at (1,-1);
    \drawShockwave{(inA)}{(outA)};
    \drawPerturbativeVertex{(c1)}
    \draw [photonTestDirected] (c1) to[out=40,in=140] (c2);
    \drawPerturbativeVertex{(c2)}
  \end{tikzpicture}
  \quad + \quad
  \begin{tikzpicture}[baseline=(anchor)]
    \coordinate (anchor) at (0,-.5);
    \coordinate (inA) at (-1,0);
    \coordinate (outA) at (1,0);
    \coordinate (v) at (0,0);
    \coordinate (c1) at (-1,-1);
    \coordinate (c2) at (1,-1);
    \drawShockwave{(inA)}{(outA)};
    \drawGravitonLine{(c1)}{(v)}
    \drawPerturbativeVertex{(c1)}
    \drawGravitonLineDirected{(v)}{(c2)}
    \drawPerturbativeVertex{(c2)}
    \draw [fill=white, thick] (v) circle (.3);
    \node at (v) {$\I\Sigma$} ;
  \end{tikzpicture}
  \, .
\end{align}
More explicitly, we define $\Sigma^{\mn\ab}(k_1,k_2)$ through the equation,
\begin{align}\label{eq:response_w_props}
  \mR_{\mn\ab}(k_1,k_2)
  =
  \frac{
    \I\mP_{\mn\ab}
    \dd(k_1-k_2)
  }{
    k_1^2+\iO (P\cdot k_1)
  }
  +
    \frac{
    \I\mP_{\mn\sigma\rho}
  }{
    k_1^2+\iO (P\cdot k_1)
  }
  \I\Sigma^{\sigma\rho\kappa\delta}(k_1,k_2)
  \frac{
    \I\mP_{\kappa\delta\ab}
  }{
    k_2^2+\iO (P\cdot k_2)
  }
  \, ,
\end{align}
where, as shown, $\Sigma^{\mn\ab}(k_1,k_2)$ is amputated.
The interpretation is as follows:  Either the graviton travels freely through space (first term) or the graviton travels freely, then hits the shockwave plane where it interacts gravitationally, before continuing freely after the interaction (second term).

Remarkably, we are able to compute the interaction response $\Sigma^{\mn\ab}(k_1,k_2)$ exactly in $G$ by carrying out an infinite sum of PM-expanded diagrams.
What comes to our aid once more, is that the null property of the shockwave momentum $P^{\mu}$ leads to a cancellation of a large class of graphs, leaving us at higher PM orders only with ladder-type contributions that we are able to resum.

\subsection{Diagrammatic expansion of the shockwave response}\label{sec:diag_simplicity_shockwave}
To obtain the shockwave response, we want to resum all perturbative diagrams generated by the Schwinger Dyson equation~\eqref{eq:SchwingerDyson}. For this purpose, we first focus on the ``self-energy'' i.e.~all one-particle-irreducible diagrams contained in the square brackets of \eqn{eq:SchwingerDyson}. 
Amazingly, and in analogy to the Aichelburg-Sexl metric computation, only a few leading-order PM diagrams are nonzero.
More specifically, as we will argue in a moment, we find
\begin{align}\label{eq:selfEnergySimplification}
  \begin{tikzpicture}[baseline=(anchor)]
    \coordinate (anchor) at (0,.1);
    %incoming/outgoing upper coordinates
    \coordinate (in2) at (-1.5,.7);
    \coordinate (out2) at (1.5,.7);
    % central vertex
    \coordinate (v) at (0,-.2);
    %outgoing gravitons
    \coordinate (x) at (-1,-.2);
    \coordinate (y) at (1,-.2);
    % upper vertices
    \coordinate (v2) at (-.7,.7);
    \coordinate (v3) at (-.1,.7);
    \coordinate (v4) at (.7,.7);
    % dot position
    \coordinate (dots) at ($(v3)!.5!(v4)$);
    %draw propagators
    \draw [photonTest] (x) -- (v);
    \draw [photonTest] (v) -- (y);
    \draw [photonTest] (v2) -- (v);
    \draw [photonTest] (v3) -- (v);
    \draw [photonTest] (v4) -- (v);
    \drawShockwave{(in2)}{(out2)}
    % draw circles
    \drawPerturbativeVertex{(v)}
    \begin{pgfonlayer}{foreground}
    \draw [fill=black, thick] (v2) circle (.055) node [above] {$1$};
    \draw [fill=black, thick] (v3) circle (.055) node [above] {$2$};
    \draw [fill=black, thick] (v4) circle (.055) node [above] {$n$};
    \end{pgfonlayer}
    \node[above=2.5pt] at (dots) {...};
  \end{tikzpicture}
  \ 
  =
  \ 
  0\quad\text{for }n\geq3 
  \, .
\end{align}
In this way, the ``self energy'' of eq.~\eqref{eq:SchwingerDyson} is 3PM-exact and the geometric series expansion of the shockwave response function reduces to
\be\label{fatgravitonexp}
  \begin{tikzpicture}[baseline={(0,-.4)}]
    \coordinate (anchor) at (0,-.5);
    \coordinate (inA) at (-.7,0);
    \coordinate (outA) at (.7,0);
    \coordinate (v) at (0,0);
    \coordinate (c1) at (-.7,-.7);
    \coordinate (c2) at (.7,-.7);
    \drawShockwave{(inA)}{(outA)};
    \drawGravitonLine{(c1)}{(v)}
    \drawGravitonLineDirected{(v)}{(c2)}
    \draw [fill=white, thick] (v) circle (.3);
    \node at (v) {$\I\Sigma$} ;
  \end{tikzpicture}
  \ \ =\ \ 
  \sum_{n=1}^{\infty} \Biggl[\,\,\,
  \begin{tikzpicture}[baseline={(0,-.4)}]
    \coordinate (currentLocation);
    \drawAbsorb
    \drawWL
    \drawEmitDirected
  \end{tikzpicture}\;
  +
  \;\begin{tikzpicture}[baseline={(0,.3)}]
    \coordinate (x) at (-.7,0);
    \coordinate (y) at (.7,0);
    \coordinate (in2) at (-.7,.7);
    \coordinate (out2) at (.7,.7);
    \coordinate (v2) at (0,.7);
    \coordinate (v1) at (0,0);
    \draw [photonTest] (x) -- (v1);
    %\draw [photonTest] (v1) -- (y);
    \drawGravitonLineDirected{(v1)}{(y)};
    \draw [photonTest] (v2) -- (v1);
    \drawShockwave{(in2)}{(out2)};
    \drawPerturbativeVertex{(v1)}
    \drawPerturbativeVertex{(v2)}
  \end{tikzpicture}\;
  +
  \frac{1}{2}
  \;\begin{tikzpicture}[baseline={(0,.3)}]
    \coordinate (x) at (-.7,0);
    \coordinate (y) at (.7,0);
    \coordinate (in2) at (-.7,.7);
    \coordinate (out2) at (.7,.7);
    \coordinate (v2) at (-0.25,.7);
    \coordinate (v1) at (0,0);
    \coordinate (v4) at (.25,.7);
    \draw [photonTest] (x) -- (v1);
    %\draw [photonTest] (v1) -- (y);
    \drawGravitonLineDirected{(v1)}{(y)};
    \draw [photonTest] (v2) -- (v1);
    \draw [photonTest] (v4) -- (v1);
    \drawShockwave{(in2)}{(out2)};
    \drawPerturbativeVertex{(v1)}
    \drawPerturbativeVertex{(v2)}
    \drawPerturbativeVertex{(v4)}
\end{tikzpicture} 
\,\,\,\Biggr] ^{n}
\, .
\ee
We can prove eq.~\eqref{eq:selfEnergySimplification} analogously to the proof of the 1PM-exactness of the one-point function $\vev{h_{\mu\nu}}$ in section \ref{Vienna}: Looking at a contribution to the ``self energy'' from a vertex with multiplicity $(n+2)$, one has an $(n-1)$-loop integral,
\be
  \int_{\ell_1,\ldots,\ell_{n-1}}\frac{\mathcal{N}^{\mu\nu\rho\sigma}_{\mu_1\nu_1\cdots\mu_{n-1}\nu_{n-1}}(k_1,k_2,\ell_1,\ldots,\ell_{n-1})}{(q - \ell_{1\cdots(n-1)})^2 [(q^+-\ell_{1\cdots (n-1)}^+)^2 + \Lambda^2]}\bigg[\prod_{i=1}^{n}\frac{\dd(P\cdot\ell_i)P^{\mu_i}P^{\nu_i}}{\ell_i^2 [{\ell_i^+}^2 + \Lambda^2]}\bigg],
\ee
where $\ell_{1\cdots(n-1)}^\mu = \sum_{i=1}^{n-1}\ell_i^\mu$ and $q^{\mu}=k_2^{\mu}-k_1^{\mu}$.
As the numerator $\mathcal{N}$ has four free indices and is quadratic in the momenta, we can at most get a non-vanishing result with six factors of $P^\mu$.
At eight or more, we invariably encounter $P^2$ or $P\cdot\ell_i$ in every term.
This leaves the cubic, quartic and quintic vertices.
Plugging in the explicit forms of these vertices, we observe a further simplification: The five-point diagram evaluates to zero in de Donder gauge. 
This leaves the non-trivial cases shown in \eqn{fatgravitonexp}.

\begin{figure}[t]
  \def\xcor{.98} % tiny rescaling in x to make diagrams fit width of page
  \centering
  \subcaptionbox{}{
    \begin{tikzpicture}[baseline=(currentLocation)]
      \coordinate (currentLocation) at (0,0);
      \drawPropagationMedium
      \drawQuartic
      \drawPropagationLong
      \drawCubic
      \drawPropagationShortDirected
    \end{tikzpicture}
  }\hfill
  \subcaptionbox{}{
    \begin{tikzpicture}[baseline=(currentLocation)]
      \coordinate (currentLocation) at (0,0);
      \drawPropagationMedium
      \drawQuartic
      \drawPropagationLong
      \drawQuartic
      \drawPropagationMediumDirected
    \end{tikzpicture}
  }\hfill
  \subcaptionbox{}{
    \begin{tikzpicture}[baseline=(currentLocation)]
      \coordinate (currentLocation) at (0,0);
      \drawPropagationShort
      \drawCubic
      \drawPropagationShort
      \drawCubic
      \drawPropagationShort
      \drawCubic
      \drawAbsorb
      \drawWL
      \drawEmitDirected
    \end{tikzpicture}
  }\hfill
  \subcaptionbox{}{
    \begin{tikzpicture}[baseline=(currentLocation)]
      \coordinate (currentLocation) at (0,0);
      \drawPropagationShort
      \drawCubic
      \drawPropagationShort
      \drawCubic
      \drawAbsorb
      \drawWL
      \drawEmit
      \drawCubic
      \drawPropagationShortDirected
    \end{tikzpicture}
  }\\[1.5em] 
  \subcaptionbox{}{
    \begin{tikzpicture}[scale=0.98,baseline=(currentLocation)]
      \coordinate (currentLocation) at (0,0);
      \drawAbsorb
      \drawWL
      \drawEmit
      \drawCubic
      \drawPropagationShort
      \drawCubic
      \drawAbsorb
      \drawWL
      \drawEmitDirected
    \end{tikzpicture}
  }\hfill
  \subcaptionbox{}{
    \begin{tikzpicture}[scale=0.98,baseline=(currentLocation)]
      \coordinate (currentLocation) at (0,0);
      \drawPropagationShort
      \drawCubic
      \drawAbsorb
      \drawWL
      \drawGirlande
      \drawWL
      \drawEmitDirected
    \end{tikzpicture}
  }\hfill
  \subcaptionbox{}{
    \begin{tikzpicture}[scale=0.98,baseline=(currentLocation)]
      \coordinate (currentLocation) at (0,0);
      \drawAbsorb
      \drawWL
      \drawGirlande
      \drawWL
      \drawGirlande
      \drawWL
      \drawEmitDirected
    \end{tikzpicture}
  }\hfill
  \caption{All these diagrams vanish off-shell purely on account of numerator algebra and the fundamental fact that $P^2 = 0$. As they vanish off-shell, we are additionally guaranteed that any diagram containing these as subdiagrams also vanishes. Importantly, this together with eq.~\eqref{eq:selfEnergySimplification} implies that only ladder diagrams contribute at 4PM and beyond.}
  \label{fig:vanishing-diagrams}
\end{figure}

Let us then turn to iterations of the self-energy and recoils, i.e.~powers of $n\ge2$ in eq.~\eqref{fatgravitonexp}.
At this stage, further simplifications arise due to $P^2=0$, summarised by the vanishing iteration diagrams in fig.~\ref{fig:vanishing-diagrams}.
These relations are in fact so strong that at each PM order beyond 3PM only one non-zero diagram appears, namely a \textit{ladder diagram}.
The resulting perturbative expansion to all orders in $G$ is shown in fig.~\ref{fig:response_PM_diagrams}, organised according to PM counting.

Intriguingly, the $(n-1)$-loop ladder diagrams of the final line of fig.~\ref{fig:response_PM_diagrams} can be computed at all orders.
They exhibit a recursive structure that exponentiates; the computational details underlying this result are presented in sec.~\ref{sec:ladder_structure}.
This observation suggest that the exact, PM-resummed response $\Sigma^{\mn\ab}$ can be decomposed into two distinct parts: (1) a contribution $\Sigma^{\mn\ab}_\mathrm{sum}$ containing infinite PM orders in a resummed form, and (2) a 3PM-exact remainder piece $\Sigma^{\mn\ab}_\mathrm{rem}$.
Accordingly, the PM-resummed off-shell response function is written as
\begin{equation}\label{eq:response_sum_rem_split}
  \Sigma^{\mu\nu\ab}(k_1,k_2)=\Sigma_\mathrm{sum}^{\mu\nu\ab}(k_1,k_2)+\Sigma^{\mu\nu\ab}_\mathrm{rem}(k_1,k_2)\, .
\end{equation}
The resummed piece $\Sigma_\mathrm{sum}^{\mu\nu\ab}(k_1,k_2)$ captures the full tower of ladder contributions and takes an impressively simple exponential form.
The remainder $\Sigma_\mathrm{rem}^{\mn\ab}(k_1,k_2)$, on the other hand, collects the low-order contributions that do not follow the pattern observed for higher-PM ladder diagrams.
From the diagrammatic expansion in fig.~\ref{fig:response_PM_diagrams}, it is clear that such terms only arise up to 3PM order, i.e.~up to the order where recoil diagrams contribute. Notably, the ladder diagrams themselves do not yet display the simple high-order structure at 1PM, 2PM, and 3PM. 
However, the full sum of diagrams (ladder and recoil) at each of these PM order exhibits considerably better properties than each individual piece.
\begin{figure}[t!]
  \centering
\begin{align}
  &\begin{tikzpicture}[baseline=(anchor)]
      \coordinate (anchor) at (0,-.5);
      \coordinate (inA) at (-.7,0);
      \coordinate (outA) at (.7,0);
      \coordinate (v) at (0,0);
      \coordinate (c1) at (-.7,-.7);
      \coordinate (c2) at (.7,-.7);
      \drawShockwave{(inA)}{(outA)};
      \drawGravitonLine{(c1)}{(v)}
      \drawGravitonLineDirected{(v)}{(c2)}
      \draw [fill=white, thick] (v) circle (.3);
      \node at (v) {$\I\Sigma$} ;
    \end{tikzpicture}\;=\;\begin{tikzpicture}[scale=1,baseline={(0,-0.5)}]
      \coordinate (currentLocation) at (0,0);
      \drawPropagationShort
      \drawCubic
      \drawPropagationShortDirected
    \end{tikzpicture}
    \;+\;
    \begin{tikzpicture}[scale=1,baseline={(0,-0.5)}]
      \coordinate (currentLocation) at (0,0);
      \drawAbsorb
      \drawWL
      \drawEmitDirected
    \end{tikzpicture}
    & \text{(1PM)}\nonumber\\[15pt]
  &+\;
  \begin{tikzpicture}[scale=1,baseline={(0,-0.5)}]
  \coordinate (currentLocation) at (0,0);
  \drawPropagationShort
  \drawCubic
  \drawPropagationMedium
  \drawCubic
  \drawPropagationShortDirected
  \end{tikzpicture}
  \;+\;\frac{1}{2}\;
  \begin{tikzpicture}[scale=1,baseline={(0,-0.5)}]
    \coordinate (currentLocation) at (0,0);
    \drawPropagationShort
    \drawQuartic
    \drawPropagationShortDirected
  \end{tikzpicture}
  \;+\;
  \begin{tikzpicture}[scale=1,baseline={(0,-0.5)}]
    \coordinate (currentLocation) at (0,0);
    \drawAbsorb
    \drawWL
    \drawGirlande
    \drawWL
    \drawEmitDirected
  \end{tikzpicture}
  & \multirow{2}{*}[-8pt]{\text{(2PM)}} \nonumber\\[8pt]
  &+\;
  \begin{tikzpicture}[scale=1,baseline={(0,-0.5)}]
    \coordinate (currentLocation) at (0,0);
    \drawPropagationShort
    \drawCubic
    \drawAbsorb
    \drawWL
    \drawEmitDirected
  \end{tikzpicture}
  \;+\;
  \begin{tikzpicture}[scale=1,baseline={(0,-0.5)}]
    \coordinate (currentLocation) at (0,0);
    \drawAbsorb
    \drawWL
    \drawEmit
    \drawCubic
    \drawPropagationShortDirected
  \end{tikzpicture}
  & \nonumber\\[15pt]
  &+\; 
  \begin{tikzpicture}[scale=1,baseline={(0,-0.5)}]
    \coordinate (currentLocation) at (0,0);
    \drawPropagationShort
    \drawCubic
    \drawPropagationShort
    \drawCubic
    \drawPropagationShort
    \drawCubic
    \drawPropagationShortDirected
  \end{tikzpicture}
  \;+\;
  \begin{tikzpicture}[scale=1,baseline={(0,-0.5)}]
    \coordinate (currentLocation) at (0,0);
    \drawPropagationShort
    \drawCubic
    \drawAbsorb
    \drawWL
    \drawEmit
    \drawCubic
    \drawPropagationShortDirected
  \end{tikzpicture}
  \;+\;
  \begin{tikzpicture}[scale=1,baseline={(0,-0.5)}]
    \coordinate (currentLocation) at (0,0);
    \drawAbsorb
    \drawWL
    \drawCubGirlande
    \drawWL
    \drawEmitDirected
  \end{tikzpicture}
  & \multirow{2}{*}[-8pt]{\text{(3PM)}} \nonumber\\[8pt]
  &+\;
  \begin{tikzpicture}[scale=1,baseline={(0,-0.5)}]
  \coordinate (currentLocation) at (0,0);
  \drawPropagationShort
  \drawCubic
  \drawPropagationShort
  \drawCubic
  \drawAbsorb
  \drawWL
  \drawEmitDirected
  \end{tikzpicture}
  \;+\;
  \begin{tikzpicture}[scale=1,baseline={(0,-0.5)}]
    \coordinate (currentLocation) at (0,0);
    \drawAbsorb
    \drawWL
    \drawEmit
    \drawCubic
    \drawPropagationShort
    \drawCubic
    \drawPropagationShortDirected
  \end{tikzpicture}
  & \nonumber\\[12pt]
  &+\;\sum_{n=4}^\infty \;\;
  \Biggl[\,\,\,\begin{tikzpicture}[scale=1,baseline={(0,-0.5)}]
      \coordinate (currentLocation) at (0,0);
      \drawPropagationShort
      \drawCubic
      \drawPropagationShortDirected
    \end{tikzpicture}
    \,\,\,\Biggr]^n 
  &(\geq\text{4PM})\nonumber
\end{align}
\vspace{-15pt}
\caption{Post-Minkowskian diagrams of the response $\Sigma^{\mu\nu\alpha\beta}$, ordered by PM counting. Only non-zero contributions are shown. Notably, at 4PM and beyond only iterations of ladders appear.
}\label{fig:response_PM_diagrams}
\end{figure}

In the following, we discuss the explicit evaluation of these two contributions to the two-point response, starting with the resummation of the ladder diagrams in sec.~\ref{sec:ladder_structure} and then including the remainder contributions in sec.~\ref{sec:response<=3PM}.

\subsection{Computation of ladder diagrams}\label{sec:ladder_structure}
We begin by examining the $n$PM ladder diagrams for all $n\ge4$ shown in the final line of fig.~\ref{fig:response_PM_diagrams}. These diagrams exhibit a remarkably simple and universal structure that enables their resummation.
As we will show, this resummation directly yields the contribution $\Sigma^{\mn\ab}_\mathrm{sum}$ appearing in eq.~\eqref{eq:response_sum_rem_split}.

We define the $n$-rung ladder $\mathcal{L}_n^{\mn\ab}$, where $n$ counts the number of shockwave-graviton vertices, as the diagram
\begin{align}
  \lad^{\mn\ab}_n &=
  \begin{tikzpicture}[baseline={(0,.2)}]
    \tikzmath{\sc=.8;\gap=1.2;};
    \coordinate (in2) at (-.9,.7);
    \coordinate (z) at (-.9,0);
    \coordinate (v2) at ($(in2)+(.6,0)$);
    \coordinate (v1) at ($(z)+(.6,0)$);
    \coordinate (v4) at ($(v2)+(\sc,0)$);
    \coordinate (v3) at ($(v1)+(\sc,0)$);
    \coordinate (v6) at ($(v4)+(\gap,0)$);
    \coordinate (v5) at ($(v3)+(\gap,0)$);
    \coordinate (v8) at ($(v6)+(\sc,0)$);
    \coordinate (v7) at ($(v5)+(\sc,0)$);
    \coordinate (out2) at ($(v8)+(.6,0)$);
    \coordinate (w) at ($(v7)+(.6,0)$);
    \node (d) at ($(v3)!.5!(v5)+(.03,.35)$){\ldots}; % (1.4,.35)
    \draw [photonTest] (z)  node [left] {$h_{\mn}(k_{1})$} -- (v1);
    \draw [photonTest] (v1) -- (v3);
    \draw [photonTest] (v3) -- (v5);
    \draw [photonTest] (v5) -- (v7);
    \drawGravitonLineDirected{(v7)}{(w)};
    \node[right] at (w) {$h_{\ab}(k_{2})$};
    \draw [photonTest] (v2) node [above=2pt] {$\ell_{1}$} -- (v1);
    \draw [photonTest] (v4) node [above=2pt] {$\ell_{2}$} -- (v3);
    \draw [photonTest] (v6) node [above=2pt] {$\ell_{n-1}$} -- (v5);
    \draw [photonTest] (v8) node [above=2pt] {$\ell_{n}$} -- (v7);
    \drawShockwave{(in2)}{(out2)};
    \drawdots{{v1,v2,v3,v4,v5,v6,v7,v8}}
  \end{tikzpicture}\,,
\end{align}
where momentum conservation fixes the last graviton momentum in terms of the rest: $\ell_n = k_2-k_1-\sum_{i=1}^{n-1}\ell_i$.
All momenta are labelled along the direction of causality flow, meaning $k_1$ is incoming while $k_2$ is outgoing.
The $n$-rung ladder forms a diagram with $(n-1)$ loops.
Its most generic analytical structure is
\be\label{eq:ladderStructure}
  \lad^{\mn\ab}_n=\frac{\kappa_{D}^{2n}}{2^n}(P\cdot k_1)^{2n} \int_{\ell_1,\ldots,\ell_{n-1}} \Omega_n^{\mu\nu\ab}(k_1,k_2,P, \ell_1,\cdots, \ell_{n-1}) \,  \integrand_{n-1}\,,
\ee
{  where $\Omega^{\mu\nu\alpha\beta}_n$ is a generic tensor structure and we have defined the $n$-loop scalar-integrand $\integrand_n$}
\begin{align}
  \integrand_{n}
  =
  \frac{\dd(P\cdot q)}{(q - \ell_{1\cdots n})^2}
  \frac{\Lambda^2}{(q^+-\ell_{1\cdots n}^+)^2 + \Lambda^2}
  \prod_{j=1}^n
  \frac{\dd(P\cdot\ell_{j})}{\ell_{j}^2[(k_1+\ell_{1\cdots j})^2+ \iO(P\cdot k_1)]}
  \frac{\Lambda^2}{{\ell_{j}^+}^2 + \Lambda^2}\,,
\end{align}
with $\ell_{1\cdots j}^\mu = \sum_{i=1}^{j}\ell_i^\mu$.
For convenience, we have also introduced the total emitted momentum from the shockwave $q^\mu = k_2^\mu - k_1^\mu$. Note that only the propagators containing $k_1$ are active, meaning that the integral depends on their pole prescription. Accordingly, we have included the $\iO(P\cdot k_1)$ prescription only in the relevant places.

\subsubsection{The $n$-rung ladder numerator}\label{sec:nladder_numerator}
The central ingredient to the $(n\geq4)$-rung ladders are their numerator structures $\Omega_n^{\mu\nu\ab}$. We read them off of the integrand in \eqn{eq:ladderStructure}, which can be computed recursively using the Feynman rules and the help of \texttt{FeynCalc} \cite{Shtabovenko:2023idz}. The extracted numerator turns out to follow a simple pattern depending on PM order $n$.
In fact, we find that the entire $n$-dependence resides in the overall factor $\kappa_D^{2n}(P\cdot k_1)^{2n}/2^n$ of \eqn{eq:ladderStructure}, while the numerator itself is independent of the loop momenta and evaluates to
\begin{align}
  \Omega_{n\ge4}^{\mu\nu\ab}(k_1,k_2,P, \ell_1,\cdots, \ell_{n-1}) &= \Omega^{\mu\nu\ab}(k_1,k_2,P)
  \nn
  \\
  &= -\I\, \Pi^{\mu\nu}{}_{\kappa\lambda}(k_1)\Pi^{\kappa\lambda\ab}(k_2)=-\I\, (\Pi_1\cdot\Pi_2)^{\mu\nu\ab}\,.
  \label{eq:vecstruc_double_copy}
\end{align}
To write it in a compact form, we have used the transverse traceless (TT) projector 
\be\label{eq:TTprojector}
  \Pi^{\mn\ab}_i=\Pi^{\mn\ab}(k_i) = \pi_i^{\mu(\rho}\pi_i^{\alpha)\beta} - \frac{1}{D-2}\pi_i^{\mu\nu}\pi_i^{\alpha\beta}.
\ee
where the spatial projector $\pi_i^{\mu\nu}$ was defined in \eqn{eq:gauge-fixed-projector}. 

Writing the numerator in terms of the TT projectors makes the following properties of the ladder numerator manifest; it is transverse to $k_1^\mu$ ($k_2^\mu$) in the first (second) pair of indices, and transverse to $P^\mu$ in both pairs. One consequence of the above is that contracting it with the de Donder numerator simply gives back the same structure: $\Omega^{\mu\nu\kappa\lambda}\mathcal{P}_{\kappa\lambda}^{~~\ab}=\Omega^{\mu\nu\ab}$.
This last property, specifically, is important to prove the expression \eqn{eq:vecstruc_double_copy} at all orders in $n$, which we do by induction.

First, one may explicitly check \eqn{eq:vecstruc_double_copy} at $n=4$. Assuming \eqn{eq:ladderStructure} for $\mathcal L^{\mn\ab}_n$, the tensor structure of $\mathcal L^{\mn\ab}_{n+1}$ follows from multiplying a three-point vertex and integrating over the former external momentum $k'_2$:
\begin{equation}\nonumber
\begin{aligned}
  &\mathcal L^{\mn\ab}_{n+1}=
  \begin{tikzpicture}[baseline={(0,.2)}]
    \tikzmath{\sc=.8;\gap=1.2;};
    \coordinate (in2) at (-.9,.7);
    \coordinate (z) at (-.9,0);
    \coordinate (v2) at ($(in2)+(.6,0)$);
    \coordinate (v1) at ($(z)+(.6,0)$);
    \coordinate (v4) at ($(v2)+(\sc,0)$);
    \coordinate (v3) at ($(v1)+(\sc,0)$);
    \coordinate (v6) at ($(v4)+(\gap,0)$);
    \coordinate (v5) at ($(v3)+(\gap,0)$);
    \coordinate (v8) at ($(v6)+(\sc,0)$);
    \coordinate (v7) at ($(v5)+(\sc,0)$);
    \coordinate (out2) at ($(v8)+(.6,0)$);
    \coordinate (w) at ($(v7)+(.6,0)$);
    \node (d) at ($(v3)!.5!(v5)+(.03,.35)$){\ldots}; 
    \draw [photonTest] (z)  node [left] {$h_{\mn}(k_{1})$} -- (v1);
    \draw [photonTest] (v1) -- (v3);
    \draw [photonTest] (v3) -- (v5);
    \draw [photonTest] (v5) -- (v7);
    \drawGravitonLineDirected{(v7)}{(w)};
    \node[right] at (w) {$h_{\ab}(k_{2})$};
    \draw [photonTest] (v2) node [above=2pt] {$\ell_{1}$} -- (v1);
    \draw [photonTest] (v4) node [above=2pt] {$\ell_{2}$} -- (v3);
    \draw [photonTest] (v6) node [above=2pt] {$\ell_{n}$} -- (v5);
    \draw [photon] (v8) node [above=2pt] {$\ell_{n+1}$} -- (v7);
    \drawdots{{v1,v2,v3,v4,v5,v6,v7,v8}}
    \drawShockwave{(in2)}{(out2)}

  \end{tikzpicture}
  \\
  &
  =
  \int_{k'_2}
  \frac{\kappa^{2n}(P\cdot k_1)^{2n}}{2^n}
  \Omega^{\mu\nu\rho\sigma}(k_1,k'_2,P)\,\integral_{n-1}
  \times
  \frac{\I \mathcal{P}_{\rho\sigma\gamma\delta}}{{k'_2}^2+\iO(P\cdot k'_2)}
  \times\!
  \begin{tikzpicture}[baseline={(0,.6)}]
    \tikzmath{\sc=.8;};
    \coordinate (a) at (-0.4,.3);
    \coordinate (b) at (0.4,.3);
    \coordinate (c) at (0,.3);
    \coordinate (d) at (0,.93);
    \node[above=2pt] at (d) {$\ell_{n+1}$};
    \draw[photonTest] (a)--(c);
    \drawGravitonLineDirected{(c)}{(b)};
    \draw[photonTest] (d)--(c);
    \drawShockwave{($(d)-(0.4,0)$)}{($(d)+(0.4,0)$)};
    \node[left] at (a) {$h_{\gamma\delta}(k'_2)$};
    \node[right] at (b) {$h_{\ab}(k_2)$};
    \drawdots{{c,d}};
  \end{tikzpicture}
  \\
  &=
  \frac{\kappa^{2(n+1)}(P\cdot k_1)^{2(n+1)}}{2^{n+1}}\Omega^{\mu\nu\ab}(k_1,k_2,P)\,\integral_{n}
  \, ,\\
  &\text{where}\quad\quad k'_2=k_1+\sum_i^{n-1}\ell_i, \quad k_2=k'_2+\ell_n\, ,\quad \ell_{n+1} = q - \sum_i^{n}\ell_i\;.
\end{aligned}
\end{equation}
In the second line, we have written the $(n+1)$-rung ladder in terms of the $n$-rung ladder and the three-point vertex interacting with a shockwave.
Using the induction hypothesis, we have expressed the $n$-rung ladder in terms of its numerator $\Omega^{\mu\nu\rho\sigma}$ and absorbed the de Donder propagator.
Multiplying with the one-rung ladder, the dependence on $k'_2$ drops out of the numerator and the remaining $k'_2$-integration simply evaluates to $\integral_n$.
In conclusion, the numerator once more follows the form of \eqn{eq:vecstruc_double_copy}, proving inductively that all ladders at $n\geq 4$ take this structure.

A crucial observation of the ladders at 4PM and beyond is that their numerator structure is independent of the loop momenta. As a result, no tensor-integral reduction is required, and the ladder diagrams are expressed directly in terms of scalar integrals. The full $n$-rung ladder contribution then takes the form
\begin{align}\label{eq:fullLadder}
  \lad_n^{\mn\ab}&=-\I \frac{\kappa_{D}^{2n}(P\cdot k_1)^{2n}}{2^n} (\Pi_1\cdot \Pi_2)^{\mu\nu\ab} \, \integral_{n-1}\,,
\end{align}
where we have defined the scalar integral $\integral_n = \int_{\ell_1,\ldots,\ell_{n}}  \integrand_{n}$. With this representation, the problem of evaluating the ladder diagrams reduces entirely to the computation of the scalar integrals $\integral_n$, to which we now turn.

We note in passing that the recursive simplicity of the $n$-rung ladders depends crucially on the gauge choice of the graviton field (which, in this work, is linear de-Donder gauge).
As a counter example, using the ``optimised gauge'' of ref.~\cite{Driesse:2024feo}, we do not observe any simple recursive structure.

\subsubsection{Integration of the $n$-rung ladder}\label{sec:n-ladder_integration}
Working in the light-cone basis, we evaluate the scalar integrals $\integral_n$ in two steps. First, we perform the integrals over the $\ell_i^+$ components by contour integration. Then, we are left with integrals over the spatial loop momenta $\belln{i}$, which we evaluate using the method of regions in the limit $\Lambda \to \infty$.

To build intuition for the integration strategy, we first consider the one-loop case
\begin{align}
  \integral_1 = \dd(P\cdot q) \int_{\ell_1}\frac{\dd(P\cdot\ell_1)}{\ell_1^2(q-\ell_1)^2[(k_1+\ell_1)^2 + \iO(P\cdot k_1)]}\frac{\Lambda^2}{{\ell_1^+}^2 + \Lambda^2}\frac{\Lambda^2}{(q^+-\ell_1^+)^2 + \Lambda^2}\,.
\end{align}
Decomposing the loop momentum in the light-cone basis 
\be
  \ell_1^\mu = \frac{\ell_1^+e^\mu + \ell_1^-\bar e^\mu}{2} + \belln{1}^\mu,\quad \d^D\ell_1 =\frac{1}{2} \d^{D-2}\belln{1} \d\ell_1^{+} \d\ell_1^{-}\,,
\ee
and using $P^\mu = Ee^\mu$, we integrate out the $\ell_1^-$ component via the delta function $\dd(P\cdot\ell_1)$, leaving 
\begin{align}
  \integral_1 
  &=
  \frac{\dd(P\cdot q)}{2 E}
  \int_{\belln{1}}
  \frac{1}{\belln{1}^2(\bperp{q}-\belln{1})^2}\\
  &\times\int_{\ell_1^+}\frac{1}{k_1^-\ell_1^+ + k_1^-k_1^+ - (\bperpn{k}{1} + \belln{1})^2 + \iO k_1^-} \frac{\Lambda^2}{{\ell_1^+}^2 + \Lambda^2}\frac{\Lambda^2}{(q^+-\ell_1^+)^2 + \Lambda^2}\,.
\end{align}
By extracting the prefactor $k_1^-$ from the linear denominator, the $\ell_1^+$-integral takes the form of a simple contour integral with five poles. 
Two of them reside in the upper half-plane, at $\ell_1^+ = \I\Lambda$ and $\ell_1^+ =q^+ + \I\Lambda$, and since the integrand falls of as $1/{\ell_1^+}^{5}$ for large $\ell_1^+$, closing the contour upward generates no arc contribution.
Note that convergence crucially relies on the presence of the shockwave regulator $\Lambda$. 
Evaluating the residues yields
\begin{align}
\cint_1 
&=
\int_{\ell_1^+}\frac{1}{\ell_1^+ +m_1 + \iO }\frac{\Lambda^2}{{\ell_1^+}^2 + \Lambda^2}\frac{\Lambda^2}{(q^+-\ell_1^+)^2 + \Lambda^2}
\\
&=\frac{\Lambda^3}{2 q^+}\Big[\frac{1}{(q^+ -2 \I \Lambda)(m_1+\I\Lambda+\iO)} +\frac{1}{(q^+ +2 \I \Lambda)(m_1+q^++\I\Lambda+\iO)}  \Big]
\nn\,,
\end{align}
where we have defined
\be
m_1 = k_1^+ - \frac{(\bperpn{k}{1} + \belln{1})^2}{k_1^-}\,.
\ee
Finally, the integral over the spatial loop momentum $\belln{1}$ remains.
Given that $m_1$ depends non-trivially on $\belln{1}$, evaluating the full integral directly is challenging and, as stated, we will address it using the method of regions in a moment.

The more general integral $\integral_{n}$ is obtained analogously. Decomposing all loop momenta into the light-cone basis and integrating out the $n$ delta functions fixes the $\ell_i^-$ components and produces a prefactor $1/(2E)^{n}$. 
Upon factoring out $k_1^-$ from all linear propagators, the remaining integral is over the $\ell_i^+$ and $\belln{i}$ components
\begin{align}\label{eq:ladderIntegralFull}
\integral_{n} =& (-1)^{n+1}\frac{\dd(P\cdot q)}{(2 k_1^+ E)^{n}} \int_{\belln{1},\cdots,\belln{n}} \frac{1}{\belln{1}^2\cdots\belln{n}^2(\bperp{q} - \belln{1}-\cdots-\belln{n})^2}\cint_n\,,
\end{align}
where $\cint_n$ denotes the integral over the $\ell_i^+$ components:
\begin{align}
\label{LambdaLimit}
  \cint_{n} = 
  \int_{{\ell}_1^+,\cdots,{\ell}_{n}^+}
  \frac{\Lambda^2}{(q^+-\ell^+_{1\cdots n})^2 + \Lambda^2}
  \prod_{j=1}^n
  \frac{1}{\ell^+_{1\cdots j} + m_j + \iO}
  \frac{\Lambda^2}{{\ell_j^+}^2 + \Lambda^2}\,.
\end{align}
This can be evaluated via contour integration to all orders in $n$ and permits the closed-form expression
\begin{align}
\cint_{n} &=\sum_{j=0}^{n}
\frac{1}{2^{n}}
\frac{\Lambda^{\,n+2}}
{\bigl(q^+ + \I (n+1-2j)\Lambda\bigr)
 \bigl(q^+ + \I (n-1-2j)\Lambda\bigr)}
\;\nn\\
&\times
\prod_{k=1}^{j}
\frac{1}{  m_k +\I k \Lambda + \iO }
\;
\prod_{l=j+1}^{n}
\frac{1}{ q^+ + m_l + \I (n+1-l)\Lambda  + \iO }\label{eq:ladderContourResult}
\end{align}
where we have defined
\begin{align}
  m_i = k_1^+ &- \frac{(\bperpn{k}{1} + \belln{1} + \cdots + \belln{i})^2}{k_1^-}\,. 
\end{align}
Substituting back into \eqn{eq:ladderIntegralFull}, we are once again left with a challenging integral over spatial loop momenta $\belln{i}$.

We tackle this with the method of regions \cite{Beneke:1997zp,Smirnov:2012gma,Becher:2014oda}. In this work, we restrict ourselves to the regions where incoming and outgoing external momenta scale as $k_1^{\mu}\sim k_2^{\mu}\sim \Lambda^0$, which is the scaling relevant for the on-shell graviton response function, capturing the scattering of a gravitational wave off the shockwave.
For each loop momentum, there are two possible scalings that lead to non-scaleless integrals:
\begin{align}\label{eq:regions_In}
  &\text{(a) External scaling:}\quad \belln{i}\sim \Lambda^0\,,\qquad
  &\text{(b) Hard scaling:}\quad \belln{i}\sim \Lambda^{1/2}\,.
\end{align}
When at least one loop momentum is hard, the $\Lambda\rightarrow \infty$ limit inside the integrand gives
\begin{align}
I_{n}^{\mathrm{hard}} \sim  \Lambda^{-1-\epsilon}\,,
\end{align}
and is thus suppressed in said limit. 
The scalar integral is therefore controlled entirely by the external region where all $\belln{i}\sim\Lambda^{0}$.
There, the contour integrals $\cint_{n}$ decouples from the loop momenta $\belln{i}$ and simplifies to
\begin{align}
\cint_{n} \xrightarrow{\Lambda\rightarrow\infty} \frac{(-\I)^n}{(n+1)!}\,.
\end{align}
A posteriori, we note that one arrives at this result upon taking the naïve $\Lambda\gg m_{i}$ limit of \eqn{LambdaLimit}.
Accordingly, the full scalar integral is determined by the external-region contribution, and evaluates to
\begin{align}
 I_{n}^{\mathrm{external}}
 &=-\frac{\I^n}{(n+1)!} \frac{\dd( P\cdot q)}{(2 P\cdot k_1)^{n}}\integralspatial_n \, ,
  \label{eq:externalIntegral}
\end{align}
where $\integralspatial_n$ denotes the scalar integral over the spatial loop momenta,
\begin{align}
    \integralspatial_n &= \int_{\belln{1},\cdots,\belln{n}}\frac{1}{
      \belln{1}^2\cdots\belln{n}^2(\bperp{q} - \belln{1}\cdots-\belln{n})^2
    }
  =
  \frac{1}{(\bperp{q}^2)^{1+n \epsilon}}
  \frac{ \Gamma (-\epsilon )^{n+1} \Gamma (n \epsilon +1)}{(4 \pi )^{n (1-\epsilon)}\Gamma (-(n+1) \epsilon )}\, .
  \label{eq:spatialIntegral}
\end{align}
For details on the explicit evaluation of the spatial integral $\integralspatial_n$ we refer the reader to ref.~\cite{Cheung:2024byb}.

\subsubsection{Resummation of ladders}\label{sec:ladder_resummation}
To resum the ladder contributions, it is convenient to pass to position space. 
The key observation is that $\integralspatial_n$ has the structure of a convolution: each loop integration contributes one copy of the Fourier transform of the Aichelburg-Sexl metric
\be
  \frac{\kappa_D^2\dd(P\cdot q)}{-\bperp{q}^2} = \dd(P\cdot q)\int\d^{D-2}\bperp{x}\,\euler^{-\I \bperp{q}\cdot \bperp{x}}[8 G f(\bperp{x}^2)], \quad Gf(\bperp{x}^2) = -\Gamma(-\epsilon)\left(\frac{\bperp{x}^2}{4\euler^\mascheroni L^2}\right)^\epsilon,
\ee
where $f(\bperp{x}^2)$ is reprinted for convenience from \eqn{eq:Gf_def}.
By the convolution theorem, the $n$-fold convolution in momentum space becomes a simple product in position space 
 \begin{align}
 \kappa_D^{2(n+1)}\dd(P\cdot q)\integralspatial_n&= \dd(P\cdot q) \int \d^{D-2} \bperp{x}\, \euler^{-\I \bperp{q}\cdot \bperp{x}} [-8Gf(\bperp{x}^2)]^{n+1}\;,
 \end{align}
leading to the following expression for the $(n\ge4)$-rung ladder
\begin{align}\label{eq:ladder_n_pos}
  \mathcal{L}^{\mn\ab}_n&=  \dd(P\cdot q)\int\d^{D-2} \bperp{x}\,\euler^{-\I \bperp{q}\cdot \bperp{x}} (\Pi_1  \cdot  \Pi_2)^{\mn\ab}\frac{2 \I^{n} }{2^{2n} n!} (P\cdot\k_1)^{n+1} [-8 Gf(\bperp{x}^2)]^n\,,
\end{align}
valid for $k_1^\mu\sim k_2^\mu\sim \Lambda^0$.

The above formula captures all $(n\ge4)$-rung ladders and, in effect, all contributions to the response function except leading order PM contributions up to third order.
It is, however, natural to expect the simple pattern at (arbitrarily) high PM orders to extend to the leading PM orders, at least in part.
Therefore we define the ``resummed contribution'' $\Sigma_{\mathrm{sum}}^{\mn\ab}$ to the shockwave response function by extending eq.~\eqref{eq:ladder_n_pos} to all $n$:
\begin{align}\label{eq:resummed_response}
  &\I\Sigma_{\mathrm{sum}}^{\mn\ab}(k_1,k_2) \equiv\sum_{n=1}^{\infty} \lad_n^{\mn\ab}\\
  &= 2 \dd(P\cdot (k_2-k_1)) \int\d^{D-2}\bperp{x}\,\euler^{-\I \bperp{q}\cdot \bperp{x}}(\Pi_1\cdot \Pi_2)^{\mn\ab}(P\cdot k_1) \Big[e^{-2 \I G (P\cdot k_1) f(\bperp{x}^2) }-1 \Big]\nn\,.
\end{align}
This gives exactly the resummed piece of \eqn{eq:response_sum_rem_split}.

\subsection{Computation of the remainder}\label{sec:response<=3PM}
To complete the computation of the shockwave response function, we must evaluate the remainder contribution $\Sigma_{\mathrm{rem}}^{\mn\ab}$ in eq.~\eqref{eq:response_sum_rem_split} which, by definition, collects the contributions of $\Sigma^{\mn\ab}$ not captured by the resummed ladder pattern $\Sigma^{\mu\nu\alpha\beta}_\mathrm{sum}$.
As discussed, such leftover remainder pieces appear only up to and including 3PM order.
In this section, we determine them by explicitly computing the response at 1PM, 2PM, and 3PM orders and taking the difference to the ladder pattern at each order.
Interestingly, as we will discuss in sec.~\ref{sec:response_discussion}, the remainder vanishes on-shell so that it is relevant only off-shell.
Further, it vanishes non-trivially at 3PM order and is thus even further restricted to the two leading PM orders.

We proceed with our calculation order by order, beginning with the structurally simplest 3PM contribution before turning to the more involved 2PM and 1PM cases.

\subsubsection{3PM remainder}\label{sec:response_3PM}
At 3PM order, the response function receives contributions from five diagrams, as shown in fig.~\ref{fig:response_PM_diagrams}. Each diagram individually does not follow the pattern $\Sigma^{\mu\nu\alpha\beta}_\mathrm{sum}|_{3\text{PM}}$ observed at 4PM and beyond. However, taking the sum of {\it all} diagrams, the full $3$PM response does in fact reproduce the pattern
\begin{align}\label{eq:response_3PM}
   \Sigma_{\mathrm{rem}}^{\mu\nu\alpha\beta}
   \Big|_{3\text{PM}}
   =
   \big(
    \Sigma^{\mu\nu\alpha\beta} - \Sigma_\mathrm{sum}^{\mu\nu\alpha\beta}
  \big)
   \Big|_{3\text{PM}}
   =
   0\,.
\end{align}
In other words, the remainder vanishes identically at 3PM order.
The cancellation happens directly by a sum of integrands of all diagrams - no integrals have to be computed, nor is tensor reduction necessary.

\subsubsection{2PM remainder}\label{sec:response_2PM}
Counterintuitively, the 2PM response is more involved than its 3PM sibling.
This is because lower PM diagrams have fewer factors of $P^\mu$, which leads to fewer cancelations.
The 2PM response is represented diagrammatically in fig. \ref{fig:response_PM_diagrams} and summing all contributions we obtain
\begin{equation}\label{eq:remainder2PM}
  \Sigma^{\mu\nu\alpha\beta}_\mathrm{rem}
  \Big|_{2\text{PM}}
  =
  \big(
    \Sigma^{\mu\nu\alpha\beta} - \Sigma_\mathrm{sum}^{\mu\nu\alpha\beta}
  \big)
    \Big|_{2\text{PM}}
  =
  { -\I}c_{2\mathrm{PM}}^{\mn\ab}(I_{011}+I_{110}+\bperp{q}^2 I_{111})\, .
\end{equation}
Here, the scalar integrals are defined by
\begin{equation}\label{eq:response_2PM_family}
  I_{ijk}
  =
  \int_{\ell}\frac{\dd(P\cdot \ell)\dd(P\cdot q)}{[\ell^2]^i[(k_1+\ell)^2+\i0^+k_1^-]^j[(q-\ell)^2]^k}\frac{\Lambda^4}{[\ell_+^2+\Lambda^2][(q_+-\ell_+)^2+\Lambda^2]}\, ,
\end{equation}
and appear with the overall coefficient
\begin{align}
c_{2\mathrm{PM}}^{\mn\ab}
  &=
  \frac{\kappa_D^4}{16} (k_1\!\cdot\! P)^2
  \bigg[ \pi^{\mn}
   (k_1)P^\alpha P^\beta 
    +
    \pi^{\ab}(k_2)P^\mu P^\nu 
  \bigg]
  \, .
\end{align}
The two terms are interchanged by swapping external legs, and the appearance of projectors $\pi^\mn(k_i)$ (from eq.~\eqref{eq:gauge-fixed-projector}) imply that the 2PM remainder vanishes on-shell. 

Integrals $I_{ijk}$ can be evaluated with the method of regions and Feynman parameters, explicitly shown in \app{app:I_ijk}. We find the combination appearing in \eqn{eq:remainder2PM} evaluates to
\begin{equation}\label{eq:problem_piece_evaluated}
  I_{011}+I_{110}+\bperp{q}^2 I_{111}
  =
 \I \frac{2  \log \left(\frac{\Lambda  k_1^-}{\bperp{q}^2}\right)-3 \pi\I -1 }{16 \pi (k_1\!\cdot\!P)}
  \dd(P\cdot q)
  +\mathcal O(\epsilon,\Lambda^{-1})\,,
\end{equation}
which is finite for $\epsilon\rightarrow 0$ but exhibits a logarithmic divergence in the limit $\Lambda\rightarrow \infty$.
We will comment on this divergence below in section~\ref{sec:response_discussion}.

\subsubsection{1PM remainder}\label{sec:response_1PM}
Finally, we treat the 1PM response which, as illustrated in fig. \ref{fig:response_PM_diagrams}, involves only two diagrams: the $1$-ladder and the simple recoil.
The tensor structure of the 1PM remainder is even more intricate than that at 2PM which we, again, ascribe to the even fewer instances of $P^\mu$ in its numerator.
Adding up both diagrams and subtracting the 1PM part of $\Sigma^{\mn\ab}_{\rm sum}$, we obtain the 1PM remainder:
\begin{align}
  &\Sigma_{\mathrm{rem}}^{\mn\ab}
  \Big|_{1 \mathrm{PM}} 
  = 
  \frac{ \kappa_D^2 \dd(P\cdot q)}{4 q^2}\Big[P^{\alpha}P^{\beta}(2 k_1^{(\mu}q^{\nu)}-\eta^{\mu\nu}) +P^{(\mu}q^{\nu)}P^{(\alpha}q^{\beta)}-(P\!\cdot\!k_1)^2\frac{\pi_1^{\mu\nu}\pi_2^{\ab}}{D-2} \nonumber \\
  &-\frac{(k_1^2 +k_2^2)P^{\mu}P^{\nu}}{2}\left(\frac{q^2 P^{\alpha}P^{\beta}}{2 (P\!\cdot\!k_1)^2}-\pi_1^{\alpha\beta}+\pi_2^{\ab}  \right)+(P\!\cdot\!k_1)^2\frac{(\pi_1^{\mu\nu}+\pi_2^{\mu\nu})(\pi_1^{\alpha\beta}+\pi_2^{\alpha\beta})}{4}   \Big]\nonumber\\
  &+(\mu\nu,k_1 \leftrightarrow \alpha\beta,-k_2)\,.
\end{align}
In contrast to the 2PM remainder, this tensor structure is not completely transverse to $P^\mu$ and, consequently, it cannot be simplified as much in terms of the $\pi$-projectors.
However, it is transverse to $P^\rho P^\sigma$ for any combination of two indices $\rho,\sigma\in\{\mu,\nu,\alpha,\beta\}$.
This is the important fact that allowed us to drop the $\delta(0)$ in section~\ref{sec:graviton-light}.
As with the 2PM remainder, the 1PM remainder vanishes on-shell.

\section{Results and Discussion of the Shockwave Response}\label{sec:response_discussion}
With the explicit calculation of the response function completed, we now collect and reorganise the result in order to better display its underlying structure.

\subsection{Summary of shockwave response results} 
Since the full, off-shell response corresponds to the graviton propagator on the shockwave background, it is naturally split into free propagation and an (amputated) interaction $\Sigma$ (eq.~\eqref{eq:sigmasplit}):
\begin{align}
  \begin{tikzpicture}[baseline=(anchor)]
    \coordinate (anchor) at (0,-.5);
    \coordinate (inA) at (-1,0);
    \coordinate (outA) at (1,0);
    \coordinate (v) at (0,0);
    \coordinate (c1) at (-1,-1);
    \coordinate (c2) at (1,-1);
    \drawShockwave{(inA)}{(outA)};
    \drawGravitonLine{(c1)}{(v)}
    \drawPerturbativeVertex{(c1)}
    \drawGravitonLineDirected{(v)}{(c2)}
    \drawPerturbativeVertex{(c2)}
    \drawResponseBlob{(v)}
  \end{tikzpicture}
  \quad = \quad
  \begin{tikzpicture}[baseline=(anchor)]
    \coordinate (anchor) at (0,-.5);
    \coordinate (inA) at (-1,0);
    \coordinate (outA) at (1,0);
    \coordinate (c1) at (-1,-1);
    \coordinate (c2) at (1,-1);
    \drawShockwave{(inA)}{(outA)};
    \drawPerturbativeVertex{(c1)}
    \draw [photonTestDirected] (c1) to[out=40,in=140] (c2);
    \drawPerturbativeVertex{(c2)}
  \end{tikzpicture}
  \quad + \quad
  \begin{tikzpicture}[baseline=(anchor)]
    \coordinate (anchor) at (0,-.5);
    \coordinate (inA) at (-1,0);
    \coordinate (outA) at (1,0);
    \coordinate (v) at (0,0);
    \coordinate (c1) at (-1,-1);
    \coordinate (c2) at (1,-1);
    \drawShockwave{(inA)}{(outA)};
    \drawGravitonLine{(c1)}{(v)}
    \drawPerturbativeVertex{(c1)}
    \drawGravitonLineDirected{(v)}{(c2)}
    \drawPerturbativeVertex{(c2)}
    \draw [fill=white, thick] (v) circle (.3);
    \node at (v) {$\I\Sigma$} ;
  \end{tikzpicture}
  \, \,,
\end{align}
where the interaction $\Sigma$ has two distinct terms:
A resummed piece, $\Sigma_{\rm sum}$, and a 2PM remainder $\Sigma_{\rm rem}$ (eq.~\eqref{eq:response_sum_rem_split}):
\begin{align}
    \Sigma^{\mu\nu\ab}(k_1,k_2)=\Sigma_\mathrm{sum}^{\mu\nu\ab}(k_1,k_2)+\Sigma^{\mu\nu\ab}_\mathrm{rem}(k_1,k_2)\, .
\end{align}

The resummed part collects contributions from an infinite series of ladder diagrams, and has the remarkably simple structure
\begin{align}\label{eq:sigmaSum}
    &\Sigma_{\mathrm{sum}}^{\mn\ab}(k_1,k_2)  =-2\I \dd(P\cdot (k_2-k_1))\,(P\cdot k_1)
    (\Pi_1 \cdot \Pi_2)^{\mn\ab} 
  \mathcal{F}(q_\perp;\eps)\,.
\end{align}
The tensor structure comprises transverse traceless (TT) projectors, which we discuss below. The response's analytic structure is encoded in the function $\mathcal{F}(q_\bot,\epsilon)$ with the resummed expression
\begin{align}\label{eq:resummedF}
    \mathcal{F}(q_\perp;\eps) = \int\d^{D-2}\bperp{x}\,\euler^{-\I\bperp{q}\!\cdot\bperp{x}}\big[\euler^{- \I W f(\bperp{x}^{2})}-1 \big]
    \, ,
\end{align}
with $f(\bperp{x}^{2})$ defined in \eqn{eq:Gf_def} and 
where the energy dependence on $k_1\cdot P=k_2\cdot P$ is encoded in the Weinberg phase
\begin{align}
    W = 2G(P\cdot k_1)\,
    .
\end{align}
In a PM-expanded form, $\mathcal{F}$ captures the infinite series of loop diagrams
\begin{align}
  \mathcal{F}(q_\perp;\eps) = \bigg(\frac{4\pi}{\bperp{q}^2}\bigg)^{1-\eps}\sum_{n=1}^\infty\frac{(\I W)^n}{n!}\frac{\Gamma(-\eps)^n}{(\bperp{q}^2\euler^\mascheroni L^2)^{n\eps}}\frac{\Gamma(1+(n-1)\eps)}{\Gamma(-n\eps)}
  \, ,
\end{align}
where $W$ controls the PM expansion.

The tensor structure of eq.~\eqref{eq:sigmaSum} is the product of two TT projectors (see \eqn{eq:TTprojector} for the definition):
\begin{align}\label{eq:tensorStruc}
    (\Pi_1 \cdot \Pi_2)^{\mu_1\nu_1\mu_2\nu_2} 
    =
    \Pi_1^{\mu_1\nu_1}{}_{\rho\sigma}
    \Pi_2^{\rho\sigma\mu_2\nu_2} \,.
\end{align}
It is hence symmetric and traceless in each pair of indices, $\mn$ and $\ab$, and is symmetric under the exchange of external legs $1\leftrightarrow2$.
Furthermore, it is transverse to $P^{\mu}$ in every index, as well as transverse to $k_1^{\mu}$ in the first pair and to $k_2^{\alpha}$ in the second pair:
\begin{align}
  (\Pi_1\cdot\Pi_2)^{\mn\ab}k_{1\mu}=(\Pi_1\cdot\Pi_2)^{\mn\ab}k_{2\alpha}=(\Pi_1\cdot\Pi_2)^{\mn\ab}P_{\mu}=(\Pi_1\cdot\Pi_2)^{\mn\ab}P_{\alpha}=0\nonumber\,.
\end{align}
Interestingly, this combination of projectors is itself a projector, fulfilling
\begin{equation}
  (\Pi_1\cdot\Pi_2)^{\mn\rho\sigma}(\Pi_1\cdot\Pi_2)_{\rho\sigma}^{~~\ab} = (\Pi_1\cdot\Pi_2)^{\mn\ab}\,,
\end{equation}
which is a consequence of the more general relation $\Pi_{1}\cdot\Pi_{2}\cdot\Pi_{3}=\Pi_{1}\cdot\Pi_{3}$.

The second piece of the shockwave response is the ``remainder'', $\Sigma^{\mu\nu\ab}_\mathrm{rem}(k_1,k_2)$, which terminates at 2PM order:
\begin{align}
  &\Sigma_{\rm rem}^{\mn\ab}(k_1,k_2) = \frac{\dd(P\!\cdot\!q)}{4}\Bigg(
    \frac{\kappa_D^2 }{q^2}\Big[P^{\alpha}P^{\beta}(2 k_1^{(\mu}q^{\nu)}-\eta^{\mu\nu}) +P^{(\mu}q^{\nu)}P^{(\alpha}q^{\beta)}-(P\!\cdot\!k_1)^2\frac{\pi_1^{\mu\nu}\pi_2^{\ab}}{D-2} \nonumber \\
  &-\frac{(k_1^2 +k_2^2)P^{\mu}P^{\nu}}{2}\left(\frac{q^2 P^{\alpha}P^{\beta}}{2 (P\!\cdot\!k_1)^2}-\pi_1^{\alpha\beta}+\pi_2^{\ab}  \right)+(P\!\cdot\!k_1)^2\frac{(\pi_1^{\mu\nu}+\pi_2^{\mu\nu})(\pi_1^{\alpha\beta}+\pi_2^{\alpha\beta})}{4}   \Big]\nonumber\\
  &+\frac{\i\kappa_D^4}{64\pi}(k_1\!\cdot\! P) \pi_1^{\mn}P^\alpha P^\beta 
 \Big[2  \log \left(\frac{\Lambda  k_1^-}{\bperp{q}^2}\right)-3 \pi\I -1 \Big]
  +(\mu\nu,k_1 \leftrightarrow \alpha\beta,-k_2)
  \Bigg) \nonumber\\
 &+\mathcal O(\epsilon,\Lambda^{-1})\,.
\end{align}
This structure vanishes on-shell and is thus only relevant off-shell.
It displays a logarithmical divergence in $\Lambda$ which 
will need to cancel in observables and, as stated above, is reminiscient of the rapidity 
divergencies observed in SCET.

Our off-shell response function $\Sigma^{\mn\ab}(k_1,k_2)$ is valid in the region where the external momenta scale as $k_1\sim k_2\sim\Lambda^0$, which is appropriate for taking the on-shell limit of the response.
Additional regions may arise off-shell, in particular when the response function is attached to a light body, as in the 1SF observables of eqs.~\eqref{eq:1sfWaveform} and \eqref{eq:z_SF_expansion}.
However, as a first nontrivial test, we have verified that no other regions contribute to the 1SF waveform in \eqn{eq:1sfWaveform} of a light Schwarzschild black hole 
scattering off the shockwave.

\subsection{On-shell response}

The on-shell limit of the response is of much interest, as it encodes the wave scattering phase shift, which is a physical observable. In the massive case, it is ofted referred to as the gravitational \textit{Compton} \cite{Bjerrum-Bohr:2025bqg,Bautista:2026qse,Bjerrum-Bohr:2026fhs} or \textit{Raman} \cite{Ivanov:2026icp} amplitude and corresponds to the transfer matrix element $\bra{k_2,2 h_2}\hat{T}\ket{k_1,2 h_1}$ with $h_{i}=\pm 1$ capturing the helicities. It is obtained by contracting incoming (outgoing) legs with polarisation tensors $\varepsilon_{1\,{\mu\nu}}^{(2h_1)}=\varepsilon_{1\,\mu}^{(h_1)}\varepsilon_{1\,\nu}^{(h_1)}$ (${\bar{\varepsilon}}_{2\,\ab}^{(2h_2)}={\bar\varepsilon}_{2\,\alpha}^{(h_2)}{\bar\varepsilon}_{2\,\beta}^{(h_2)}$) that, for each $i=1,2$, fulfill transversality $\varepsilon^{(h_i)}_{i\,\mu}k_i^{\mu}=\bar{\varepsilon}^{(h_i)}_{i\,\mu}k_i^{\mu}=0$ and tracelessness $\varepsilon_{i\,\mu}^{(h_i)}\varepsilon_{i}^{(h_i)\mu}=\bar{\varepsilon}_{i\,\mu}^{(h_i)}\bar{\varepsilon}_{i}^{(h_i)\mu}=0$. Thereafter imposing the on-shell conditions $k_i^2=0$, we obtain the on-shell response function
\begin{align}
  T^{(2h_1,2h_2)}(k_1,k_2)
  =
  \bra{k_2,2 h_2}\hat{T}\ket{k_1,2 h_1}
  =
  {\varepsilon}_{1\,\mn}^{(2h_1)}\Sigma^{\mu\nu\ab}(k_1,k_2){\bar{\varepsilon}}_{2\,\ab}^{(2h_2)}\big\vert_\text{on-shell}\,.
\end{align}
where $\Sigma=\Sigma_\text{sum}+\Sigma_\text{rem}$, as above. Strikingly, we observe that the remainder, $\Sigma_\text{rem}$, vanishes completely on-shell
\begin{align}
  \varepsilon_{1\,\mn}^{(2h_1)}\Sigma^{\mn\ab}_{\rm rem}(k_1,k_2)\bar{\varepsilon}_{2\,\ab}^{(2h_2)}\big\vert_\text{on-shell} &= 0\, .
\end{align}
The on-shell response $T^{(2h_1,2h_2)}(k_1,k_2)$ is thus fully determined by the resummed contribution, $\Sigma_\text{sum}$. It takes the very compact, PM-resummed form of an exponential in position space
\begin{align}\label{eq:response-on-shell}
  T(k_1,k_2)= -2\I(P\cdot k_1)\dd(P\cdot q)A^{(2h_1,2h_2)}
  \mathcal{F}(q_\bot,\epsilon)
\end{align}
The associated vector structure is denoted $A^{(2h_1,2h_2)}$, which derives from the TT projectors $\Pi_1\cdot \Pi_2$ of \eqn{eq:tensorStruc}, contracted with polarisation tensors and put on-shell. It takes the gauge-invariant form
\begin{equation}\label{eq:def_onshellStructure}
  A^{(2h_1,2h_2)} \equiv {\varepsilon}_{1\,\mn}^{(2h_1)}(\Pi_1\cdot\Pi_2)^{\mu\nu\ab}\bar{\varepsilon}_{2\,\ab}^{(2h_2)}\big\vert_\text{on-shell} = \frac{(P\cdot F_1^{(h_1)}\!\cdot F_2^{(h_2)}\!\cdot P)^2}{(P\cdot k_1)^4}\,,
\end{equation}
where we have written the resulting expression in terms of linearised field strength tensors $F_i^{(h_i)\mu\nu} = 2k_i^{[\mu}\varepsilon_i^{(h_i)\nu]}$.
In the TT-gauge, where we choose the polarisation vectors transverse to $P^\mu$, the structure further simplifies to
\begin{equation}
  A^{(2h_1,2h_2)}=(\varepsilon_1^{(h_1)}\cdot \bar{\varepsilon}_2^{(h_2)})^2\, ,\qquad \text{for}\qquad P\cdot \varepsilon^{(h_i)}_i=P\cdot \bar{\varepsilon}^{(h_i)}_i=0\, .
\end{equation}
Notably, the polarisation structure, $A^{(2h_1,2h_2)}$, conforms to the double-copy form \cite{Kawai:1985xq,Bern:2008qj,Bern:2010ue} familiar from the 1PM on-shell response of a massive BH (ie. the Compton amplitude). Here, we have shown that for a massless black hole, this polarisation structure remarkably extends to all orders in perturbation theory.

\subsection{Four-dimensional limit and infrared structure}\label{sec:weinberg_phase}

We now take the four dimendional limit of the on-shell response and expose its
infrared structure.
Our starting point is the PM-expanded form of the on-shell response from eq.~\eqref{eq:response-on-shell}, with the objective being to obtain the 4-dimensional limit of the sum,
\begin{align}\label{eq:F_4d_lim_def}
  \mathcal{F}(q_\perp;\eps) = \bigg(\frac{4\pi}{\bperp{q}^2}\bigg)^{1-\eps}\sum_{n=1}^\infty\frac{(\I W g(q_\perp))^n}{n!}\frac{\Gamma(1+(n-1)\eps)}{\Gamma(-n\eps)}\,,
\end{align}
where $W$ is defined as
\begin{equation}
  W = 2G(P\cdot k_1)\,,
\end{equation}
and, in analogy to the position space function $f(\bperp{x}^{2})$, we take
\begin{equation}\label{eq:g-definition}
  g(q_\perp) = \Gamma(-\eps)(\bperp{q}^2L^2\euler^\mascheroni)^{-\eps} = -\frac{1}{\eps} + \log(\bperp{q}^2L^2) + \mathcal{O}(\eps)\,.
\end{equation}
It is expedient to begin by slightly rewriting \eqn{eq:F_4d_lim_def} as
\begin{equation}
  \mathcal{F}(q_\perp; \eps) = -\I W\eps g(q_\perp)\bigg(\frac{4\pi}{\bperp{q}^2}\bigg)^{1-\eps}\sum_{m=0}^\infty\frac{(\I W g(q_\perp))^m}{m!}\frac{\Gamma(1+m\eps)}{\Gamma(1-(m+1)\eps)}\,.
\end{equation}
Then, to do the sum, we leverage the integral representations of the gamma function and its reciprocal,
\begin{equation}
  \Gamma(z) = \int_0^\infty\d t\,t^{z-1}\euler^{-z}, \qquad \frac1{\Gamma(z)} = \frac{\I}{2\pi}\int_H\d t\,(-t)^{-z}\euler^{-t}\,,
\end{equation}
which, when combined, yield
\begin{equation}
  \frac{\Gamma(1+m\eps)}{\Gamma(1-(m+1)\eps)} = \frac{\I}{2\pi}\int_\mathcal{D}\d^2t\,\frac{(-t_1t_2)^{m\eps}\euler^{-t_1-t_2}}{(-t_1)^{1-\eps}}\,,
\end{equation}
where $\mathcal{D} = H\times[0,\infty]$ and $H$ is a Hankel contour pointed toward positive real infinity. Inserting this into the sum and interchanging the summation and integration, we get
\begin{align}
  \mathcal{F}(q_\perp; \eps) &=\bigg(\frac{4\pi}{\bperp{q}^2}\bigg)^{1-\eps} \frac{\I}{2\pi}\int_\mathcal{D}\d^2t\,\frac{\euler^{-t_1-t_2}}{(-t_1)^{1-\eps}}\sum_{m=0}^\infty\frac{(\I W g(q_\perp)(-t_1t_2)^\eps)^m}{m!}(-\I W\eps g(q_\perp)) \notag\\
  &=\bigg(\frac{4\pi}{\bperp{q}^2}\bigg)^{1-\eps} \frac{\I}{2\pi}\int_\mathcal{D}\d^2t\,\frac{\euler^{-t_1-t_2}\euler^{\I Wg(q_\perp)(-t_1t_2)^\eps}}{(-t_1)^{1-\eps}}(-\I W\eps g(q_\perp))\,.
\end{align}
Remarkably, expanding the above expression in the dimensional regulator and using eq.~\eqref{eq:g-definition}, we find that the Weinberg phase cleanly separates from the rest:
\begin{align}
  \mathcal{F}(q_\perp; \eps) &= \euler^{-\I W/\eps}\frac{4\pi}{\bperp{q}^2}\frac{\I}{2\pi}\int_\mathcal{D}\d^2t\,\frac{(-t_1t_2)^{-\I W}\euler^{-t_1-t_2}}{-t_1}\I W(\bperp{q}^2L^2)^{\I W} + \mathcal{O}(\eps) \notag\\
  &= 
  \bigg[
    \euler^{-\I W/\eps}\frac{\Gamma(1-\I W)}{\Gamma(1+\I W)}\frac{\I}{(\bperp{q}^2L^2)^{-\I W}}
  \bigg]
  \frac{4\pi W}{\bperp{q}^2} + \mathcal{O}(\eps)\,,
\end{align}
where we reassembled the integral representation in the second step. 
Note that the factor in square brackets forms a pure phase and the reggeised propagator structure. 
Thus, we find that the finite on-shell response in four dimensions is given by 
\begin{align}
  \lim_{\eps\to0}T_\text{fin.}^{(2h_1,2h_2)} &= \lim_{\eps\to0}\euler^{\I W/\eps}T^{(2h_1,2h_2)}  \nn\\
  &=\frac{4\pi L^{2}W^{2}}{G} A^{(2h_1,2h_2)}
  \frac{\Gamma(1-\I W)}{\Gamma(1+\I W)}\frac{\dd(P\cdot q)}{(\bperp{q}^2L^2)^{1-\I W}}\, .
\end{align}
with  $W = 2G(P\cdot k_1)$ and $q=k_{2}-k_{1}$. 
Put differently, we find that the Born amplitude is corrected to all orders in $G$ by a pure phase factor, as
\be
T^{(2h_1,2h_2)}=  \frac{\Gamma(1-\I W)}{\Gamma(1+\I W)}\frac{\euler^{-\I W/\eps}}{(\bperp{q}^2L^2)^{-\I W}}\, T^{(2h_1,2h_2)}_{\text{Born}}\, .
\ee
We see that all the $\epsilon$-divergences are captured by the Weinberg phase $\euler^{-\I W/\eps}$.
The finite part of the response has a pleasingly simple form, directly analogous to the
’t Hooft ultra-high energy graviton dominance amplitude \cite{tHooft:1987vrq} as well as the
divergent ``Newtonian'' part of the gravitational wave scattering problem \cite{Dolan:2007ut,Bjerrum-Bohr:2026fhs}. Indeed, it also resembles the Coulomb phase 
(in the double-copy sense) for the electromagnetic wave scattering problem \cite{Gordon:1928,Mott:1930}, see also the recent review \cite{Mizera:2023tfe}.

Importantly, while we have focused on the on-shell response here, the analysis fully generalises to the off-shell case. The Weinberg phase cleanly factorises in the same way, and the finite part of the response is given by the same expression as above, but with the off-shell tensor structure instead of the on-shell one. Thus, the off-shell response is equally well-defined in the four-dimensional limit. 

\subsection{The  Magnus $\hat N$-operator}\label{sec:magnusian}
The Magnus $\hat N$-operator has been observed to capture the physical content of the on-shell response function in a concise manner~\cite{Bautista:2026qse}.
First, it has been observed to take a structurally simpler form: It is real and free of divergencies in $\epsilon$.
Second, in the massive case it maps directly to the phase shift of the wave scattering event.

The Magnus $\hat N$-operator appears in the exponential representation of the scattering matrix \cite{Magnus:1954zz,Blanes:2008xlr}
\begin{equation}
  \hat S=e^{i\hat N}\, .
\end{equation}
Importantly, it serves as a compact generator of scattering observables \cite{Gonzo:2024zxo,Kim:2024svw,Kim:2025hpn,Alessio:2025flu,Kim:2025olv,Kim:2025gis,Haddad:2025cmw}, and is related to the on-shell two-point response by
\begin{equation}\label{eq:response_magnusian_relation}
\I T^{(2h_1,2 h_2)}(k_1,k_2)=\bra{k_2,h_2}\hat S-\mathbb{1}\ket{k_1,h_1}=\bra{k_2,h_2}e^{i\hat N}-\mathbb{1}\ket{k_1,h_1}\, .
\end{equation}
Matrix elements of the Magnus operator, $\bra{k_2,h_2}\hat N\ket{k_1,h_1}$, may be extracted from \eqn{eq:response_magnusian_relation} by matching PM expansions order by order. Carrying out this matching, at leading PM order one finds
  \begin{align}\label{eq:magnusian}
  \bra{k_2,h_2}&\hat N\ket{k_1,h_1}=-2\i \dd( P\cdot (k_1-k_2))
 \, (P\cdot k_1) \\\nonumber
  &\times\bigg(
A^{(2h_1,2h_2)}
  \int\d^{D-2}\bperp{x}
  \,
  \euler^{-\I (\mathbf{k}_2-\mathbf{k}_1)_{\bot}\cdot \bperp{x}}
  \Big[-2 \I G (P\cdot k_1) f(\bperp{x}^2) \Big]
  \bigg)\, ,
\end{align}
where $A^{(2h_1,2h_2)}$ has been defined in \eqn{eq:def_onshellStructure}. As we will now show, \eqn{eq:magnusian} is in fact the full, PM-exact Magnus operator matrix element. In this sense, the Magnus operator is completely determined by tree-level data.

To prove this, we demonstrate that the matrix element of \eqn{eq:magnusian} indeed reproduces the full two-point, on-shell response function. We plug it back into \eqn{eq:response_magnusian_relation}, checking that the correct PM-resummed on-shell response of \eqn{eq:response-on-shell} is recovered. One may expand the exponential on the right hand side of \eqn{eq:response_magnusian_relation} as
\begin{align}\label{eq:m+1_expansion_N}
    \bra{k_2,h_2}e^{i\hat N}-\mathbb{1}\ket{k_1,h_1} &= \sum_{n=1}^{\infty}\frac{1}{n!}
  \bra{k_2,h_2}(\i\hat N)^n\ket{k_1,h_1}\, .
\end{align}
We now show that matrix elements of the  ${\hat N}^n$-operator take the form
\begin{align}\label{eq:Nnoperator}
  \bra{k_2,h_2} (\i\hat N)^n \ket{k_1,h_1}=& 2 \dd(P\cdot(k_2-k_1))(P\!\cdot\! k_1)
A^{(2h_1,2h_2)}(k_1,k_2)\\
  &\times \int\d^{D-2}\bperp{x}
  \,
  \euler^{-\I (\bperpn{k}{2}-\bperpn{k}{1})\cdot\bperp{x}}
  \Big[-2 \I G (P\!\cdot\!k_1) f(\bperp{x}^2) \Big]^n\, .\nonumber
\end{align}
The proof follows by induction, since we know it to be true for $n=1$. Assuming eq.~\eqref{eq:Nnoperator} holds for $n$, one wishes to compute $\bra{k_2,h_2} (\i\hat N)^{n+1} \ket{k_1,h_1}$.
This may be achieved by inserting a complete set of classical, on-shell single-graviton states with positive energy
\begin{align}\label{eq:complete_set_states}
  \mathbb{1}
  =
  \sum_{h}\int_\ell \dd(\ell^2)\theta(\ell^0)\,\ket{\ell,h}\bra{\ell,h}\, .
\end{align}
It is important to note that $\hat N$ here only lives in the single BH Hilbert space but also has multi-graviton 
matrix elements. Yet, including these  states in the completeness relation above
 would lead to quantum effects which we neglect in
our analysis.
Furthermore, for the same reason as we can work with the Mandelstam-Leibbrandt prescription in the retarded propagator, we may replace the positive energy condition $\theta(\ell^0)$ with $\theta(P\cdot\ell)$, which is more natural in the shockwave context.
Upon this replacement the matrix element $\bra{k_2,h_2} (\i\hat N)^{n+1} \ket{k_1,h_1}$ reads
\begin{align}\nonumber
  &\bra{k_2,h_2} (\i\hat N)^{n+1} \ket{k_1,h_1} = \sum_h \int_{\ell}\dd(\ell^2)\theta(P\cdot \ell)\bra{k_2,h_2}(\i\hat N)^{n} \ket{\ell,h}\bra{\ell,h} \i\hat N \ket{k_1,h_1} \\\nonumber
  & = 4 \sum_h \int_{\ell}\dd(\ell^2)\theta(P\cdot \ell)(P\!\cdot\!k_1)^2 \dd(P\!\cdot\! (k_1-\ell))\dd( P\!\cdot\! (\ell-k_2))A^{(2h_1,2h)}(k_1,\ell)A^{(2h,2h_2)}(\ell,k_2)\\
  &\times \int\d^{D-2}\bperp{x}\int\d^{D-2}\bperp{y}\euler^{-\I(\bell - \bperpn{k}{1})\cdot \bperp{x} -\I (\bperpn{k}{2}-\bell)\cdot \bperp{y}}[-2 \I G(P\!\cdot\!k_1)]^{n+1}f(\bperp{x}^2)^n f(\bperp{y}^2)
  \label{eq:N_n+1_expanded}
\end{align}
after inserting eqs.~\eqref{eq:magnusian}~and~\eqref{eq:Nnoperator} for the matrix elements of $\hat N$ and $\hat N^n$. 
The crucial simplification arises from the observation that
\begin{align}\label{eq:completeness_relation_A}
   \sum_h A^{(2h_1,2h)}(k_1,\ell)A^{(2h,2h_2)}(\ell,k_2)=A^{(2h_1,2h_2)}(k_1,k_2)
\end{align}
in presence of the on-shell and energy-conserving terms $\dd(\ell^2)\dd(P\cdot(k_1-\ell))\dd(P\cdot(\ell-k_2))$ appearing in \eqn{eq:N_n+1_expanded}. This follows directly from the completeness relation for graviton polarizations, 
\begin{equation}\label{eq:grav_pol_sum}
\sum_{h}\varepsilon^{(2h)\mn}\bar{\varepsilon}^{(2h)\ab}= \Pi^{\mn\ab}\,.
\end{equation}
Inserting \eqn{eq:completeness_relation_A} in \eqn{eq:N_n+1_expanded}, the tensor structure $A^{(2h_1,2h_2)}(k_1,k_2)$ factorises from the integration. The remaining integrals over $\ell^+$ and $\ell^-$ are trivial on account of the $\dd$-distributions. Similarly, the integration over $\bperp{\ell}$ simply yields $\delta(\bperp y-\bperp x)$ which trivialises the integration over $\bperp{y}$. At this point, the positive energy requirement has been replaced with $\theta(P\cdot k_1)$. Since $k_1$ is taken on-shell, we may again replace $\theta(P\cdot k_1)\rightarrow \theta(k_1^0)$, which can be dropped since external states manifestly have positive energy. Thus, the expression reduces to
\begin{align}
  \bra{k_2,h_2} (\i\hat N)^{n+1} \ket{k_1,h_1}=& 2\dd(P\cdot(k_1-k_2))(P\cdot k_1)
A^{(2h_1,2h_2)}(k_1,k_2)\\
  &\times \int\d^{D-2}\bperp{x}
  \,
  \euler^{-\I (\bperpn{k}{2}-\bperpn{k}{1})\cdot\bperp{x}}
  \Big[-2 \I G (P\!\cdot\!k_1) f(\bperp{x}^2) \Big]^{n+1}\, .\nonumber
\end{align}
This concludes the inductive proof of \eqn{eq:Nnoperator} for all $n\in\mathbb N$. Consequently, substituting this expression into eq.~\eqref{eq:m+1_expansion_N}, we recover exactly the full response function $T^{(2 h_1,2 h_2)}$. This shows that the 1PM matrix element of \eqn{eq:magnusian} is PM-exact.

In particular, the matrix element of the Magnus operator is exactly given by the 1PM response and has the diagrammatic representation
\begin{equation}\label{eq:magnusian_diagrams}
  \bra{k_2,h_2}\i \hat N\ket{k_1,h_1}
  =\I
  T^{(2 h_1,2 h_2)}_\mathrm{1PM}(k_1,k_2)
  =
  \begin{tikzpicture}[baseline=(anchor)]
    \coordinate (anchor) at (0,-.42);
    \coordinate (currentLocation) at (0,0);
    \drawPropagationShort
    \drawCubic
    \drawPropagationShortDirected
  \end{tikzpicture}
  \;+\;
  \begin{tikzpicture}[baseline=(anchor)]
    \coordinate (anchor) at (0,-.42);
    \coordinate (currentLocation) at (0,0);
    \drawAbsorb
    \drawWL
    \drawEmitDirected
  \end{tikzpicture}\,,
\end{equation}
where in TT gauge the second diagram vanishes and one is left only with the ladder.
We may leverage this fact to show that the on-shell response exponentiates 1PM diagrams:
Expand \eqn{eq:response_magnusian_relation} to $n$'th PM order, and insert $n-1$ complete sets of states $\ket{\ell_j,h_j}$ labelled by $j=1,...,n-1$. One finds
\begin{align}\nonumber
  \i T_\text{$n$PM}^{(2 h_1,2 h_2)} &= \frac{1}{n!}\Big(\prod_{j=1}^{n-1}\sum_{h'_j}\int_{\ell_j}\dd(\ell_j^2)\theta(P\!\cdot\!\ell_j)\bra{\ell_{j+1},h'_{j+1}}\i \hat{N}\ket{\ell_j,h'_j}\Big)\bra{\ell_{1},h'_{1}}\i \hat{N}\ket{k_1,h_1}\\
  &=\frac{1}{n!}\left(
  \begin{tikzpicture}[baseline=(anchor)]
    \coordinate (anchor) at (0,-.42);
    \coordinate (currentLocation) at (0,0);
    \drawPropagationShort
    \drawCubic
    \drawPropagationShortDirected
  \end{tikzpicture}
  \;+\;
  \begin{tikzpicture}[baseline=(anchor)]
    \coordinate (anchor) at (0,-.42);
    \coordinate (currentLocation) at (0,0);
    \drawAbsorb
    \drawWL
    \drawEmitDirected
  \end{tikzpicture}
  \right)^n
  \label{eq:response_on-shell_nPM_diagram}
\end{align}
where we have denoted the final state $(\ell_n,h_n)=(k_2,h_2)$ and inserted \eqn{eq:magnusian_diagrams} in the last line. The nPM response is thus a product of 1PM diagrams; internal lines correspond to cut propagators $\dd(\ell^2)\theta(P\cdot \ell)$ and are integrated over momentum $\ell$ \footnote{The tensor structure $\Pi_\ell^{\mu\nu\alpha\beta}$ arises in \eqn{eq:response_on-shell_nPM_diagram} by evaluating the sum over helicities $h$ using \eqn{eq:grav_pol_sum}.}. The $n$PM response contribution of \eqn{eq:response_on-shell_nPM_diagram} may now be (re)summed over $n$, upon which
\begin{equation}
  \i T^{(2h_1,2h_2)}=
  \sum_{n=1}^\infty \i T_\text{$n$PM}^{(2 h_1,2 h_2)} 
  =\exp\left(
  \begin{tikzpicture}[baseline=(anchor)]
    \coordinate (anchor) at (0,-.42);
    \coordinate (currentLocation) at (0,0);
    \drawPropagationShort
    \drawCubic
    \drawPropagationShortDirected
  \end{tikzpicture}
  \;+\;
  \begin{tikzpicture}[baseline=(anchor)]
    \coordinate (anchor) at (0,-.42);
    \coordinate (currentLocation) at (0,0);
    \drawAbsorb
    \drawWL
    \drawEmitDirected
  \end{tikzpicture}
    \right)-1\,.
\end{equation}
We conclude that the PM-resummed shockwave response is simply an exponential of on-shell 1PM diagrams.

\section{Conclusions}\label{sec:conclusions}

In this work, we have formulated \emph{black hole response theory} as a WQFT-based framework to diagrammatically compute 
self-force dynamics for binary black-hole systems. Starting from the WQFT action of an isolated heavy black hole, we constructed  an effective black hole response action that is given through an infinite tower of response functions.
These response functions are defined as in-in correlation functions of $(n-1)$ incoming and one outgoing graviton, upon integrating out the
heavy black hole’s deviation from a straight line trajectory to all orders in the post-Minkowskian (PM) expansion. The two-point response function plays the role of the propagator in the background of a 
heavy black hole, including recoil, while the higher-point responses act as effective vertices in a non-flat background. Coupling the response functions to a compact object then yields a systematic diagrammatic self-force (SF) expansion of observables.  In particular, an $n$SF computation requires the knowledge of 
response functions only up to $(n+1)$-point order.

As an application, we considered a massless, high-energy particle as the ``heavy’’ object, sourcing the Aichelburg–Sexl shockwave geometry. In this setting, many PM diagrams vanish due to the null properties of the shockwave, making exact calculations in $G$ feasible. Within this theory, we derived both the Aichelburg–Sexl metric and the associated geodesic motion from a diagrammatic perspective. Importantly, the singular nature of the shockwave required the introduction of an additional finite-width regulator, on top of the dimensional regulator, in order to make the unregulated light-cone momentum integrals well-defined.

A central result of this work is the PM-exact computation of the two-point shockwave response function. Due to the vanishing of large classes of iterated diagrams, only ladder diagrams contribute at high PM orders. These may be computed at all orders, albeit so far only in the region for external graviton momenta being independent of the shockwave regulator scale $\Lambda$. Resumming these ladder contributions to all PM orders leads to an exponentiation in position space. The final two-point response consists of this exponential structure and a 2PM-exact remainder.
The infrared-divergent part factorises into a Weinberg phase, while the finite part takes
an fascinatingly simple Coulomb phase form relative to the Born term. 
Furthermore, extracting the matrix element of the 
$\hat N$-operator reproduces the 1PM contribution to the response function, making the
exponentiation structure manifest.

Having established the black hole response EFT using the WQFT formalism, there are several future research directions to pursue. The exact two-point response and the resummed probe vertices provide the necessary ingredients to compute physical observables, such as the 1SF waveform and impulse for the scattering of a massive particle off the high-energy
shockwave in an $m/E$ expansion. While additional regions in the integral could in principle contribute, we have verified that they do not affect the waveform, paving the way for an exact in $G$ calculation.

More generally, the two-point response function serves as a building block for higher-point response functions and thus for higher-order $m/E$ or SF type corrections. In particular, the required three-point response functions can be constructed from the two-point response, enabling the computation of 2SF observables, such as the impulse and waveform. These exact in $G$ results will be valuable for comparing to the high-energy limits of the
massive scattering scenario, which are divergent starting at 4PM and known at present up to 5PM order.
In addition they will shed light on the intricate question of the convergence of the PM expansion.

It would also be interesting to compare the presented results to a calculation in black hole perturbation theory. This would mean the calculation of quasi normal modes of the shockwave background and would help fixing the fiducial length scale~$L$.

Finally, the central future challenge is to take our formalism to the massive scattering 
scenario, with the grand goal of computing SF observables 
in physically relevant binary systems exactly in $G$. 

\section*{Acknowledgments}
We especially thank Vittorio del Duca, Riccardo Gonzo, Gustav Mogull and Emanuele Rosi for
very insightful discussions and important exchanges. Moreover, we
are thankful to Fabian Bautista, Thibault Damour, Kays Haddad, Gregor K\"alin, Dimitrios Kosmopoulos, Paolo Di Vecchia, Maarten van de Meent, Niels Warburton and Mao Zheng for helpful comments in the course of this work.
We greatly benefitted from interactions at the ``Amplitudes, Strong-Field Gravity and Resummation’’ workshop at Nordita where this work was first presented.
This work was funded by the Deutsche Forschungsgemeinschaft
(DFG, German Research Foundation)
Projektnummer 417533893/GRK2575 ``Rethinking Quantum Field Theory'' (LB,CE) 
and by the European Union through the 
European Research Council under grant ERC Advanced Grant 101097219 (GraWFTy) (JH,JP).
Views and opinions expressed are, however, those of the authors only and do not necessarily reflect those of the European Union or European Research Council Executive Agency. Neither the European Union nor the granting authority can be held responsible for them.

\appendix
\counterwithout*{equation}{subsection} % "undefine" previous equation counter reset
\counterwithin*{equation}{section} % reset equation counter for each section
\renewcommand{\theequation}{\Alph{section}.\arabic{equation}} % eq = (A.#)
\clearpage
\section{Integrals of the 2PM response}\label{app:I_ijk}
In this appendix, we provide the details of the 2PM response computation, specifically the evaluation of its integral family, \eqn{eq:response_2PM_family},
\begin{equation}\label{eq:response_2PM_family_appendix}
  I_{ijk}\equiv \int_{\ell}\frac{\dd(P\cdot l)\dd(P\cdot q)}{[\ell^2]^i[(k_1+\ell)^2+\i0^+k^-_1]^j[(q-\ell)^2]^k}\frac{\Lambda^4}{[(\ell^+)^2+\Lambda^2][(q^+-\ell^+)^2+\Lambda^2]}\, .
\end{equation}
The combination of integrals appearing in the 2PM response remainder $\Sigma_\mathrm{rem}^{\mu\nu\alpha\beta}\big|_{2\text{PM}}$ evaluates to
\begin{align}\label{eq:response_2PM_tensor_decomp_appendix}
  I_{011}+I_{110}+\bperp{q}^2 I_{111}=\I\frac{2\log \left(\frac{\Lambda  k_1^-}{\bperp q^2}\right)-1-3\i\pi }{16 \pi  (P\cdot k_1)}
  \dd(P\cdot q)
  +\mathcal O(\epsilon,\Lambda^{-1})
\end{align}
to leading order in $\Lambda$ and $\epsilon$. 
We show this by explicitly computing the three integral topologies.

The first integral, $I_{011}$, reads
\begin{equation}\label{eq:I011_def}
  I_{011}=\int_{\ell}\frac{\dd(P\cdot \ell)\dd(P\cdot q)}{[(k_1+\ell)^2+\i0^+k_1^-][(q-\ell)^2]}\frac{\Lambda^4}{[(\ell^+)^2+\Lambda^2][(q^+-\ell^+)^2+\Lambda^2]}\, .
\end{equation}
Performing the trivial integral over $\ell^-$, one finds
\begin{align}\label{eq:I011_l-_done}
  I_{011}=&-\frac{1}{2E}\int_{\bell\ell^+}\frac{\dd(P\cdot q)}{[k_1^-(k_1^++\ell^++\i0^+)-(\bperpn{k}{1}+\bell)^2](\bperp q-\bell)^2}\nonumber\\
  &\times\frac{\Lambda^4}{[(\ell^+)^2+\Lambda^2][(q^+-\ell^+)^2+\Lambda^2]}\, .
\end{align}
The $\ell^+$-integral is subsequently evaluated by contour integration, and is given by $\mathcal C_1$ of \eqn{eq:ladderContourResult}. For completeness, we repeat some details. The integrand of \eqn{eq:I011_l-_done} has five simple poles, located at
\begin{equation}
  \ell^+=\pm \i \Lambda, \quad\quad \ell^+=q^+\pm \i\Lambda, \quad\quad \ell^+=(\bperpn{k}{1}+\bell)^2/k_1^--k_1^+-\i0^+\, .
\end{equation}
Closing the contour in the upper half-plane, only two poles are picked up, and we find
\begin{align}\nonumber
  I_{011}=
  \dd(P\cdot q)
  \frac{\Lambda ^3}{4E}\Big[
  \frac{1}{(q^+)^2-2 \i \Lambda  q^+}
  &
  \int_{\bell}
  \frac{1}{(\bell-\bperp q)^2 ((\bperpn{k}{1}+\bell)^2-k_1^- (k_1^++\i \Lambda ))}\\
  +
  \frac{1}{(q^+)^2+2 \i \Lambda q^+ }
  &
  \int_{\bell}
  \frac{1}{(\bell-\bperp q)^2 ((\bperpn{k}{1}+\bell)^2-k_1^- (\i \Lambda +k_1^++q^+))}
  \Big]\,.
  \label{eq:I011_pf}
\end{align}
The remaining integrals may be evaluated with the method of regions, similar to the discussion in sec. \ref{sec:n-ladder_integration} following \eqn{eq:regions_In}. Letting external momenta scale independent of $\Lambda$,
\begin{equation}
  k_1\sim q\sim \Lambda^0\;,
\end{equation}
only the hard region (see \eqn{eq:regions_In}) is not scaleless,
\begin{equation}
  \bell\sim \Lambda^{1/2}\, .
\end{equation}
Correspondingly, expanding \eqn{eq:I011_pf} we find
\begin{align}\nonumber
  I_{011}&=
  \dd(P\cdot q)
  \Lambda^{-\epsilon}\int_{\bell}\frac{\left(\bell^2-2 \i k_1^-\right)}{8E \bell^2 \left(\bell^2-\i k_1^-\right)^2}
  +
  \mathcal O(\Lambda^{-1-\epsilon})\\
  &=
  \dd(P\cdot q)
  \Lambda^{-\epsilon}(I^{(1)}_{011}+I^{(2)}_{011})
  +
  \mathcal O(\Lambda^{-1-\epsilon})\, .
\end{align}
Integrals $I^{(1)}_{011},I^{(2)}_{011}$ are given by \footnote{Note that the $x$-integral of $I^{(2)}_{011}$ only converges for $-2<\epsilon<0$. However, all poles of \eqn{eq:I2} are isolated in the complex $\epsilon$-plane. Analytic continuation of \eqn{eq:I2} to $\epsilon\in \mathbb C$, including $\epsilon>0$, is thus possible.}
\begin{subequations}
\begin{equation}\label{eq:I1}
  I^{(1)}_{011}\equiv
  \int_{\bell}\frac{1}{8E \left(\bell^2-\i k_1^-\right)^2}
  =
  -\frac{2^{2 \epsilon -5} \pi ^{\epsilon -1} \epsilon \Lambda ^{-\epsilon } \Gamma (2-\epsilon ) \Gamma (\epsilon -1)}{E \Gamma (1-\epsilon )}(-\i k_1^-)^{-\epsilon -1}\, ,
\end{equation}
\begin{align}\nonumber
  I^{(2)}_{011}
  &\equiv
  \frac{-\i k_1^-}{4E}
  \int_{\bell}\frac{1}{\bell^2 \left(\bell^2-\i k_1^-\right)^2}
  =
  \frac{-\i k_1^-}{4E}
  \int_0^1\!\d x\,\int_{\bell}
  \frac{2(1- x)}{\left(\bell^2-\i k_1^- (1-x)\right)^3}\\\nonumber
  &=\frac{-\i k_1^-}{4E}
  \int_0^1\!\d x\,
  \frac{(x-1) (4 \pi )^{\epsilon -1} \epsilon  (\epsilon +1) \Gamma (2-\epsilon ) \Gamma (\epsilon -1) \left(\frac{\i}{k_1^- (1-x)}\right)^{\epsilon +2}}{\Gamma (1-\epsilon )}\\
  &=
\frac{\i k_1^-}{4E}
  \frac{(4 \pi )^{\epsilon -1} \Gamma (2-\epsilon ) \Gamma (\epsilon -1)}{(k_1^-)^2 \epsilon  \Gamma (-\epsilon -1)}(-\i k_1^-)^{-\epsilon }\, ,
  \label{eq:I2}
\end{align}
\end{subequations}
and taking the sum of eqs. \eqref{eq:I1} and \eqref{eq:I2}, $I_{011}$ at leading order in $\Lambda$ reads
\begin{equation}\label{eq:I011_exp}
  I_{011}
  =
  -\frac{2^{2 \epsilon -5} \pi ^{\epsilon -1} (\epsilon +2) \Gamma (\epsilon )}{E}
  \dd(P\cdot q)
  (-\i k_1^-)^{-\epsilon -1}
  \Lambda ^{-\epsilon }
  +
  \mathcal O(\Lambda^{-1-\epsilon})\, .
\end{equation}

Moving on to the integral $I_{110}$ we may relate it to $I_{011}$ of \eqn{eq:I011_def} by letting $q\rightarrow0$ after factoring out the $\dd(P\cdot q)$ term. From \eqn{eq:I011_exp}, one therefore has
\begin{align}
  I_{110}=
    \begin{aligned}[t]
      I_{011}
      +\mathcal O(\Lambda^{-1-\epsilon})\, .
    \end{aligned}
  \label{eq:I110}
\end{align}
Again, we implicitly assume $k_1\sim q\sim \Lambda^0$.

The remaining integral, $I_{111}$, was computed in the main text, eqs. \eqref{eq:externalIntegral} and \eqref{eq:spatialIntegral} with $n=1$. It reads
\begin{equation}\label{eq:I111_exp}
  I_{111}=
  -\frac{\I}{2} \frac{\dd( P\cdot q)}{(2 P\cdot k_1)}
  \frac{1}{(\bperp{q}^2)^{1+ \epsilon}}
  \frac{ \Gamma (-\epsilon )^{2} \Gamma (\epsilon +1)}{(4 \pi )^{ (1-\epsilon)}\Gamma (-2\epsilon )}
\end{equation}
The integrals eqs. \eqref{eq:I011_exp}, \eqref{eq:I110} and \eqref{eq:I111_exp} are plugged into \eqref{eq:response_2PM_tensor_decomp_appendix}, and expanded in $\epsilon$ to obtain \eqn{eq:response_2PM_tensor_decomp_appendix}.

\clearpage

\bibliographystyle{JHEP}
\bibliography{comptonshock.bib}

\providecommand{\href}[2]{#2}\begingroup\raggedright\begin{thebibliography}{100}

\bibitem{LIGOScientific:2016aoc}
{\scshape LIGO Scientific, Virgo} collaboration, B.~P. Abbott et~al., \emph{{Observation of Gravitational Waves from a Binary Black Hole Merger}}, \href{https://doi.org/10.1103/PhysRevLett.116.061102}{\emph{Phys. Rev. Lett.} {\bfseries 116} (2016) 061102} [\href{https://arxiv.org/abs/1602.03837}{{\ttfamily 1602.03837}}].

\bibitem{LIGOScientific:2017vwq}
{\scshape LIGO Scientific, Virgo} collaboration, B.~P. Abbott et~al., \emph{{GW170817: Observation of Gravitational Waves from a Binary Neutron Star Inspiral}}, \href{https://doi.org/10.1103/PhysRevLett.119.161101}{\emph{Phys. Rev. Lett.} {\bfseries 119} (2017) 161101} [\href{https://arxiv.org/abs/1710.05832}{{\ttfamily 1710.05832}}].

\bibitem{KAGRA:2021vkt}
{\scshape KAGRA, VIRGO, LIGO Scientific} collaboration, R.~Abbott et~al., \emph{{GWTC-3: Compact Binary Coalescences Observed by LIGO and Virgo during the Second Part of the Third Observing Run}}, \href{https://doi.org/10.1103/PhysRevX.13.041039}{\emph{Phys. Rev. X} {\bfseries 13} (2023) 041039} [\href{https://arxiv.org/abs/2111.03606}{{\ttfamily 2111.03606}}].

\bibitem{LIGOScientific:2025slb}
{\scshape LIGO Scientific, VIRGO, KAGRA} collaboration, A.~G. Abac et~al., \emph{{GWTC-4.0: Updating the Gravitational-Wave Transient Catalog with Observations from the First Part of the Fourth LIGO-Virgo-KAGRA Observing Run}},  \href{https://arxiv.org/abs/2508.18082}{{\ttfamily 2508.18082}}.

\bibitem{LISA:2017pwj}
{\scshape LISA} collaboration, P.~Amaro-Seoane et~al., \emph{{Laser Interferometer Space Antenna}},  \href{https://arxiv.org/abs/1702.00786}{{\ttfamily 1702.00786}}.

\bibitem{Punturo:2010zz}
M.~Punturo et~al., \emph{{The Einstein Telescope: A third-generation gravitational wave observatory}}, \href{https://doi.org/10.1088/0264-9381/27/19/194002}{\emph{Class. Quant. Grav.} {\bfseries 27} (2010) 194002}.

\bibitem{Ballmer:2022uxx}
S.~W. Ballmer et~al., \emph{{Snowmass2021 Cosmic Frontier White Paper: Future Gravitational-Wave Detector Facilities}},  in \emph{{Snowmass 2021}}, 3, 2022, \href{https://arxiv.org/abs/2203.08228}{{\ttfamily 2203.08228}}.

\bibitem{ET:2025xjr}
{\scshape ET} collaboration, A.~Abac et~al., \emph{{The Science of the Einstein Telescope}},  \href{https://arxiv.org/abs/2503.12263}{{\ttfamily 2503.12263}}.

\bibitem{Blanchet:2013haa}
L.~Blanchet, \emph{{Gravitational Radiation from Post-Newtonian Sources and Inspiralling Compact Binaries}}, \href{https://doi.org/10.12942/lrr-2014-2}{\emph{Living Rev. Rel.} {\bfseries 17} (2014) 2} [\href{https://arxiv.org/abs/1310.1528}{{\ttfamily 1310.1528}}].

\bibitem{Porto:2016pyg}
R.~A. Porto, \emph{{The effective field theorist\textquoteright{}s approach to gravitational dynamics}}, \href{https://doi.org/10.1016/j.physrep.2016.04.003}{\emph{Phys. Rept.} {\bfseries 633} (2016) 1} [\href{https://arxiv.org/abs/1601.04914}{{\ttfamily 1601.04914}}].

\bibitem{Levi:2018nxp}
M.~Levi, \emph{{Effective Field Theories of Post-Newtonian Gravity: A comprehensive review}}, \href{https://doi.org/10.1088/1361-6633/ab12bc}{\emph{Rept. Prog. Phys.} {\bfseries 83} (2020) 075901} [\href{https://arxiv.org/abs/1807.01699}{{\ttfamily 1807.01699}}].

\bibitem{Kosower:2022yvp}
D.~A. Kosower, R.~Monteiro and D.~O'Connell, \emph{{The SAGEX review on scattering amplitudes Chapter 14: Classical gravity from scattering amplitudes}}, \href{https://doi.org/10.1088/1751-8121/ac8846}{\emph{J. Phys. A} {\bfseries 55} (2022) 443015} [\href{https://arxiv.org/abs/2203.13025}{{\ttfamily 2203.13025}}].

\bibitem{Bjerrum-Bohr:2022blt}
N.~E.~J. Bjerrum-Bohr, P.~H. Damgaard, L.~Plante and P.~Vanhove, \emph{{The SAGEX review on scattering amplitudes Chapter 13: Post-Minkowskian expansion from scattering amplitudes}}, \href{https://doi.org/10.1088/1751-8121/ac7a78}{\emph{J. Phys. A} {\bfseries 55} (2022) 443014} [\href{https://arxiv.org/abs/2203.13024}{{\ttfamily 2203.13024}}].

\bibitem{Buonanno:2022pgc}
A.~Buonanno, M.~Khalil, D.~O'Connell, R.~Roiban, M.~P. Solon and M.~Zeng, \emph{{Snowmass White Paper: Gravitational Waves and Scattering Amplitudes}},  in \emph{{Snowmass 2021}}, 4, 2022, \href{https://arxiv.org/abs/2204.05194}{{\ttfamily 2204.05194}}.

\bibitem{DiVecchia:2023frv}
P.~Di~Vecchia, C.~Heissenberg, R.~Russo and G.~Veneziano, \emph{{The gravitational eikonal: From particle, string and brane collisions to black-hole encounters}}, \href{https://doi.org/10.1016/j.physrep.2024.06.002}{\emph{Phys. Rept.} {\bfseries 1083} (2024) 1} [\href{https://arxiv.org/abs/2306.16488}{{\ttfamily 2306.16488}}].

\bibitem{Jakobsen:2023oow}
G.~U. Jakobsen, \emph{{Gravitational Scattering of Compact Bodies from Worldline Quantum Field Theory}}, Ph.D. thesis, Humboldt U., Berlin, Humboldt U., Berlin (main), 2023.
\newblock \href{https://arxiv.org/abs/2308.04388}{{\ttfamily 2308.04388}}.
\newblock 10.18452/27075.

\bibitem{Kovacs:1978eu}
S.~J. Kovacs and K.~S. Thorne, \emph{{The Generation of Gravitational Waves. 4. Bremsstrahlung}}, \href{https://doi.org/10.1086/156350}{\emph{Astrophys. J.} {\bfseries 224} (1978) 62}.

\bibitem{Westpfahl:1979gu}
K.~Westpfahl and M.~Goller, \emph{{Gravitational scattering of two relativistic particles in postlinear approximation}}, \href{https://doi.org/10.1007/BF02817047}{\emph{Lett. Nuovo Cim.} {\bfseries 26} (1979) 573}.

\bibitem{Bel:1981be}
L.~Bel, T.~Damour, N.~Deruelle, J.~Ibanez and J.~Martin, \emph{{Poincar\'e-invariant gravitational field and equations of motion of two pointlike objects: The postlinear approximation of general relativity}}, \href{https://doi.org/10.1007/BF00756073}{\emph{Gen. Rel. Grav.} {\bfseries 13} (1981) 963}.

\bibitem{Damour:2017zjx}
T.~Damour, \emph{{High-energy gravitational scattering and the general relativistic two-body problem}}, \href{https://doi.org/10.1103/PhysRevD.97.044038}{\emph{Phys. Rev. D} {\bfseries 97} (2018) 044038} [\href{https://arxiv.org/abs/1710.10599}{{\ttfamily 1710.10599}}].

\bibitem{Hopper:2022rwo}
S.~Hopper, A.~Nagar and P.~Rettegno, \emph{{Strong-field scattering of two spinning black holes: Numerics versus analytics}}, \href{https://doi.org/10.1103/PhysRevD.107.124034}{\emph{Phys. Rev. D} {\bfseries 107} (2023) 124034} [\href{https://arxiv.org/abs/2204.10299}{{\ttfamily 2204.10299}}].

\bibitem{Kalin:2020mvi}
G.~K\"alin and R.~A. Porto, \emph{{Post-Minkowskian Effective Field Theory for Conservative Binary Dynamics}}, \href{https://doi.org/10.1007/JHEP11(2020)106}{\emph{JHEP} {\bfseries 11} (2020) 106} [\href{https://arxiv.org/abs/2006.01184}{{\ttfamily 2006.01184}}].

\bibitem{Kalin:2020fhe}
G.~K\"alin, Z.~Liu and R.~A. Porto, \emph{{Conservative Dynamics of Binary Systems to Third Post-Minkowskian Order from the Effective Field Theory Approach}}, \href{https://doi.org/10.1103/PhysRevLett.125.261103}{\emph{Phys. Rev. Lett.} {\bfseries 125} (2020) 261103} [\href{https://arxiv.org/abs/2007.04977}{{\ttfamily 2007.04977}}].

\bibitem{Kalin:2020lmz}
G.~K\"alin, Z.~Liu and R.~A. Porto, \emph{{Conservative Tidal Effects in Compact Binary Systems to Next-to-Leading Post-Minkowskian Order}}, \href{https://doi.org/10.1103/PhysRevD.102.124025}{\emph{Phys. Rev. D} {\bfseries 102} (2020) 124025} [\href{https://arxiv.org/abs/2008.06047}{{\ttfamily 2008.06047}}].

\bibitem{Mogull:2020sak}
G.~Mogull, J.~Plefka and J.~Steinhoff, \emph{{Classical black hole scattering from a worldline quantum field theory}}, \href{https://doi.org/10.1007/JHEP02(2021)048}{\emph{JHEP} {\bfseries 02} (2021) 048} [\href{https://arxiv.org/abs/2010.02865}{{\ttfamily 2010.02865}}].

\bibitem{Jakobsen:2021smu}
G.~U. Jakobsen, G.~Mogull, J.~Plefka and J.~Steinhoff, \emph{{Classical Gravitational Bremsstrahlung from a Worldline Quantum Field Theory}}, \href{https://doi.org/10.1103/PhysRevLett.126.201103}{\emph{Phys. Rev. Lett.} {\bfseries 126} (2021) 201103} [\href{https://arxiv.org/abs/2101.12688}{{\ttfamily 2101.12688}}].

\bibitem{Dlapa:2021npj}
C.~Dlapa, G.~K\"alin, Z.~Liu and R.~A. Porto, \emph{{Dynamics of binary systems to fourth Post-Minkowskian order from the effective field theory approach}}, \href{https://doi.org/10.1016/j.physletb.2022.137203}{\emph{Phys. Lett. B} {\bfseries 831} (2022) 137203} [\href{https://arxiv.org/abs/2106.08276}{{\ttfamily 2106.08276}}].

\bibitem{Dlapa:2021vgp}
C.~Dlapa, G.~K\"alin, Z.~Liu and R.~A. Porto, \emph{{Conservative Dynamics of Binary Systems at Fourth Post-Minkowskian Order in the Large-Eccentricity Expansion}}, \href{https://doi.org/10.1103/PhysRevLett.128.161104}{\emph{Phys. Rev. Lett.} {\bfseries 128} (2022) 161104} [\href{https://arxiv.org/abs/2112.11296}{{\ttfamily 2112.11296}}].

\bibitem{Mougiakakos:2021ckm}
S.~Mougiakakos, M.~M. Riva and F.~Vernizzi, \emph{{Gravitational Bremsstrahlung in the post-Minkowskian effective field theory}}, \href{https://doi.org/10.1103/PhysRevD.104.024041}{\emph{Phys. Rev. D} {\bfseries 104} (2021) 024041} [\href{https://arxiv.org/abs/2102.08339}{{\ttfamily 2102.08339}}].

\bibitem{Riva:2021vnj}
M.~M. Riva and F.~Vernizzi, \emph{{Radiated momentum in the post-Minkowskian worldline approach via reverse unitarity}}, \href{https://doi.org/10.1007/JHEP11(2021)228}{\emph{JHEP} {\bfseries 11} (2021) 228} [\href{https://arxiv.org/abs/2110.10140}{{\ttfamily 2110.10140}}].

\bibitem{Dlapa:2022lmu}
C.~Dlapa, G.~K\"alin, Z.~Liu, J.~Neef and R.~A. Porto, \emph{{Radiation Reaction and Gravitational Waves at Fourth Post-Minkowskian Order}}, \href{https://doi.org/10.1103/PhysRevLett.130.101401}{\emph{Phys. Rev. Lett.} {\bfseries 130} (2023) 101401} [\href{https://arxiv.org/abs/2210.05541}{{\ttfamily 2210.05541}}].

\bibitem{Dlapa:2023hsl}
C.~Dlapa, G.~K\"alin, Z.~Liu and R.~A. Porto, \emph{{Bootstrapping the relativistic two-body problem}}, \href{https://doi.org/10.1007/JHEP08(2023)109}{\emph{JHEP} {\bfseries 08} (2023) 109} [\href{https://arxiv.org/abs/2304.01275}{{\ttfamily 2304.01275}}].

\bibitem{Liu:2021zxr}
Z.~Liu, R.~A. Porto and Z.~Yang, \emph{{Spin Effects in the Effective Field Theory Approach to Post-Minkowskian Conservative Dynamics}}, \href{https://doi.org/10.1007/JHEP06(2021)012}{\emph{JHEP} {\bfseries 06} (2021) 012} [\href{https://arxiv.org/abs/2102.10059}{{\ttfamily 2102.10059}}].

\bibitem{Mougiakakos:2022sic}
S.~Mougiakakos, M.~M. Riva and F.~Vernizzi, \emph{{Gravitational Bremsstrahlung with Tidal Effects in the Post-Minkowskian Expansion}}, \href{https://doi.org/10.1103/PhysRevLett.129.121101}{\emph{Phys. Rev. Lett.} {\bfseries 129} (2022) 121101} [\href{https://arxiv.org/abs/2204.06556}{{\ttfamily 2204.06556}}].

\bibitem{Riva:2022fru}
M.~M. Riva, F.~Vernizzi and L.~K. Wong, \emph{{Gravitational bremsstrahlung from spinning binaries in the post-Minkowskian expansion}}, \href{https://doi.org/10.1103/PhysRevD.106.044013}{\emph{Phys. Rev. D} {\bfseries 106} (2022) 044013} [\href{https://arxiv.org/abs/2205.15295}{{\ttfamily 2205.15295}}].

\bibitem{Jakobsen:2021lvp}
G.~U. Jakobsen, G.~Mogull, J.~Plefka and J.~Steinhoff, \emph{{Gravitational Bremsstrahlung and Hidden Supersymmetry of Spinning Bodies}}, \href{https://doi.org/10.1103/PhysRevLett.128.011101}{\emph{Phys. Rev. Lett.} {\bfseries 128} (2022) 011101} [\href{https://arxiv.org/abs/2106.10256}{{\ttfamily 2106.10256}}].

\bibitem{Jakobsen:2021zvh}
G.~U. Jakobsen, G.~Mogull, J.~Plefka and J.~Steinhoff, \emph{{SUSY in the sky with gravitons}}, \href{https://doi.org/10.1007/JHEP01(2022)027}{\emph{JHEP} {\bfseries 01} (2022) 027} [\href{https://arxiv.org/abs/2109.04465}{{\ttfamily 2109.04465}}].

\bibitem{Jakobsen:2022fcj}
G.~U. Jakobsen and G.~Mogull, \emph{{Conservative and Radiative Dynamics of Spinning Bodies at Third Post-Minkowskian Order Using Worldline Quantum Field Theory}}, \href{https://doi.org/10.1103/PhysRevLett.128.141102}{\emph{Phys. Rev. Lett.} {\bfseries 128} (2022) 141102} [\href{https://arxiv.org/abs/2201.07778}{{\ttfamily 2201.07778}}].

\bibitem{Jakobsen:2022zsx}
G.~U. Jakobsen and G.~Mogull, \emph{{Linear response, Hamiltonian, and radiative spinning two-body dynamics}}, \href{https://doi.org/10.1103/PhysRevD.107.044033}{\emph{Phys. Rev. D} {\bfseries 107} (2023) 044033} [\href{https://arxiv.org/abs/2210.06451}{{\ttfamily 2210.06451}}].

\bibitem{Jakobsen:2022psy}
G.~U. Jakobsen, G.~Mogull, J.~Plefka and B.~Sauer, \emph{{All things retarded: radiation-reaction in worldline quantum field theory}}, \href{https://doi.org/10.1007/JHEP10(2022)128}{\emph{JHEP} {\bfseries 10} (2022) 128} [\href{https://arxiv.org/abs/2207.00569}{{\ttfamily 2207.00569}}].

\bibitem{Shi:2021qsb}
C.~Shi and J.~Plefka, \emph{{Classical double copy of worldline quantum field theory}}, \href{https://doi.org/10.1103/PhysRevD.105.026007}{\emph{Phys. Rev. D} {\bfseries 105} (2022) 026007} [\href{https://arxiv.org/abs/2109.10345}{{\ttfamily 2109.10345}}].

\bibitem{Bastianelli:2021nbs}
F.~Bastianelli, F.~Comberiati and L.~de~la Cruz, \emph{{Light bending from eikonal in worldline quantum field theory}}, \href{https://doi.org/10.1007/JHEP02(2022)209}{\emph{JHEP} {\bfseries 02} (2022) 209} [\href{https://arxiv.org/abs/2112.05013}{{\ttfamily 2112.05013}}].

\bibitem{Comberiati:2022cpm}
F.~Comberiati and C.~Shi, \emph{{Classical Double Copy of Spinning Worldline Quantum Field Theory}}, \href{https://doi.org/10.1007/JHEP04(2023)008}{\emph{JHEP} {\bfseries 04} (2023) 008} [\href{https://arxiv.org/abs/2212.13855}{{\ttfamily 2212.13855}}].

\bibitem{Wang:2022ntx}
T.~Wang, \emph{{Binary dynamics from worldline QFT for scalar QED}}, \href{https://doi.org/10.1103/PhysRevD.107.085011}{\emph{Phys. Rev. D} {\bfseries 107} (2023) 085011} [\href{https://arxiv.org/abs/2205.15753}{{\ttfamily 2205.15753}}].

\bibitem{Ben-Shahar:2023djm}
M.~Ben-Shahar, \emph{{Scattering of spinning compact objects from a worldline EFT}}, \href{https://doi.org/10.1007/JHEP03(2024)108}{\emph{JHEP} {\bfseries 03} (2024) 108} [\href{https://arxiv.org/abs/2311.01430}{{\ttfamily 2311.01430}}].

\bibitem{Bhattacharyya:2024aeq}
A.~Bhattacharyya, D.~Ghosh, S.~Ghosh and S.~Pal, \emph{{Observables from classical black hole scattering in Scalar-Tensor theory of gravity from worldline quantum field theory}}, \href{https://doi.org/10.1007/JHEP04(2024)015}{\emph{JHEP} {\bfseries 04} (2024) 015} [\href{https://arxiv.org/abs/2401.05492}{{\ttfamily 2401.05492}}].

\bibitem{Jakobsen:2023ndj}
G.~U. Jakobsen, G.~Mogull, J.~Plefka, B.~Sauer and Y.~Xu, \emph{{Conservative Scattering of Spinning Black Holes at Fourth Post-Minkowskian Order}}, \href{https://doi.org/10.1103/PhysRevLett.131.151401}{\emph{Phys. Rev. Lett.} {\bfseries 131} (2023) 151401} [\href{https://arxiv.org/abs/2306.01714}{{\ttfamily 2306.01714}}].

\bibitem{Jakobsen:2023hig}
G.~U. Jakobsen, G.~Mogull, J.~Plefka and B.~Sauer, \emph{{Dissipative Scattering of Spinning Black Holes at Fourth Post-Minkowskian Order}}, \href{https://doi.org/10.1103/PhysRevLett.131.241402}{\emph{Phys. Rev. Lett.} {\bfseries 131} (2023) 241402} [\href{https://arxiv.org/abs/2308.11514}{{\ttfamily 2308.11514}}].

\bibitem{Jakobsen:2023pvx}
G.~U. Jakobsen, G.~Mogull, J.~Plefka and B.~Sauer, \emph{{Tidal effects and renormalization at fourth post-Minkowskian order}}, \href{https://doi.org/10.1103/PhysRevD.109.L041504}{\emph{Phys. Rev. D} {\bfseries 109} (2024) L041504} [\href{https://arxiv.org/abs/2312.00719}{{\ttfamily 2312.00719}}].

\bibitem{Neill:2013wsa}
D.~Neill and I.~Z. Rothstein, \emph{{Classical Space-Times from the S Matrix}}, \href{https://doi.org/10.1016/j.nuclphysb.2013.09.007}{\emph{Nucl. Phys. B} {\bfseries 877} (2013) 177} [\href{https://arxiv.org/abs/1304.7263}{{\ttfamily 1304.7263}}].

\bibitem{Luna:2017dtq}
A.~Luna, I.~Nicholson, D.~O'Connell and C.~D. White, \emph{{Inelastic Black Hole Scattering from Charged Scalar Amplitudes}}, \href{https://doi.org/10.1007/JHEP03(2018)044}{\emph{JHEP} {\bfseries 03} (2018) 044} [\href{https://arxiv.org/abs/1711.03901}{{\ttfamily 1711.03901}}].

\bibitem{Kosower:2018adc}
D.~A. Kosower, B.~Maybee and D.~O'Connell, \emph{{Amplitudes, Observables, and Classical Scattering}}, \href{https://doi.org/10.1007/JHEP02(2019)137}{\emph{JHEP} {\bfseries 02} (2019) 137} [\href{https://arxiv.org/abs/1811.10950}{{\ttfamily 1811.10950}}].

\bibitem{Cristofoli:2021vyo}
A.~Cristofoli, R.~Gonzo, D.~A. Kosower and D.~O'Connell, \emph{{Waveforms from amplitudes}}, \href{https://doi.org/10.1103/PhysRevD.106.056007}{\emph{Phys. Rev. D} {\bfseries 106} (2022) 056007} [\href{https://arxiv.org/abs/2107.10193}{{\ttfamily 2107.10193}}].

\bibitem{Bjerrum-Bohr:2013bxa}
N.~E.~J. Bjerrum-Bohr, J.~F. Donoghue and P.~Vanhove, \emph{{On-shell Techniques and Universal Results in Quantum Gravity}}, \href{https://doi.org/10.1007/JHEP02(2014)111}{\emph{JHEP} {\bfseries 02} (2014) 111} [\href{https://arxiv.org/abs/1309.0804}{{\ttfamily 1309.0804}}].

\bibitem{Bjerrum-Bohr:2018xdl}
N.~E.~J. Bjerrum-Bohr, P.~H. Damgaard, G.~Festuccia, L.~Plant\'e and P.~Vanhove, \emph{{General Relativity from Scattering Amplitudes}}, \href{https://doi.org/10.1103/PhysRevLett.121.171601}{\emph{Phys. Rev. Lett.} {\bfseries 121} (2018) 171601} [\href{https://arxiv.org/abs/1806.04920}{{\ttfamily 1806.04920}}].

\bibitem{Bern:2019nnu}
Z.~Bern, C.~Cheung, R.~Roiban, C.-H. Shen, M.~P. Solon and M.~Zeng, \emph{{Scattering Amplitudes and the Conservative Hamiltonian for Binary Systems at Third Post-Minkowskian Order}}, \href{https://doi.org/10.1103/PhysRevLett.122.201603}{\emph{Phys. Rev. Lett.} {\bfseries 122} (2019) 201603} [\href{https://arxiv.org/abs/1901.04424}{{\ttfamily 1901.04424}}].

\bibitem{Bern:2019crd}
Z.~Bern, C.~Cheung, R.~Roiban, C.-H. Shen, M.~P. Solon and M.~Zeng, \emph{{Black Hole Binary Dynamics from the Double Copy and Effective Theory}}, \href{https://doi.org/10.1007/JHEP10(2019)206}{\emph{JHEP} {\bfseries 10} (2019) 206} [\href{https://arxiv.org/abs/1908.01493}{{\ttfamily 1908.01493}}].

\bibitem{Bjerrum-Bohr:2021wwt}
N.~E.~J. Bjerrum-Bohr, L.~Plant\'e and P.~Vanhove, \emph{{Post-Minkowskian radial action from soft limits and velocity cuts}}, \href{https://doi.org/10.1007/JHEP03(2022)071}{\emph{JHEP} {\bfseries 03} (2022) 071} [\href{https://arxiv.org/abs/2111.02976}{{\ttfamily 2111.02976}}].

\bibitem{Cheung:2020gyp}
C.~Cheung and M.~P. Solon, \emph{{Classical gravitational scattering at $ \mathcal{O} $(G$^{3}$) from Feynman diagrams}}, \href{https://doi.org/10.1007/JHEP06(2020)144}{\emph{JHEP} {\bfseries 06} (2020) 144} [\href{https://arxiv.org/abs/2003.08351}{{\ttfamily 2003.08351}}].

\bibitem{Bjerrum-Bohr:2021din}
N.~E.~J. Bjerrum-Bohr, P.~H. Damgaard, L.~Plant\'e and P.~Vanhove, \emph{{The amplitude for classical gravitational scattering at third Post-Minkowskian order}}, \href{https://doi.org/10.1007/JHEP08(2021)172}{\emph{JHEP} {\bfseries 08} (2021) 172} [\href{https://arxiv.org/abs/2105.05218}{{\ttfamily 2105.05218}}].

\bibitem{DiVecchia:2020ymx}
P.~Di~Vecchia, C.~Heissenberg, R.~Russo and G.~Veneziano, \emph{{Universality of ultra-relativistic gravitational scattering}}, \href{https://doi.org/10.1016/j.physletb.2020.135924}{\emph{Phys. Lett. B} {\bfseries 811} (2020) 135924} [\href{https://arxiv.org/abs/2008.12743}{{\ttfamily 2008.12743}}].

\bibitem{DiVecchia:2021bdo}
P.~Di~Vecchia, C.~Heissenberg, R.~Russo and G.~Veneziano, \emph{{The eikonal approach to gravitational scattering and radiation at $ \mathcal{O} $(G$^{3}$)}}, \href{https://doi.org/10.1007/JHEP07(2021)169}{\emph{JHEP} {\bfseries 07} (2021) 169} [\href{https://arxiv.org/abs/2104.03256}{{\ttfamily 2104.03256}}].

\bibitem{DiVecchia:2021ndb}
P.~Di~Vecchia, C.~Heissenberg, R.~Russo and G.~Veneziano, \emph{{Radiation Reaction from Soft Theorems}}, \href{https://doi.org/10.1016/j.physletb.2021.136379}{\emph{Phys. Lett. B} {\bfseries 818} (2021) 136379} [\href{https://arxiv.org/abs/2101.05772}{{\ttfamily 2101.05772}}].

\bibitem{DiVecchia:2022piu}
P.~Di~Vecchia, C.~Heissenberg, R.~Russo and G.~Veneziano, \emph{{Classical gravitational observables from the Eikonal operator}}, \href{https://doi.org/10.1016/j.physletb.2023.138049}{\emph{Phys. Lett. B} {\bfseries 843} (2023) 138049} [\href{https://arxiv.org/abs/2210.12118}{{\ttfamily 2210.12118}}].

\bibitem{Heissenberg:2022tsn}
C.~Heissenberg, \emph{{Angular Momentum Loss due to Tidal Effects in the Post-Minkowskian Expansion}}, \href{https://doi.org/10.1103/PhysRevLett.131.011603}{\emph{Phys. Rev. Lett.} {\bfseries 131} (2023) 011603} [\href{https://arxiv.org/abs/2210.15689}{{\ttfamily 2210.15689}}].

\bibitem{Damour:2020tta}
T.~Damour, \emph{{Radiative contribution to classical gravitational scattering at the third order in $G$}}, \href{https://doi.org/10.1103/PhysRevD.102.124008}{\emph{Phys. Rev. D} {\bfseries 102} (2020) 124008} [\href{https://arxiv.org/abs/2010.01641}{{\ttfamily 2010.01641}}].

\bibitem{Herrmann:2021tct}
E.~Herrmann, J.~Parra-Martinez, M.~S. Ruf and M.~Zeng, \emph{{Radiative classical gravitational observables at $ \mathcal{O} $(G$^{3}$) from scattering amplitudes}}, \href{https://doi.org/10.1007/JHEP10(2021)148}{\emph{JHEP} {\bfseries 10} (2021) 148} [\href{https://arxiv.org/abs/2104.03957}{{\ttfamily 2104.03957}}].

\bibitem{Damgaard:2019lfh}
P.~H. Damgaard, K.~Haddad and A.~Helset, \emph{{Heavy Black Hole Effective Theory}}, \href{https://doi.org/10.1007/JHEP11(2019)070}{\emph{JHEP} {\bfseries 11} (2019) 070} [\href{https://arxiv.org/abs/1908.10308}{{\ttfamily 1908.10308}}].

\bibitem{Damgaard:2021ipf}
P.~H. Damgaard, L.~Plante and P.~Vanhove, \emph{{On an exponential representation of the gravitational S-matrix}}, \href{https://doi.org/10.1007/JHEP11(2021)213}{\emph{JHEP} {\bfseries 11} (2021) 213} [\href{https://arxiv.org/abs/2107.12891}{{\ttfamily 2107.12891}}].

\bibitem{Damgaard:2023vnx}
P.~H. Damgaard, E.~R. Hansen, L.~Plant\'e and P.~Vanhove, \emph{{The relation between KMOC and worldline formalisms for classical gravity}}, \href{https://doi.org/10.1007/JHEP09(2023)059}{\emph{JHEP} {\bfseries 09} (2023) 059} [\href{https://arxiv.org/abs/2306.11454}{{\ttfamily 2306.11454}}].

\bibitem{Aoude:2020onz}
R.~Aoude, K.~Haddad and A.~Helset, \emph{{On-shell heavy particle effective theories}}, \href{https://doi.org/10.1007/JHEP05(2020)051}{\emph{JHEP} {\bfseries 05} (2020) 051} [\href{https://arxiv.org/abs/2001.09164}{{\ttfamily 2001.09164}}].

\bibitem{AccettulliHuber:2020dal}
M.~Accettulli~Huber, A.~Brandhuber, S.~De~Angelis and G.~Travaglini, \emph{{From amplitudes to gravitational radiation with cubic interactions and tidal effects}}, \href{https://doi.org/10.1103/PhysRevD.103.045015}{\emph{Phys. Rev. D} {\bfseries 103} (2021) 045015} [\href{https://arxiv.org/abs/2012.06548}{{\ttfamily 2012.06548}}].

\bibitem{Brandhuber:2021eyq}
A.~Brandhuber, G.~Chen, G.~Travaglini and C.~Wen, \emph{{Classical gravitational scattering from a gauge-invariant double copy}}, \href{https://doi.org/10.1007/JHEP10(2021)118}{\emph{JHEP} {\bfseries 10} (2021) 118} [\href{https://arxiv.org/abs/2108.04216}{{\ttfamily 2108.04216}}].

\bibitem{Bern:2021dqo}
Z.~Bern, J.~Parra-Martinez, R.~Roiban, M.~S. Ruf, C.-H. Shen, M.~P. Solon et~al., \emph{{Scattering Amplitudes and Conservative Binary Dynamics at ${\cal O}(G^4)$}}, \href{https://doi.org/10.1103/PhysRevLett.126.171601}{\emph{Phys. Rev. Lett.} {\bfseries 126} (2021) 171601} [\href{https://arxiv.org/abs/2101.07254}{{\ttfamily 2101.07254}}].

\bibitem{Bern:2021yeh}
Z.~Bern, J.~Parra-Martinez, R.~Roiban, M.~S. Ruf, C.-H. Shen, M.~P. Solon et~al., \emph{{Scattering Amplitudes, the Tail Effect, and Conservative Binary Dynamics at O(G4)}}, \href{https://doi.org/10.1103/PhysRevLett.128.161103}{\emph{Phys. Rev. Lett.} {\bfseries 128} (2022) 161103} [\href{https://arxiv.org/abs/2112.10750}{{\ttfamily 2112.10750}}].

\bibitem{Bern:2022kto}
Z.~Bern, D.~Kosmopoulos, A.~Luna, R.~Roiban and F.~Teng, \emph{{Binary Dynamics through the Fifth Power of Spin at O(G2)}}, \href{https://doi.org/10.1103/PhysRevLett.130.201402}{\emph{Phys. Rev. Lett.} {\bfseries 130} (2023) 201402} [\href{https://arxiv.org/abs/2203.06202}{{\ttfamily 2203.06202}}].

\bibitem{Bern:2023ity}
Z.~Bern, D.~Kosmopoulos, A.~Luna, R.~Roiban, T.~Scheopner, F.~Teng et~al., \emph{{Quantum field theory, worldline theory, and spin magnitude change in orbital evolution}}, \href{https://doi.org/10.1103/PhysRevD.109.045011}{\emph{Phys. Rev. D} {\bfseries 109} (2024) 045011} [\href{https://arxiv.org/abs/2308.14176}{{\ttfamily 2308.14176}}].

\bibitem{Damgaard:2023ttc}
P.~H. Damgaard, E.~R. Hansen, L.~Plant\'e and P.~Vanhove, \emph{{Classical observables from the exponential representation of the gravitational S-matrix}}, \href{https://doi.org/10.1007/JHEP09(2023)183}{\emph{JHEP} {\bfseries 09} (2023) 183} [\href{https://arxiv.org/abs/2307.04746}{{\ttfamily 2307.04746}}].

\bibitem{Brandhuber:2023hhy}
A.~Brandhuber, G.~R. Brown, G.~Chen, S.~De~Angelis, J.~Gowdy and G.~Travaglini, \emph{{One-loop gravitational bremsstrahlung and waveforms from a heavy-mass effective field theory}}, \href{https://doi.org/10.1007/JHEP06(2023)048}{\emph{JHEP} {\bfseries 06} (2023) 048} [\href{https://arxiv.org/abs/2303.06111}{{\ttfamily 2303.06111}}].

\bibitem{Brandhuber:2023hhl}
A.~Brandhuber, G.~R. Brown, G.~Chen, J.~Gowdy and G.~Travaglini, \emph{{Resummed spinning waveforms from five-point amplitudes}}, \href{https://doi.org/10.1007/JHEP02(2024)026}{\emph{JHEP} {\bfseries 02} (2024) 026} [\href{https://arxiv.org/abs/2310.04405}{{\ttfamily 2310.04405}}].

\bibitem{DeAngelis:2023lvf}
S.~De~Angelis, P.~P. Novichkov and R.~Gonzo, \emph{{Spinning waveforms from the Kosower-Maybee-O\textquoteright{}Connell formalism at leading order}}, \href{https://doi.org/10.1103/PhysRevD.110.L041502}{\emph{Phys. Rev. D} {\bfseries 110} (2024) L041502} [\href{https://arxiv.org/abs/2309.17429}{{\ttfamily 2309.17429}}].

\bibitem{Herderschee:2023fxh}
A.~Herderschee, R.~Roiban and F.~Teng, \emph{{The sub-leading scattering waveform from amplitudes}}, \href{https://doi.org/10.1007/JHEP06(2023)004}{\emph{JHEP} {\bfseries 06} (2023) 004} [\href{https://arxiv.org/abs/2303.06112}{{\ttfamily 2303.06112}}].

\bibitem{Caron-Huot:2023vxl}
S.~Caron-Huot, M.~Giroux, H.~S. Hannesdottir and S.~Mizera, \emph{{What can be measured asymptotically?}}, \href{https://doi.org/10.1007/JHEP01(2024)139}{\emph{JHEP} {\bfseries 01} (2024) 139} [\href{https://arxiv.org/abs/2308.02125}{{\ttfamily 2308.02125}}].

\bibitem{FebresCordero:2022jts}
F.~Febres~Cordero, M.~Kraus, G.~Lin, M.~S. Ruf and M.~Zeng, \emph{{Conservative Binary Dynamics with a Spinning Black Hole at O(G3) from Scattering Amplitudes}}, \href{https://doi.org/10.1103/PhysRevLett.130.021601}{\emph{Phys. Rev. Lett.} {\bfseries 130} (2023) 021601} [\href{https://arxiv.org/abs/2205.07357}{{\ttfamily 2205.07357}}].

\bibitem{Bohnenblust:2023qmy}
L.~Bohnenblust, H.~Ita, M.~Kraus and J.~Schlenk, \emph{{Gravitational Bremsstrahlung in black-hole scattering at $ \mathcal{O}\left({G}^3\right) $: linear-in-spin effects}}, \href{https://doi.org/10.1007/JHEP11(2024)109}{\emph{JHEP} {\bfseries 11} (2024) 109} [\href{https://arxiv.org/abs/2312.14859}{{\ttfamily 2312.14859}}].

\bibitem{Bern:2025zno}
Z.~Bern, E.~Herrmann, R.~Roiban, M.~S. Ruf, A.~V. Smirnov, V.~A. Smirnov et~al., \emph{{Second-order self-force potential-region binary dynamics at $O(G^5)$ in supergravity}},  \href{https://arxiv.org/abs/2509.17412}{{\ttfamily 2509.17412}}.

\bibitem{Bern:2025wyd}
Z.~Bern, E.~Herrmann, R.~Roiban, M.~S. Ruf, A.~V. Smirnov, S.~Smith et~al., \emph{{Scattering Amplitudes and Conservative Binary Dynamics at $O(G^5)$ without Self-Force Truncation}},  \href{https://arxiv.org/abs/2512.23654}{{\ttfamily 2512.23654}}.

\bibitem{Goldberger:2004jt}
W.~D. Goldberger and I.~Z. Rothstein, \emph{{An Effective field theory of gravity for extended objects}}, \href{https://doi.org/10.1103/PhysRevD.73.104029}{\emph{Phys. Rev. D} {\bfseries 73} (2006) 104029} [\href{https://arxiv.org/abs/hep-th/0409156}{{\ttfamily hep-th/0409156}}].

\bibitem{Haddad:2024ebn}
K.~Haddad, G.~U. Jakobsen, G.~Mogull and J.~Plefka, \emph{{Spinning bodies in general relativity from bosonic worldline oscillators}}, \href{https://doi.org/10.1007/JHEP02(2025)019}{\emph{JHEP} {\bfseries 02} (2025) 019} [\href{https://arxiv.org/abs/2411.08176}{{\ttfamily 2411.08176}}].

\bibitem{Driesse:2024xad}
M.~Driesse, G.~U. Jakobsen, G.~Mogull, J.~Plefka, B.~Sauer and J.~Usovitsch, \emph{{Conservative Black Hole Scattering at Fifth Post-Minkowskian and First Self-Force Order}}, \href{https://doi.org/10.1103/PhysRevLett.132.241402}{\emph{Phys. Rev. Lett.} {\bfseries 132} (2024) 241402} [\href{https://arxiv.org/abs/2403.07781}{{\ttfamily 2403.07781}}].

\bibitem{Driesse:2024feo}
M.~Driesse, G.~U. Jakobsen, A.~Klemm, G.~Mogull, C.~Nega, J.~Plefka et~al., \emph{{Emergence of Calabi{\textendash}Yau manifolds in high-precision black-hole scattering}}, \href{https://doi.org/10.1038/s41586-025-08984-2}{\emph{Nature} {\bfseries 641} (2025) 603} [\href{https://arxiv.org/abs/2411.11846}{{\ttfamily 2411.11846}}].

\bibitem{Driesse:2026qiz}
M.~Driesse, G.~U. Jakobsen, G.~Mogull, C.~Nega, J.~Plefka, B.~Sauer et~al., \emph{{Conservative Black Hole Scattering at Fifth Post-Minkowskian and Second Self-Force Order}},  \href{https://arxiv.org/abs/2601.16256}{{\ttfamily 2601.16256}}.

\bibitem{Bohnenblust:2025gir}
L.~Bohnenblust, H.~Ita, M.~Kraus and J.~Schlenk, \emph{{Gravitational Bremsstrahlung in black-hole scattering at $ \mathcal{O}\left({G}^3\right) $: quadratic-in-spin effects}}, \href{https://doi.org/10.1007/JHEP12(2025)100}{\emph{JHEP} {\bfseries 12} (2025) 100} [\href{https://arxiv.org/abs/2505.15724}{{\ttfamily 2505.15724}}].

\bibitem{Kim:2024svw}
J.-H. Kim, J.-W. Kim, S.~Kim and S.~Lee, \emph{{Classical eikonal from Magnus expansion}}, \href{https://doi.org/10.1007/JHEP01(2025)111}{\emph{JHEP} {\bfseries 01} (2025) 111} [\href{https://arxiv.org/abs/2410.22988}{{\ttfamily 2410.22988}}].

\bibitem{Kim:2025gis}
J.-W. Kim, R.~Patil, T.~Scheopner and J.~Steinhoff, \emph{{Magnusian: relating the eikonal phase, the on-shell action, and the scattering generator}}, \href{https://doi.org/10.1007/JHEP03(2026)241}{\emph{JHEP} {\bfseries 03} (2026) 241} [\href{https://arxiv.org/abs/2511.05649}{{\ttfamily 2511.05649}}].

\bibitem{Gonzo:2026yha}
R.~Gonzo and G.~Mogull, \emph{{Canonical Quantisation of Bound and Unbound WQFT}},  \href{https://arxiv.org/abs/2603.05237}{{\ttfamily 2603.05237}}.

\bibitem{Foffa:2016rgu}
S.~Foffa, P.~Mastrolia, R.~Sturani and C.~Sturm, \emph{{Effective field theory approach to the gravitational two-body dynamics, at fourth post-Newtonian order and quintic in the Newton constant}}, \href{https://doi.org/10.1103/PhysRevD.95.104009}{\emph{Phys. Rev. D} {\bfseries 95} (2017) 104009} [\href{https://arxiv.org/abs/1612.00482}{{\ttfamily 1612.00482}}].

\bibitem{Foffa:2019rdf}
S.~Foffa and R.~Sturani, \emph{{Conservative dynamics of binary systems to fourth Post-Newtonian order in the EFT approach I: Regularized Lagrangian}}, \href{https://doi.org/10.1103/PhysRevD.100.024047}{\emph{Phys. Rev. D} {\bfseries 100} (2019) 024047} [\href{https://arxiv.org/abs/1903.05113}{{\ttfamily 1903.05113}}].

\bibitem{Foffa:2019yfl}
S.~Foffa, R.~A. Porto, I.~Rothstein and R.~Sturani, \emph{{Conservative dynamics of binary systems to fourth Post-Newtonian order in the EFT approach II: Renormalized Lagrangian}}, \href{https://doi.org/10.1103/PhysRevD.100.024048}{\emph{Phys. Rev. D} {\bfseries 100} (2019) 024048} [\href{https://arxiv.org/abs/1903.05118}{{\ttfamily 1903.05118}}].

\bibitem{Foffa:2019hrb}
S.~Foffa, P.~Mastrolia, R.~Sturani, C.~Sturm and W.~J. Torres~Bobadilla, \emph{{Static two-body potential at fifth post-Newtonian order}}, \href{https://doi.org/10.1103/PhysRevLett.122.241605}{\emph{Phys. Rev. Lett.} {\bfseries 122} (2019) 241605} [\href{https://arxiv.org/abs/1902.10571}{{\ttfamily 1902.10571}}].

\bibitem{Blumlein:2019zku}
J.~Bl{\"u}mlein, A.~Maier and P.~Marquard, \emph{{Five-Loop Static Contribution to the Gravitational Interaction Potential of Two Point Masses}}, \href{https://doi.org/10.1016/j.physletb.2019.135100}{\emph{Phys. Lett. B} {\bfseries 800} (2020) 135100} [\href{https://arxiv.org/abs/1902.11180}{{\ttfamily 1902.11180}}].

\bibitem{Foffa:2020nqe}
S.~Foffa, R.~Sturani and W.~J. Torres~Bobadilla, \emph{{Efficient resummation of high post-Newtonian contributions to the binding energy}}, \href{https://doi.org/10.1007/JHEP02(2021)165}{\emph{JHEP} {\bfseries 02} (2021) 165} [\href{https://arxiv.org/abs/2010.13730}{{\ttfamily 2010.13730}}].

\bibitem{Blumlein:2020pyo}
J.~Bl{\"u}mlein, A.~Maier, P.~Marquard and G.~Sch{\"a}fer, \emph{{The fifth-order post-Newtonian Hamiltonian dynamics of two-body systems from an effective field theory approach: potential contributions}}, \href{https://doi.org/10.1016/j.nuclphysb.2021.115352}{\emph{Nucl. Phys. B} {\bfseries 965} (2021) 115352} [\href{https://arxiv.org/abs/2010.13672}{{\ttfamily 2010.13672}}].

\bibitem{Blumlein:2021txe}
J.~Bl\"umlein, A.~Maier, P.~Marquard and G.~Sch\"afer, \emph{{The fifth-order post-Newtonian Hamiltonian dynamics of two-body systems from an effective field theory approach}}, \href{https://doi.org/10.1016/j.nuclphysb.2022.115900}{\emph{Nucl. Phys. B} {\bfseries 983} (2022) 115900} [\href{https://arxiv.org/abs/2110.13822}{{\ttfamily 2110.13822}}].

\bibitem{Porto:2024cwd}
R.~A. Porto, M.~M. Riva and Z.~Yang, \emph{{Nonlinear gravitational radiation reaction: failed tail, memories {\&} squares}}, \href{https://doi.org/10.1007/JHEP04(2025)050}{\emph{JHEP} {\bfseries 04} (2025) 050} [\href{https://arxiv.org/abs/2409.05860}{{\ttfamily 2409.05860}}].

\bibitem{Porto:2026fsd}
R.~A. Porto and M.~M. Riva, \emph{{Black Hole Dynamics at Fifth Post-Newtonian Order}},  \href{https://arxiv.org/abs/2604.09545}{{\ttfamily 2604.09545}}.

\bibitem{Blumlein:2021txj}
J.~Bl{\"u}mlein, A.~Maier, P.~Marquard and G.~Sch{\"a}fer, \emph{{The 6th post-Newtonian potential terms at $O(G_N^4)$}}, \href{https://doi.org/10.1016/j.physletb.2021.136260}{\emph{Phys. Lett. B} {\bfseries 816} (2021) 136260} [\href{https://arxiv.org/abs/2101.08630}{{\ttfamily 2101.08630}}].

\bibitem{Almeida:2026clf}
G.~L. Almeida, A.~M{\"u}ller, S.~Foffa and R.~Sturani, \emph{{Angular momentum tail contributions to compact binary dynamics}},  \href{https://arxiv.org/abs/2603.20096}{{\ttfamily 2603.20096}}.

\bibitem{Brunello:2025gpf}
G.~Brunello, M.~K. Mandal, P.~Mastrolia, R.~Patil, M.~Pegorin, J.~Ronca et~al., \emph{{Six-loop gravitational interactions at the sixth post-Newtonian order}},  \href{https://arxiv.org/abs/2512.19498}{{\ttfamily 2512.19498}}.

\bibitem{Brunello:2026anu}
G.~Brunello, M.~K. Mandal, P.~Mastrolia, R.~Patil, M.~Pegorin, S.~Smith et~al., \emph{{All-order structure of static gravitational interactions and the seventh post-Newtonian potential}},  \href{https://arxiv.org/abs/2604.14134}{{\ttfamily 2604.14134}}.

\bibitem{Schwarzschild:1916uq}
K.~Schwarzschild, \emph{{{\"U}ber das Gravitationsfeld eines Massenpunktes nach der Einstein'schen Theorie}}, {\emph{Sitzungsberichte der K{\"o}niglich Preussischen Akademie der Wissenschaften zu Berlin (Phys.-Math.~Klasse),} {\bfseries 1916} (1916) 189} [\href{https://arxiv.org/abs/physics/9905030}{{\ttfamily physics/9905030}}].

\bibitem{Kerr:1963ud}
R.~P. Kerr, \emph{{Gravitational field of a spinning mass as an example of algebraically special metrics}}, \href{https://doi.org/10.1103/PhysRevLett.11.237}{\emph{Phys. Rev. Lett.} {\bfseries 11} (1963) 237}.

\bibitem{Aichelburg:1970dh}
P.~C. Aichelburg and R.~U. Sexl, \emph{{On the Gravitational field of a massless particle}}, \href{https://doi.org/10.1007/BF00758149}{\emph{Gen. Rel. Grav.} {\bfseries 2} (1971) 303}.

\bibitem{Mino:1996nk}
Y.~Mino, M.~Sasaki and T.~Tanaka, \emph{{Gravitational radiation reaction to a particle motion}}, \href{https://doi.org/10.1103/PhysRevD.55.3457}{\emph{Phys. Rev. D} {\bfseries 55} (1997) 3457} [\href{https://arxiv.org/abs/gr-qc/9606018}{{\ttfamily gr-qc/9606018}}].

\bibitem{Poisson:2011nh}
E.~Poisson, A.~Pound and I.~Vega, \emph{{The Motion of point particles in curved spacetime}}, \href{https://doi.org/10.12942/lrr-2011-7}{\emph{Living Rev. Rel.} {\bfseries 14} (2011) 7} [\href{https://arxiv.org/abs/1102.0529}{{\ttfamily 1102.0529}}].

\bibitem{Barack:2018yvs}
L.~Barack and A.~Pound, \emph{{Self-force and radiation reaction in general relativity}}, \href{https://doi.org/10.1088/1361-6633/aae552}{\emph{Rept. Prog. Phys.} {\bfseries 82} (2019) 016904} [\href{https://arxiv.org/abs/1805.10385}{{\ttfamily 1805.10385}}].

\bibitem{Gralla:2021qaf}
S.~E. Gralla and K.~Lobo, \emph{{Self-force effects in post-Minkowskian scattering}}, \href{https://doi.org/10.1088/1361-6382/ac5d88}{\emph{Class. Quant. Grav.} {\bfseries 39} (2022) 095001} [\href{https://arxiv.org/abs/2110.08681}{{\ttfamily 2110.08681}}].

\bibitem{Pound:2019lzj}
A.~Pound, B.~Wardell, N.~Warburton and J.~Miller, \emph{{Second-Order Self-Force Calculation of Gravitational Binding Energy in Compact Binaries}}, \href{https://doi.org/10.1103/PhysRevLett.124.021101}{\emph{Phys. Rev. Lett.} {\bfseries 124} (2020) 021101} [\href{https://arxiv.org/abs/1908.07419}{{\ttfamily 1908.07419}}].

\bibitem{Kavanagh:2015lva}
C.~Kavanagh, A.~C. Ottewill and B.~Wardell, \emph{{Analytical high-order post-Newtonian expansions for extreme mass ratio binaries}}, \href{https://doi.org/10.1103/PhysRevD.92.084025}{\emph{Phys. Rev. D} {\bfseries 92} (2015) 084025} [\href{https://arxiv.org/abs/1503.02334}{{\ttfamily 1503.02334}}].

\bibitem{vandeMeent:2017bcc}
M.~van~de Meent, \emph{{Gravitational self-force on generic bound geodesics in Kerr spacetime}}, \href{https://doi.org/10.1103/PhysRevD.97.104033}{\emph{Phys. Rev. D} {\bfseries 97} (2018) 104033} [\href{https://arxiv.org/abs/1711.09607}{{\ttfamily 1711.09607}}].

\bibitem{Pound:2021qin}
A.~Pound and B.~Wardell, \emph{{Black hole perturbation theory and gravitational self-force}},  \href{https://arxiv.org/abs/2101.04592}{{\ttfamily 2101.04592}}.

\bibitem{Wardell:2021fyy}
B.~Wardell, A.~Pound, N.~Warburton, J.~Miller, L.~Durkan and A.~Le~Tiec, \emph{{Gravitational Waveforms for Compact Binaries from Second-Order Self-Force Theory}}, \href{https://doi.org/10.1103/PhysRevLett.130.241402}{\emph{Phys. Rev. Lett.} {\bfseries 130} (2023) 241402} [\href{https://arxiv.org/abs/2112.12265}{{\ttfamily 2112.12265}}].

\bibitem{Lynch:2023gpu}
P.~Lynch, M.~van~de Meent and N.~Warburton, \emph{{Self-forced inspirals with spin-orbit precession}}, \href{https://doi.org/10.1103/PhysRevD.109.084072}{\emph{Phys. Rev. D} {\bfseries 109} (2024) 084072} [\href{https://arxiv.org/abs/2305.10533}{{\ttfamily 2305.10533}}].

\bibitem{Kuchler:2024esj}
L.~K{\"u}chler, G.~Comp{\`e}re, L.~Durkan and A.~Pound, \emph{{Self-force framework for transition-to-plunge waveforms}}, \href{https://doi.org/10.21468/SciPostPhys.17.2.056}{\emph{SciPost Phys.} {\bfseries 17} (2024) 056} [\href{https://arxiv.org/abs/2405.00170}{{\ttfamily 2405.00170}}].

\bibitem{Honet:2025lmk}
L.~Honet, J.~Mathews, G.~Comp{\`e}re, A.~Pound, B.~Wardell, G.~A. Piovano et~al., \emph{{Spin-aligned inspiral waveforms from self-force and post-Newtonian theory}},  \href{https://arxiv.org/abs/2510.16112}{{\ttfamily 2510.16112}}.

\bibitem{Kuchler:2025hwx}
L.~K{\"u}chler, G.~Comp{\`e}re and A.~Pound, \emph{{Self-force framework for merger-ringdown waveforms}}, \href{https://doi.org/10.1088/1361-6382/ae2b44}{\emph{Class. Quant. Grav.} {\bfseries 43} (2026) 015018} [\href{https://arxiv.org/abs/2506.02189}{{\ttfamily 2506.02189}}].

\bibitem{Pani:2013pma}
P.~Pani, \emph{{Advanced Methods in Black-Hole Perturbation Theory}}, \href{https://doi.org/10.1142/S0217751X13400186}{\emph{Int. J. Mod. Phys. A} {\bfseries 28} (2013) 1340018} [\href{https://arxiv.org/abs/1305.6759}{{\ttfamily 1305.6759}}].

\bibitem{Bini:2018qvd}
D.~Bini and A.~Geralico, \emph{{Black Hole Perturbations: A Review of Recent Analytical Results}}, \href{https://doi.org/10.1007/s10701-018-0187-7}{\emph{Found. Phys.} {\bfseries 48} (2018) 1349}.

\bibitem{Sasaki:2003xr}
M.~Sasaki and H.~Tagoshi, \emph{{Analytic black hole perturbation approach to gravitational radiation}}, \href{https://doi.org/10.12942/lrr-2003-6}{\emph{Living Rev. Rel.} {\bfseries 6} (2003) 6} [\href{https://arxiv.org/abs/gr-qc/0306120}{{\ttfamily gr-qc/0306120}}].

\bibitem{Berti:2025hly}
J.~Abedi et~al., \emph{{Black hole spectroscopy: from theory to experiment}},  \href{https://arxiv.org/abs/2505.23895}{{\ttfamily 2505.23895}}.

\bibitem{Duff:1973zz}
M.~J. Duff, \emph{{Quantum Tree Graphs and the Schwarzschild Solution}}, \href{https://doi.org/10.1103/PhysRevD.7.2317}{\emph{Phys. Rev. D} {\bfseries 7} (1973) 2317}.

\bibitem{Damgaard:2024fqj}
P.~H. Damgaard and K.~Lee, \emph{{Schwarzschild Black Hole from Perturbation Theory to All Orders}}, \href{https://doi.org/10.1103/PhysRevLett.132.251603}{\emph{Phys. Rev. Lett.} {\bfseries 132} (2024) 251603} [\href{https://arxiv.org/abs/2403.13216}{{\ttfamily 2403.13216}}].

\bibitem{Mougiakakos:2024nku}
S.~Mougiakakos and P.~Vanhove, \emph{{Schwarzschild Metric from Scattering Amplitudes to All Orders in GN}}, \href{https://doi.org/10.1103/PhysRevLett.133.111601}{\emph{Phys. Rev. Lett.} {\bfseries 133} (2024) 111601} [\href{https://arxiv.org/abs/2405.14421}{{\ttfamily 2405.14421}}].

\bibitem{Mougiakakos:2024lif}
S.~Mougiakakos and P.~Vanhove, \emph{{Schwarzschild geodesics from scattering amplitudes to all orders in G$_{N}$}}, \href{https://doi.org/10.1007/JHEP10(2024)152}{\emph{JHEP} {\bfseries 10} (2024) 152} [\href{https://arxiv.org/abs/2407.09448}{{\ttfamily 2407.09448}}].

\bibitem{Damgaard:2026kqg}
P.~H. Damgaard, H.~Lee, K.~Lee and T.~Rahnuma, \emph{{Gravitational Metric of a Star}},  \href{https://arxiv.org/abs/2603.16493}{{\ttfamily 2603.16493}}.

\bibitem{Dray:1984ha}
T.~Dray and G.~'t~Hooft, \emph{{The Gravitational Shock Wave of a Massless Particle}}, \href{https://doi.org/10.1016/0550-3213(85)90525-5}{\emph{Nucl. Phys. B} {\bfseries 253} (1985) 173}.

\bibitem{Steinbauer:1997dw}
R.~Steinbauer, \emph{{Geodesics and geodesic deviation for impulsive gravitational waves}}, \href{https://doi.org/10.1063/1.532283}{\emph{J. Math. Phys.} {\bfseries 39} (1998) 2201} [\href{https://arxiv.org/abs/gr-qc/9710119}{{\ttfamily gr-qc/9710119}}].

\bibitem{Cheung:2023lnj}
C.~Cheung, J.~Parra-Martinez, I.~Z. Rothstein, N.~Shah and J.~Wilson-Gerow, \emph{{Effective Field Theory for Extreme Mass Ratio Binaries}}, \href{https://doi.org/10.1103/PhysRevLett.132.091402}{\emph{Phys. Rev. Lett.} {\bfseries 132} (2024) 091402} [\href{https://arxiv.org/abs/2308.14832}{{\ttfamily 2308.14832}}].

\bibitem{Cheung:2024byb}
C.~Cheung, J.~Parra-Martinez, I.~Z. Rothstein, N.~Shah and J.~Wilson-Gerow, \emph{{Gravitational scattering and beyond from extreme mass ratio effective field theory}}, \href{https://doi.org/10.1007/JHEP10(2024)005}{\emph{JHEP} {\bfseries 10} (2024) 005} [\href{https://arxiv.org/abs/2406.14770}{{\ttfamily 2406.14770}}].

\bibitem{Bjerrum-Bohr:2025bqg}
N.~E.~J. Bjerrum-Bohr, G.~Chen, C.~J. Eriksen and N.~Shah, \emph{{The gravitational Compton amplitude from flat and curved spacetimes at second post-Minkowskian order}}, \href{https://doi.org/10.1007/JHEP10(2025)235}{\emph{JHEP} {\bfseries 10} (2025) 235} [\href{https://arxiv.org/abs/2506.19705}{{\ttfamily 2506.19705}}].

\bibitem{Kosmopoulos:2023bwc}
D.~Kosmopoulos and M.~P. Solon, \emph{{Gravitational self force from scattering amplitudes in curved space}}, \href{https://doi.org/10.1007/JHEP03(2024)125}{\emph{JHEP} {\bfseries 03} (2024) 125} [\href{https://arxiv.org/abs/2308.15304}{{\ttfamily 2308.15304}}].

\bibitem{tHooft:1987vrq}
G.~'t~Hooft, \emph{{Graviton Dominance in Ultrahigh-Energy Scattering}}, \href{https://doi.org/10.1016/0370-2693(87)90159-6}{\emph{Phys. Lett. B} {\bfseries 198} (1987) 61}.

\bibitem{Muzinich:1987in}
I.~J. Muzinich and M.~Soldate, \emph{{High-Energy Unitarity of Gravitation and Strings}}, \href{https://doi.org/10.1103/PhysRevD.37.359}{\emph{Phys. Rev. D} {\bfseries 37} (1988) 359}.

\bibitem{Verlinde:1991iu}
H.~L. Verlinde and E.~P. Verlinde, \emph{{Scattering at Planckian energies}}, \href{https://doi.org/10.1016/0550-3213(92)90236-5}{\emph{Nucl. Phys. B} {\bfseries 371} (1992) 246} [\href{https://arxiv.org/abs/hep-th/9110017}{{\ttfamily hep-th/9110017}}].

\bibitem{Kabat:1992tb}
D.~N. Kabat and M.~Ortiz, \emph{{Eikonal quantum gravity and Planckian scattering}}, \href{https://doi.org/10.1016/0550-3213(92)90627-N}{\emph{Nucl. Phys. B} {\bfseries 388} (1992) 570} [\href{https://arxiv.org/abs/hep-th/9203082}{{\ttfamily hep-th/9203082}}].

\bibitem{Amati:1992zb}
D.~Amati, M.~Ciafaloni and G.~Veneziano, \emph{{Planckian scattering beyond the semiclassical approximation}}, \href{https://doi.org/10.1016/0370-2693(92)91366-H}{\emph{Phys. Lett. B} {\bfseries 289} (1992) 87}.

\bibitem{Camanho:2014apa}
X.~O. Camanho, J.~D. Edelstein, J.~Maldacena and A.~Zhiboedov, \emph{{Causality Constraints on Corrections to the Graviton Three-Point Coupling}}, \href{https://doi.org/10.1007/JHEP02(2016)020}{\emph{JHEP} {\bfseries 02} (2016) 020} [\href{https://arxiv.org/abs/1407.5597}{{\ttfamily 1407.5597}}].

\bibitem{Amati:1987wq}
D.~Amati, M.~Ciafaloni and G.~Veneziano, \emph{{Superstring Collisions at Planckian Energies}}, \href{https://doi.org/10.1016/0370-2693(87)90346-7}{\emph{Phys. Lett. B} {\bfseries 197} (1987) 81}.

\bibitem{Amati:1987uf}
D.~Amati, M.~Ciafaloni and G.~Veneziano, \emph{{Classical and Quantum Gravity Effects from Planckian Energy Superstring Collisions}}, \href{https://doi.org/10.1142/S0217751X88000710}{\emph{Int. J. Mod. Phys. A} {\bfseries 3} (1988) 1615}.

\bibitem{Eardley:2002re}
D.~M. Eardley and S.~B. Giddings, \emph{{Classical black hole production in high-energy collisions}}, \href{https://doi.org/10.1103/PhysRevD.66.044011}{\emph{Phys. Rev. D} {\bfseries 66} (2002) 044011} [\href{https://arxiv.org/abs/gr-qc/0201034}{{\ttfamily gr-qc/0201034}}].

\bibitem{Cristofoli:2020hnk}
A.~Cristofoli, \emph{{Gravitational shock waves and scattering amplitudes}}, \href{https://doi.org/10.1007/JHEP11(2020)160}{\emph{JHEP} {\bfseries 11} (2020) 160} [\href{https://arxiv.org/abs/2006.08283}{{\ttfamily 2006.08283}}].

\bibitem{Adamo:2022qci}
T.~Adamo, A.~Cristofoli, A.~Ilderton and S.~Klisch, \emph{{All Order Gravitational Waveforms from Scattering Amplitudes}}, \href{https://doi.org/10.1103/PhysRevLett.131.011601}{\emph{Phys. Rev. Lett.} {\bfseries 131} (2023) 011601} [\href{https://arxiv.org/abs/2210.04696}{{\ttfamily 2210.04696}}].

\bibitem{Aoki:2026eos}
K.~Aoki and A.~Cristofoli, \emph{{Resumming Scattering Amplitudes for Waveforms}},  \href{https://arxiv.org/abs/2601.08252}{{\ttfamily 2601.08252}}.

\bibitem{Adamo:2022rob}
T.~Adamo, A.~Cristofoli and P.~Tourkine, \emph{{The ultrarelativistic limit of Kerr}}, \href{https://doi.org/10.1007/JHEP02(2023)107}{\emph{JHEP} {\bfseries 02} (2023) 107} [\href{https://arxiv.org/abs/2209.05730}{{\ttfamily 2209.05730}}].

\bibitem{Raj:2023iqn}
H.~Raj and R.~Venugopalan, \emph{{Gravitational wave double copy of radiation from gluon shockwave collisions}}, \href{https://doi.org/10.1016/j.physletb.2024.138669}{\emph{Phys. Lett. B} {\bfseries 853} (2024) 138669} [\href{https://arxiv.org/abs/2312.03507}{{\ttfamily 2312.03507}}].

\bibitem{Raj:2024xsi}
H.~Raj and R.~Venugopalan, \emph{{QCD-gravity double-copy in the Regge regime: Shock wave propagators}}, \href{https://doi.org/10.1103/PhysRevD.110.056010}{\emph{Phys. Rev. D} {\bfseries 110} (2024) 056010} [\href{https://arxiv.org/abs/2406.10483}{{\ttfamily 2406.10483}}].

\bibitem{Weinberg:1965nx}
S.~Weinberg, \emph{{Infrared photons and gravitons}}, \href{https://doi.org/10.1103/PhysRev.140.B516}{\emph{Phys. Rev.} {\bfseries 140} (1965) B516}.

\bibitem{Bjerrum-Bohr:2026fhs}
N.~E.~J. Bjerrum-Bohr, G.~Chen, C.~Jordan~Eriksen and N.~Shah, \emph{{The gravitational Compton amplitude at third post-Minkowskian order}},  \href{https://arxiv.org/abs/2602.06947}{{\ttfamily 2602.06947}}.

\bibitem{Bautista:2026qse}
Y.~F. Bautista, M.~Driesse, K.~Haddad and G.~U. Jakobsen, \emph{{Gravitational Wave Scattering in Spinless WQFT}},  \href{https://arxiv.org/abs/2602.06125}{{\ttfamily 2602.06125}}.

\bibitem{Ivanov:2026icp}
M.~M. Ivanov, Y.-Z. Li, J.~Parra-Martinez and Z.~Zhou, \emph{{Gravitational Raman Scattering: a Systematic Toolkit for Tidal Effects in General Relativity}},  \href{https://arxiv.org/abs/2602.06951}{{\ttfamily 2602.06951}}.

\bibitem{Ivanov:2024sds}
M.~M. Ivanov, Y.-Z. Li, J.~Parra-Martinez and Z.~Zhou, \emph{{Gravitational Raman Scattering in Effective Field Theory: A Scalar Tidal Matching at O(G3)}}, \href{https://doi.org/10.1103/PhysRevLett.132.131401}{\emph{Phys. Rev. Lett.} {\bfseries 132} (2024) 131401} [\href{https://arxiv.org/abs/2401.08752}{{\ttfamily 2401.08752}}].

\bibitem{Correia:2024jgr}
M.~Correia and G.~Isabella, \emph{{The Born regime of gravitational amplitudes}}, \href{https://doi.org/10.1007/JHEP03(2025)144}{\emph{JHEP} {\bfseries 03} (2025) 144} [\href{https://arxiv.org/abs/2406.13737}{{\ttfamily 2406.13737}}].

\bibitem{Caron-Huot:2025tlq}
S.~Caron-Huot, M.~Correia, G.~Isabella and M.~Solon, \emph{{Gravitational Wave Scattering via the Born Series: Scalar Tidal Matching to O(G7) and Beyond}}, \href{https://doi.org/10.1103/qd3c-nfz6}{\emph{Phys. Rev. Lett.} {\bfseries 135} (2025) 191601} [\href{https://arxiv.org/abs/2503.13593}{{\ttfamily 2503.13593}}].

\bibitem{Correia:2025enx}
M.~Correia, T.~Gopalka, G.~Isabella and A.~M. Wolz, \emph{{Analyticity of the Black Hole S-Matrix}},  \href{https://arxiv.org/abs/2511.11794}{{\ttfamily 2511.11794}}.

\bibitem{Christodoulou:1979eu}
D.~Christodoulou and B.~G. Schmidt, \emph{{Convergent and asymptotic iteration methods in General Relativity}}, {\emph{Commun.Math. Phys.} {\bfseries 68} (1979) 275}.

\bibitem{Damour:1990rm}
T.~Damour and B.~G. Schmidt, \emph{{Reliability of Perturbation Theory in General Relativity}}, \href{https://doi.org/10.1063/1.528850}{\emph{J. Math. Phys.} {\bfseries 31} (1990) 2441}.

\bibitem{Brink:1976uf}
L.~Brink, P.~Di~Vecchia and P.~S. Howe, \emph{{A Lagrangian Formulation of the Classical and Quantum Dynamics of Spinning Particles}}, \href{https://doi.org/10.1016/0550-3213(77)90364-9}{\emph{Nucl. Phys. B} {\bfseries 118} (1977) 76}.

\bibitem{Kalin:2022hph}
G.~K\"alin, J.~Neef and R.~A. Porto, \emph{{Radiation-reaction in the Effective Field Theory approach to Post-Minkowskian dynamics}}, \href{https://doi.org/10.1007/JHEP01(2023)140}{\emph{JHEP} {\bfseries 01} (2023) 140} [\href{https://arxiv.org/abs/2207.00580}{{\ttfamily 2207.00580}}].

\bibitem{Galley:2009px}
C.~R. Galley and M.~Tiglio, \emph{{Radiation reaction and gravitational waves in the effective field theory approach}}, \href{https://doi.org/10.1103/PhysRevD.79.124027}{\emph{Phys. Rev. D} {\bfseries 79} (2009) 124027} [\href{https://arxiv.org/abs/0903.1122}{{\ttfamily 0903.1122}}].

\bibitem{Jakobsen:2020ksu}
G.~U. Jakobsen, \emph{{Schwarzschild-Tangherlini Metric from Scattering Amplitudes}}, \href{https://doi.org/10.1103/PhysRevD.102.104065}{\emph{Phys. Rev. D} {\bfseries 102} (2020) 104065} [\href{https://arxiv.org/abs/2006.01734}{{\ttfamily 2006.01734}}].

\bibitem{Hoogeveen:2025tew}
J.~Hoogeveen, G.~U. Jakobsen and J.~Plefka, \emph{{Spinning the probe in Kerr with WQFT}}, \href{https://doi.org/10.1007/JHEP10(2025)201}{\emph{JHEP} {\bfseries 10} (2025) 201} [\href{https://arxiv.org/abs/2506.14626}{{\ttfamily 2506.14626}}].

\bibitem{Almeida:2024lbv}
G.~L. Almeida, A.~M{\"u}ller, S.~Foffa and R.~Sturani, \emph{{Gravitational memory contributions to waveform and effective action}}, \href{https://doi.org/10.1103/1ncd-r238}{\emph{Phys. Rev. D} {\bfseries 112} (2025) 084077} [\href{https://arxiv.org/abs/2410.10565}{{\ttfamily 2410.10565}}].

\bibitem{Mandelstam:1982cb}
S.~Mandelstam, \emph{{Light Cone Superspace and the Ultraviolet Finiteness of the N=4 Model}}, \href{https://doi.org/10.1016/0550-3213(83)90179-7}{\emph{Nucl. Phys. B} {\bfseries 213} (1983) 149}.

\bibitem{Leibbrandt:1983pj}
G.~Leibbrandt, \emph{{The Light Cone Gauge in Yang-Mills Theory}}, \href{https://doi.org/10.1103/PhysRevD.29.1699}{\emph{Phys. Rev. D} {\bfseries 29} (1984) 1699}.

\bibitem{Chiu:2011qc}
J.-y. Chiu, A.~Jain, D.~Neill and I.~Z. Rothstein, \emph{{The Rapidity Renormalization Group}}, \href{https://doi.org/10.1103/PhysRevLett.108.151601}{\emph{Phys. Rev. Lett.} {\bfseries 108} (2012) 151601} [\href{https://arxiv.org/abs/1104.0881}{{\ttfamily 1104.0881}}].

\bibitem{Mogull:2025cfn}
G.~Mogull, J.~Plefka and K.~Stoldt, \emph{{Radiated angular momentum from spinning black hole scattering trajectories}}, \href{https://doi.org/10.1103/my38-14k5}{\emph{Phys. Rev. D} {\bfseries 112} (2025) 124076} [\href{https://arxiv.org/abs/2506.20643}{{\ttfamily 2506.20643}}].

\bibitem{Balasin:1996mq}
H.~Balasin, \emph{{Geodesics for impulsive gravitational waves and the multiplication of distributions}}, \href{https://doi.org/10.1088/0264-9381/14/2/018}{\emph{Class. Quant. Grav.} {\bfseries 14} (1997) 455} [\href{https://arxiv.org/abs/gr-qc/9607076}{{\ttfamily gr-qc/9607076}}].

\bibitem{Shtabovenko:2023idz}
V.~Shtabovenko, R.~Mertig and F.~Orellana, \emph{{FeynCalc 10: Do multiloop integrals dream of computer codes?}}, \href{https://doi.org/10.1016/j.cpc.2024.109357}{\emph{Comput. Phys. Commun.} {\bfseries 306} (2025) 109357} [\href{https://arxiv.org/abs/2312.14089}{{\ttfamily 2312.14089}}].

\bibitem{Beneke:1997zp}
M.~Beneke and V.~A. Smirnov, \emph{{Asymptotic expansion of Feynman integrals near threshold}}, \href{https://doi.org/10.1016/S0550-3213(98)00138-2}{\emph{Nucl. Phys. B} {\bfseries 522} (1998) 321} [\href{https://arxiv.org/abs/hep-ph/9711391}{{\ttfamily hep-ph/9711391}}].

\bibitem{Smirnov:2012gma}
V.~A. Smirnov, \emph{{Analytic tools for Feynman integrals}}, vol.~250. 2012, \href{https://doi.org/10.1007/978-3-642-34886-0}{10.1007/978-3-642-34886-0}.

\bibitem{Becher:2014oda}
T.~Becher, A.~Broggio and A.~Ferroglia, \emph{{Introduction to Soft-Collinear Effective Theory}}, vol.~896. Springer, 2015, \href{https://doi.org/10.1007/978-3-319-14848-9}{10.1007/978-3-319-14848-9}, [\href{https://arxiv.org/abs/1410.1892}{{\ttfamily 1410.1892}}].

\bibitem{Kawai:1985xq}
H.~Kawai, D.~C. Lewellen and S.~H.~H. Tye, \emph{{A Relation Between Tree Amplitudes of Closed and Open Strings}}, \href{https://doi.org/10.1016/0550-3213(86)90362-7}{\emph{Nucl. Phys. B} {\bfseries 269} (1986) 1}.

\bibitem{Bern:2008qj}
Z.~Bern, J.~J.~M. Carrasco and H.~Johansson, \emph{{New Relations for Gauge-Theory Amplitudes}}, \href{https://doi.org/10.1103/PhysRevD.78.085011}{\emph{Phys. Rev. D} {\bfseries 78} (2008) 085011} [\href{https://arxiv.org/abs/0805.3993}{{\ttfamily 0805.3993}}].

\bibitem{Bern:2010ue}
Z.~Bern, J.~J.~M. Carrasco and H.~Johansson, \emph{{Perturbative Quantum Gravity as a Double Copy of Gauge Theory}}, \href{https://doi.org/10.1103/PhysRevLett.105.061602}{\emph{Phys. Rev. Lett.} {\bfseries 105} (2010) 061602} [\href{https://arxiv.org/abs/1004.0476}{{\ttfamily 1004.0476}}].

\bibitem{Dolan:2007ut}
S.~R. Dolan, \emph{{Scattering of long-wavelength gravitational waves}}, \href{https://doi.org/10.1103/PhysRevD.77.044004}{\emph{Phys. Rev. D} {\bfseries 77} (2008) 044004} [\href{https://arxiv.org/abs/0710.4252}{{\ttfamily 0710.4252}}].

\bibitem{Gordon:1928}
W.~Gordon, \emph{{{\"U}ber den Sto{\ss} zweier Punktladungen nach der Wellenmechanik}}, \href{https://doi.org/10.1007/BF01351302}{\emph{Z. Physik} {\bfseries 48} (1928) 180}.

\bibitem{Mott:1930}
N.~F. Mott, \emph{{The Collision between Two Electrons}}, \href{https://doi.org/10.1098/rspa.1930.0006}{\emph{Proceedings of the Royal Society of London Series A} {\bfseries 126} (1930) 259}.

\bibitem{Mizera:2023tfe}
S.~Mizera, \emph{{Physics of the analytic S-matrix}}, \href{https://doi.org/10.1016/j.physrep.2023.10.006}{\emph{Phys. Rept.} {\bfseries 1047} (2024) 1} [\href{https://arxiv.org/abs/2306.05395}{{\ttfamily 2306.05395}}].

\bibitem{Magnus:1954zz}
W.~Magnus, \emph{{On the exponential solution of differential equations for a linear operator}}, \href{https://doi.org/10.1002/cpa.3160070404}{\emph{Commun. Pure Appl. Math.} {\bfseries 7} (1954) 649}.

\bibitem{Blanes:2008xlr}
S.~Blanes, F.~Casas, J.~A. Oteo and J.~Ros, \emph{{The Magnus expansion and some of its applications}}, \href{https://doi.org/10.1016/j.physrep.2008.11.001}{\emph{Phys. Rept.} {\bfseries 470} (2009) 151}.

\bibitem{Gonzo:2024zxo}
R.~Gonzo and C.~Shi, \emph{{Scattering and Bound Observables for Spinning Particles in Kerr Spacetime with Generic Spin Orientations}}, \href{https://doi.org/10.1103/PhysRevLett.133.221401}{\emph{Phys. Rev. Lett.} {\bfseries 133} (2024) 221401} [\href{https://arxiv.org/abs/2405.09687}{{\ttfamily 2405.09687}}].

\bibitem{Kim:2025hpn}
J.-W. Kim, \emph{{Radiation eikonal for post-Minkowskian observables}}, \href{https://doi.org/10.1103/PhysRevD.111.L121702}{\emph{Phys. Rev. D} {\bfseries 111} (2025) L121702} [\href{https://arxiv.org/abs/2501.07372}{{\ttfamily 2501.07372}}].

\bibitem{Alessio:2025flu}
F.~Alessio, R.~Gonzo and C.~Shi, \emph{{Dirac brackets for classical radiative observables}}, \href{https://doi.org/10.1103/ykgq-jqd5}{\emph{Phys. Rev. D} {\bfseries 112} (2025) 104060} [\href{https://arxiv.org/abs/2506.03249}{{\ttfamily 2506.03249}}].

\bibitem{Kim:2025olv}
S.~Kim, H.~Lee and S.~Lee, \emph{{Classical eikonal in relativistic scattering}}, \href{https://doi.org/10.1007/JHEP11(2025)032}{\emph{JHEP} {\bfseries 11} (2025) 032} [\href{https://arxiv.org/abs/2509.01922}{{\ttfamily 2509.01922}}].

\bibitem{Haddad:2025cmw}
K.~Haddad, G.~U. Jakobsen, G.~Mogull and J.~Plefka, \emph{{Unitarity and the On-Shell Action of Worldline Quantum Field Theory}}, \href{https://doi.org/10.1007/JHEP02(2026)008}{\emph{JHEP} {\bfseries 02} (2026) 008} [\href{https://arxiv.org/abs/2510.00988}{{\ttfamily 2510.00988}}].

\end{thebibliography}\endgroup

\end{document}